\documentclass[twocolumn]{aastex61}
\usepackage{enumitem}
\usepackage{amsmath}
\usepackage{tablefootnote}
\usepackage{hyperref}
\usepackage{cleveref}
\usepackage{natbib}

\newcommand{\Ni}{$^{56}$Ni}
\newcommand{\Co}{$^{56}$Co}
\newcommand{\Ti}{$^{44}$Ti}
\newcommand{\Ms}{M$_{\odot}$}
\newcommand{\Ls}{L$_{\odot}$}

\newcommand{\til}{$\sim$}

\newcommand{\mic}{$\mu$m}

\received{2018 February 25}
\revised{2018 March 30}
\accepted{2018 April 18}

\begin{document}

\title{DELAYED SHOCK-INDUCED DUST FORMATION IN THE DENSE CIRCUMSTELLAR SHELL SURROUNDING THE TYPE IIn SUPERNOVA SN~2010JL}
\shorttitle{Dust formation in  SN 2010jl}
\author{Arkaprabha Sarangi}
\affiliation{Observational Cosmology Lab, NASA Goddard Space Flight Center, Mail Code 665, Greenbelt, MD 20771, USA}
\affiliation{CRESST II/CUA/GSFC}
\email{arkaprabha.sarangi@nasa.gov}

\author{Eli Dwek}
\affiliation{Observational Cosmology Lab, NASA Goddard Space Flight Center, Mail Code 665, Greenbelt, MD 20771, USA}
\email{eli.dwek@nasa.gov}

\author{Richard G. Arendt}
\affiliation{Observational Cosmology Lab, NASA Goddard Space Flight Center, Mail Code 665, Greenbelt, MD 20771, USA}
\affiliation{CRESST II/UMBC/GSFC}

\begin{abstract}

The light curves of Type IIn supernovae are dominated by the radiative energy released through the interaction of the supernova shockwaves with their dense circumstellar medium (CSM).  The ultraluminous Type IIn supernova SN~2010jl exhibits an infrared emission component that is in excess of the extrapolated UV-optical spectrum as early as a few weeks post-explosion. This emission has been attributed by some as evidence for rapid formation of dust in the cooling postshock CSM. We investigate the physical processes that may inhibit or facilitate the formation of dust in the CSM. When only radiative cooling is considered, the temperature of the dense shocked gas rapidly drops below the dust condensation temperature. 
However, by accounting for the heating of the postshock gas by the downstream radiation from the shock, we show that dust formation is inhibited until the radiation from the shock weakens, as the shock propagates into the less dense outer regions of the CSM. In SN~2010jl dust formation can therefore only commence after day $\sim$ 380. Only the IR emission since that epoch can be attributed to the newly formed CSM dust.
Observations on day 460 and later show that the IR luminosity exceeds the UV-optical luminosity. The post-shock dust cannot extinct the radiation emitted by the expanding SN shock. Therefore, its IR emission must be powered by an interior source, which we identify with the reverse shock propagating through the SN ejecta. IR emission before day ~380 must therefore be an IR echo from preexisting CSM dust.

\end{abstract}

\section{Introduction}

Core-collapse supernovae (CCSNe) are the final fate of a massive stars larger than 8 \Ms\ at main-sequence. They play significant roles in the chemical enrichment of the host galaxy through the synthesis of metals and dust. Prior to the explosion, the progenitor stars undergo phases of mass loss in the form of winds. In most cases, the mass loss rates are estimated to vary between 10$^{-7}$ to 10$^{-4}$ \Ms yr$^{-1}$ depending on the metallicity and the mass \citep{mey15}. However in case of some progenitors, the mass loss rate can be $>$ 10$^{-2}$ \Ms\ \citep{kie12,mor14} leading to the formation of a dense circumstellar medium (CSM) surrounding the pre-explosion star \citep{chu94}. In spite of being only a small fraction ($<$ 10 \%) of the all observed CCSNe, unique features of Type IIn supernovae help us understand important aspects of the pre- and post-explosion phases in massive stars \citep{chu03,fas01,fox11,wha13}.

The transfer of radiative energy thorough the surrounding medium enables us to understand the nature of the progenitor, the explosion, and the properties of the CSM \citep{chu04,des10}. Depending on the shape of the light curve and spectral type, CCSNe has been categorized into several subclasses \citep{fil97}. Supernovae Type IIn, introduced by \cite{sch90}, are characterized by the presence of narrow (\til\ 100 km s$^{-1}$), intermediate (1-4 $\times$ 10$^3$ km s$^{-1}$) and broad (10-15 $\times$ 10$^3$ km s$^{-1}$) velocity width components. The narrow component originates from the slow moving CSM before being traversed by the supernova shock, the source of intermediate width component is the post-shock region of CSM, whereas the broad component arises from the fast expanding SN ejecta \citep{smi08}.  

Of the Type IIn supernova that have been observed over the last decade \citep{tad13}, SN~2010jl in UGC5189  \citep{sto11} has been extensively probed at X-ray wavelength with the the Chandra, NuSTAR, and Swift satellites \citep{ofe14b,cha15}, and at UV, optical, and near-IR wavelengths with the \textit{Hubble Space Telescope} (\textit{HST}), VLT/X, and Subaro observatories, and the Spitzer satellite \citep{mae13,fra14,gal14}. The observations have provided important information on the  progenitor mass, the composition and morphology of its circumstellar environment \citep{and11a, wil15, fox13, fox17}.

The luminosity of the UV-optical (UVO) light curve of SN~2010jl is characterized by a slow $t^{-0.4}$ decline from an initially observed value of $L$ $\sim$ 4 $\times$ 10$^9$ \Ls\ on day 20 to a value of $\sim$ 2 $\times$ 10$^9$ \Ls\ on day 300. It thereafter decreases at a rapid rate to a value of 5 $\times$ 10$^7$ \Ls\ by day 900. On the other hand, the infrared (IR) light curve exhibits a slow increase from 4 $\times$ 10$^8$ to 8 $\times$ 10$^8$ \Ls\ betwen days 20 to 300, followed by a more rapid increase to $\sim$ 1.5 $\times$ 10$^9$ \Ls\ by day $\sim$ 500 \citep{gal14, fra14, jen16}. 

A decline in UVO luminosity combined with a rise in IR indicates the presence of dust in the environment which absorbs and reprocesses the UVO photons to IR. The presence of dust is also supported by the gradual blue-shifting of hydrogen and oxygen intermediate velocity emission lines \citep{gal14, smi12}. Moreover, the gradual increase in blue-red asymmetry over time suggests that the receding part of the ejecta is increasingly being blocked by the presence of new dust which is lying interior to the line-emitting shocked gas \citep{smi08}.

The origin of IR emission in Type IIn SN can be attributed to (a) an echo from the pre-existing dust \citep{dwe83,and11a,bod80} (b) emission from the newly formed dust in the post-shock cool dense shell \citep{poz04, chu09, smi08, gal14} (c) emission from dust formed in the SN ejecta \citep{kot05,sza13,and11b}, or a combination of these. 

The pre-existing dust, which has survived the shock breakout, resides outside of the vaporization radius of the explosion \citep{dra79}. The luminosity of the shock-CSM interaction-front is generally much lower than the instantaneous flash of energy at the time of outburst \citep{dwe08,sod08}. Therefore, the temperature of the surviving dust grains emitting in the form of IR-echo should be much smaller than the vaporization temperature of the grains. However, analysis of the IR excess indicates dust temperatures to be close to its vaporization temperatures at early epochs \citep{fra14}. 

 Given the high dust temperature, and the asymmetric extinction of red and blue wings of the emission lines, \cite{gal14} rules out IR echo as a possible scenario in this case. \cite{and11a} argues in favor of IR-echo. However, the shock breakout luminosity assumed in the study is much lower compared to the one estimated by theoretical models \citep{bli00} or IR observations \citep{dwe08}. 
 

The CCSN ejecta is a well known site for dust synthesis. However, due to the presence of radioactive elements (such as $^{56}$Ni, $^{56}$Co, $^{44}$Ti), $\gamma$-rays and energetic Compton electrons \citep{che09, sar13}, the earliest epoch of dust synthesis in the ejecta is not before day 250. Therefore, the dense shell formed in the post-shock CSM has been considered as the most potent source of dust causing the near-IR excess at day 60 onwards \citep{gal14}. Nevertheless, dust formation in such environments is controlled by a complex chain of processes, such as the cooling rate of the shocked gas, radiative heating by the forward shock, the reverse/reflected shock traversing inwards to the ejecta, and the radioactive processes of the inner ejecta expanding into the CSM.

The detection of near-IR excess as early as day 60, like in the case of SN~2010jl poses new challenges to our understanding of dust formation scenario in any circumstellar environment. The dynamics of such high-density post-shock gas, controlled by strong ionizing radiation and rapid cooling, have never been studied before. We formulate the model in order to addresses the following issues: 
\begin{itemize}[noitemsep]
\item[--] What is the earliest epoch of new dust formation in the CSM? Can the IR echo from the pre-existing dust contribute to the light curve?
  \item[--] What are the physical and chemical processes that facilitate or impede the synthesis of dust in post-shock gas? 
\item[--] What are the primary heating sources of the dust, present at various region of the stellar system, that give rise to the IR emission? 
\end{itemize}


The paper is arranged in the following order. In Section \ref{optnir} we study the X-ray, UVO and the IR signatures of SN~2010jl light curve and their manifestation on the physical model. Following that, in Section \ref{morphology} we present a schematic diagram of a typical type IIn SN in light of its pre- and post-explosion morphology. Thereafter,  Additionally, we also focus on the probable dust compositions that can be derived from the observations. Following that, we address the scenarios for all the possible regions where pre-existing dust can survive or new dust can form in Section \ref{origin}. Dust formation in the post-shock gas being the main focus of this paper, Section \ref{constraints} deals with the constraints imposed on the pre-explosion CSM and the post-explosion shock dynamics of SN~2010jl. In Section \ref{postshock} we discuss the evolution of the post-shock gas leading to the conditions favorable to form dust. Following that in Section \ref{dustcondtions} we briefly summarize the physical and chemical processes and aid the formation and growth of dust grains in such environments. Lastly, in Section \ref{heating} we describe the energetics of the environment and the role played by the newly formed dust grain in explaining the IR emission in SN~2010jl. We summarize our results on SN~2010jl in Section \ref{summary} and its global implications on any Type IIn SN.



\begin{figure*}
\centering
\includegraphics[width=2.2in]{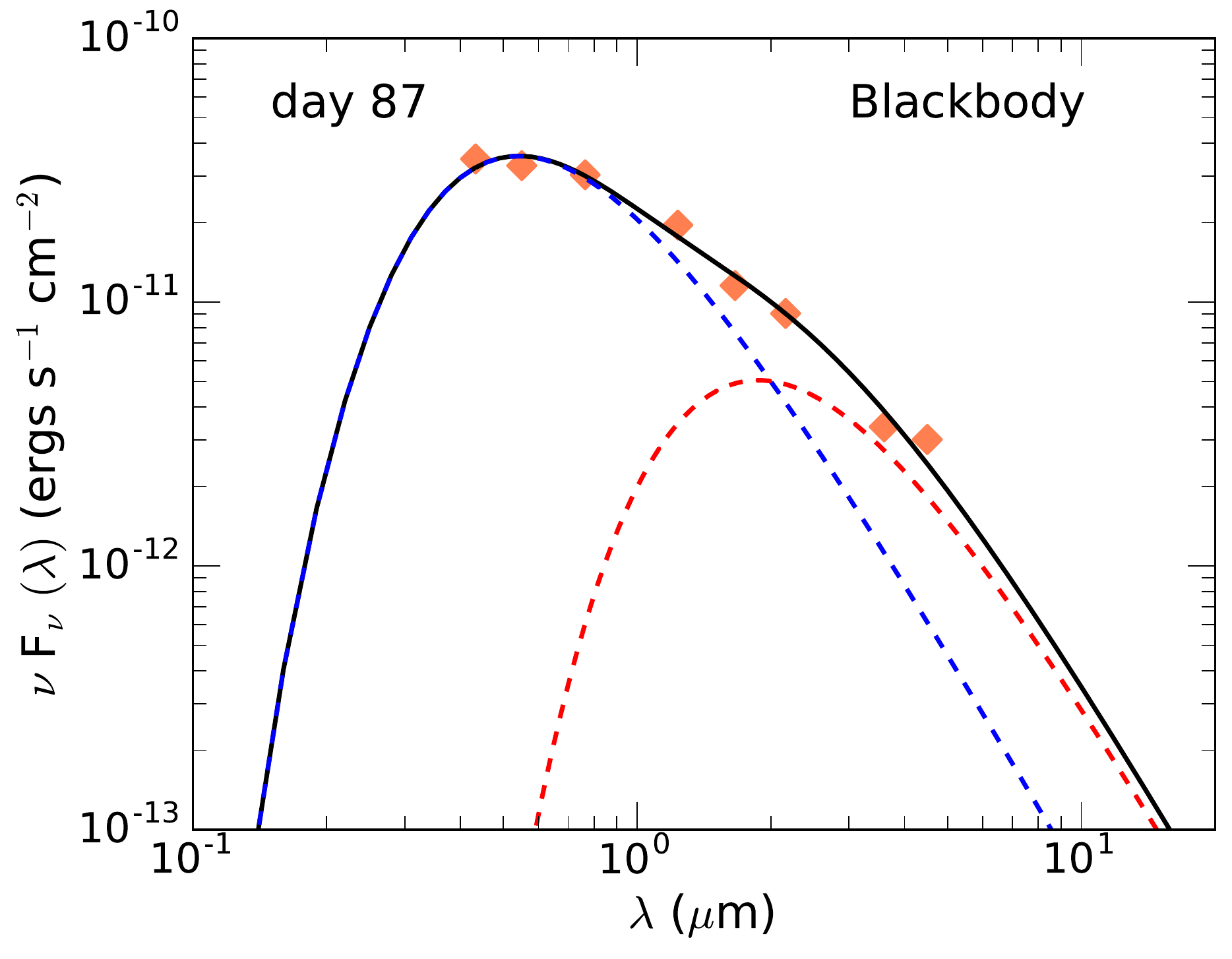}
\includegraphics[width=2.2in]{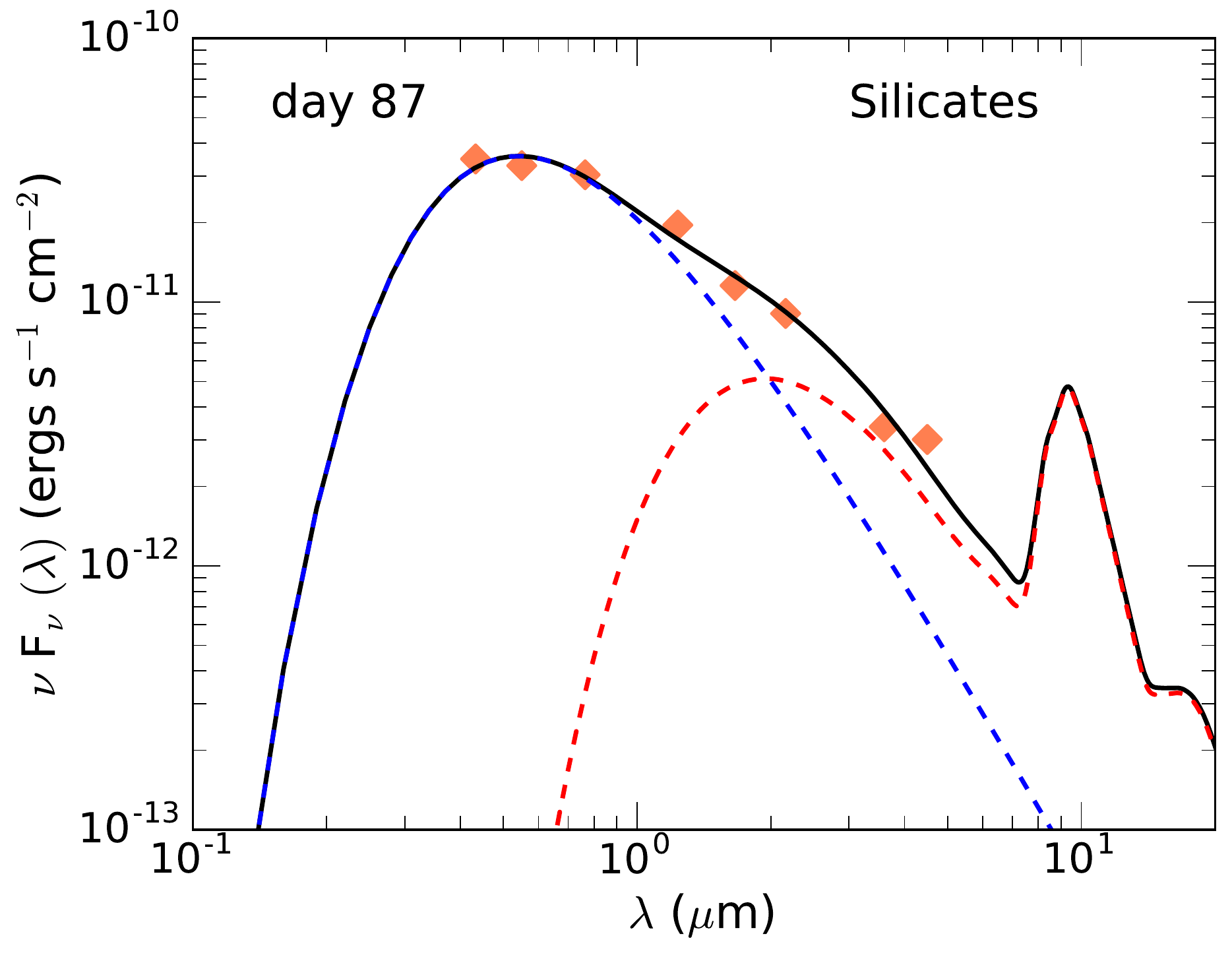}
\includegraphics[width=2.2in]{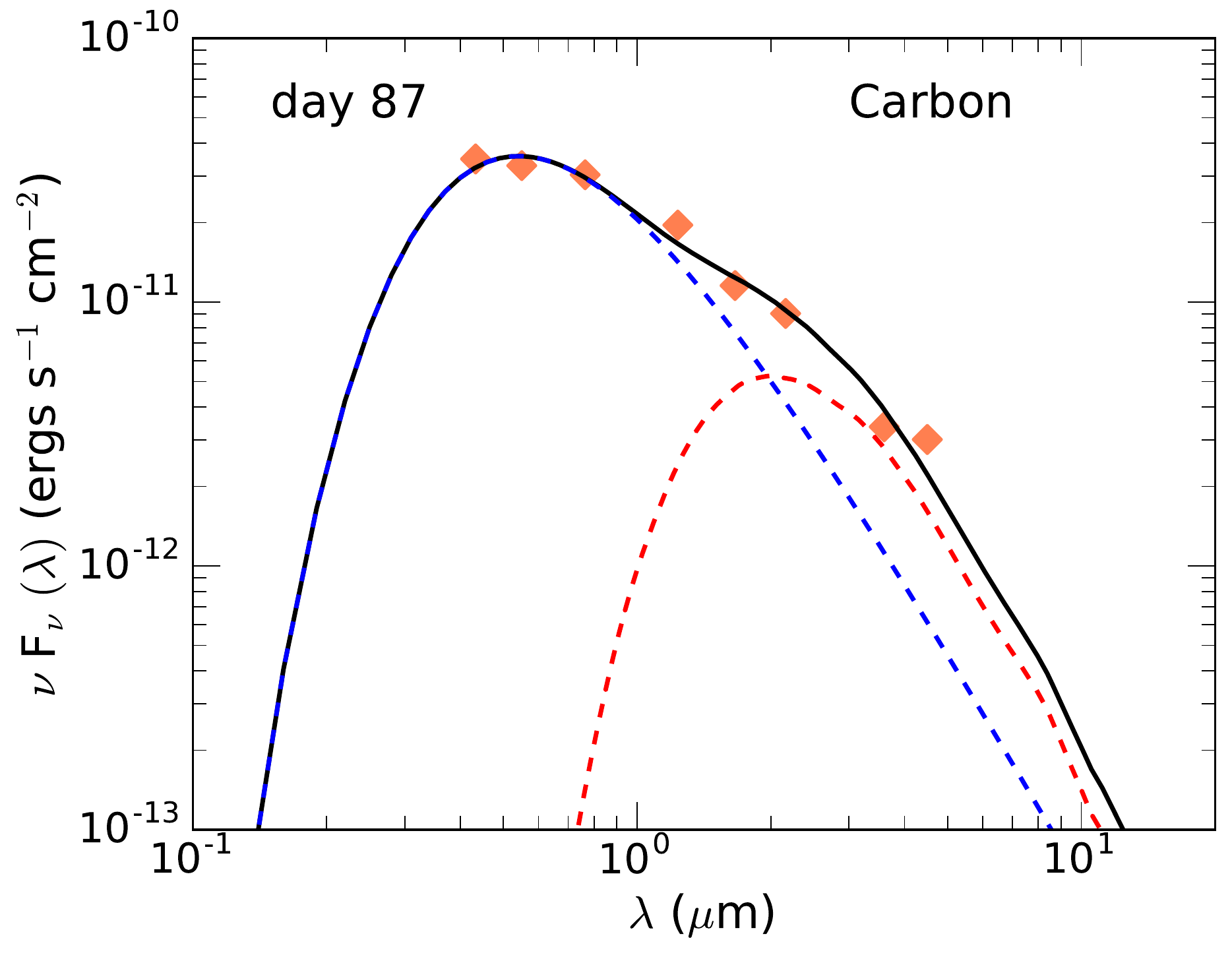}
\includegraphics[width=2.2in]{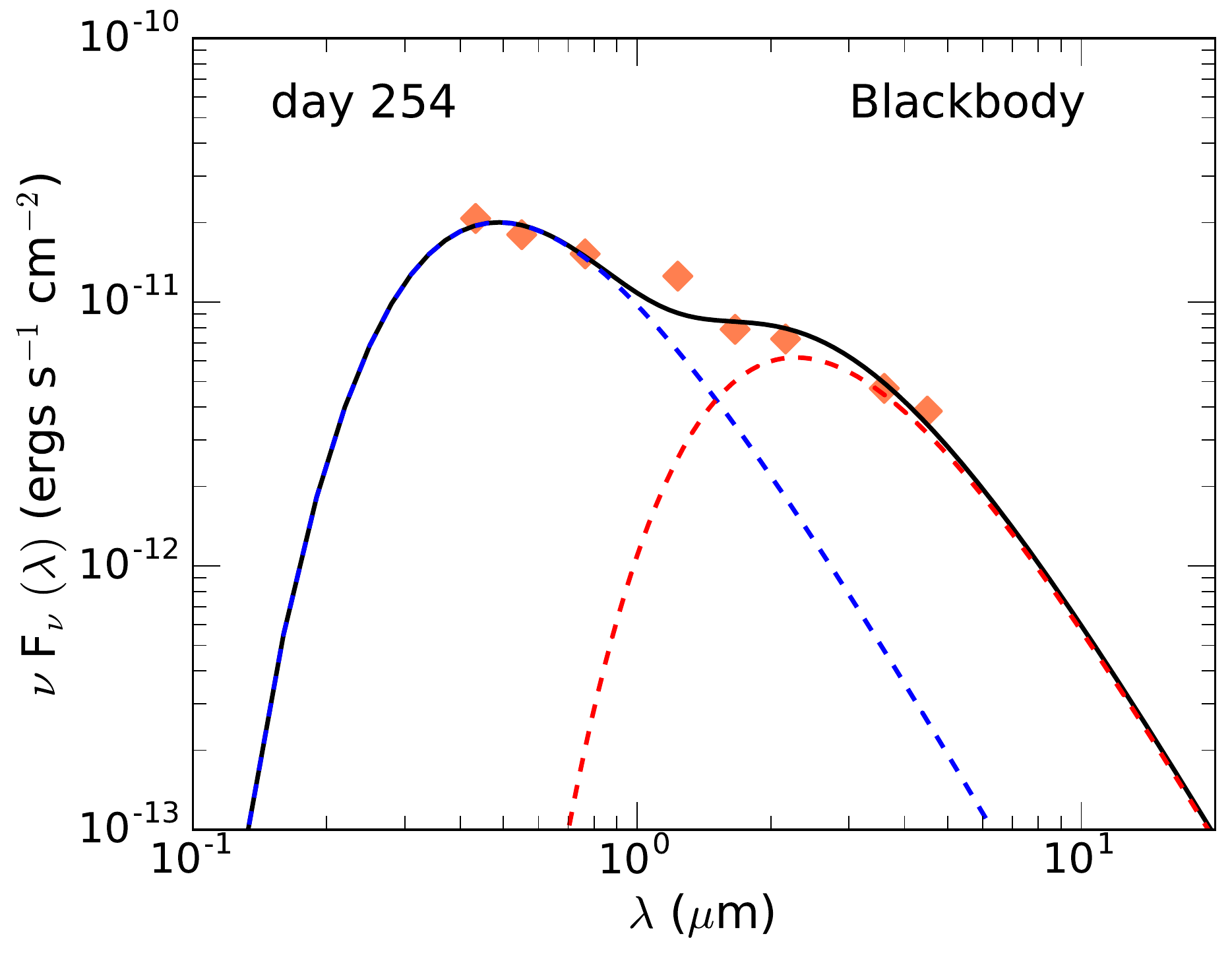}
\includegraphics[width=2.2in]{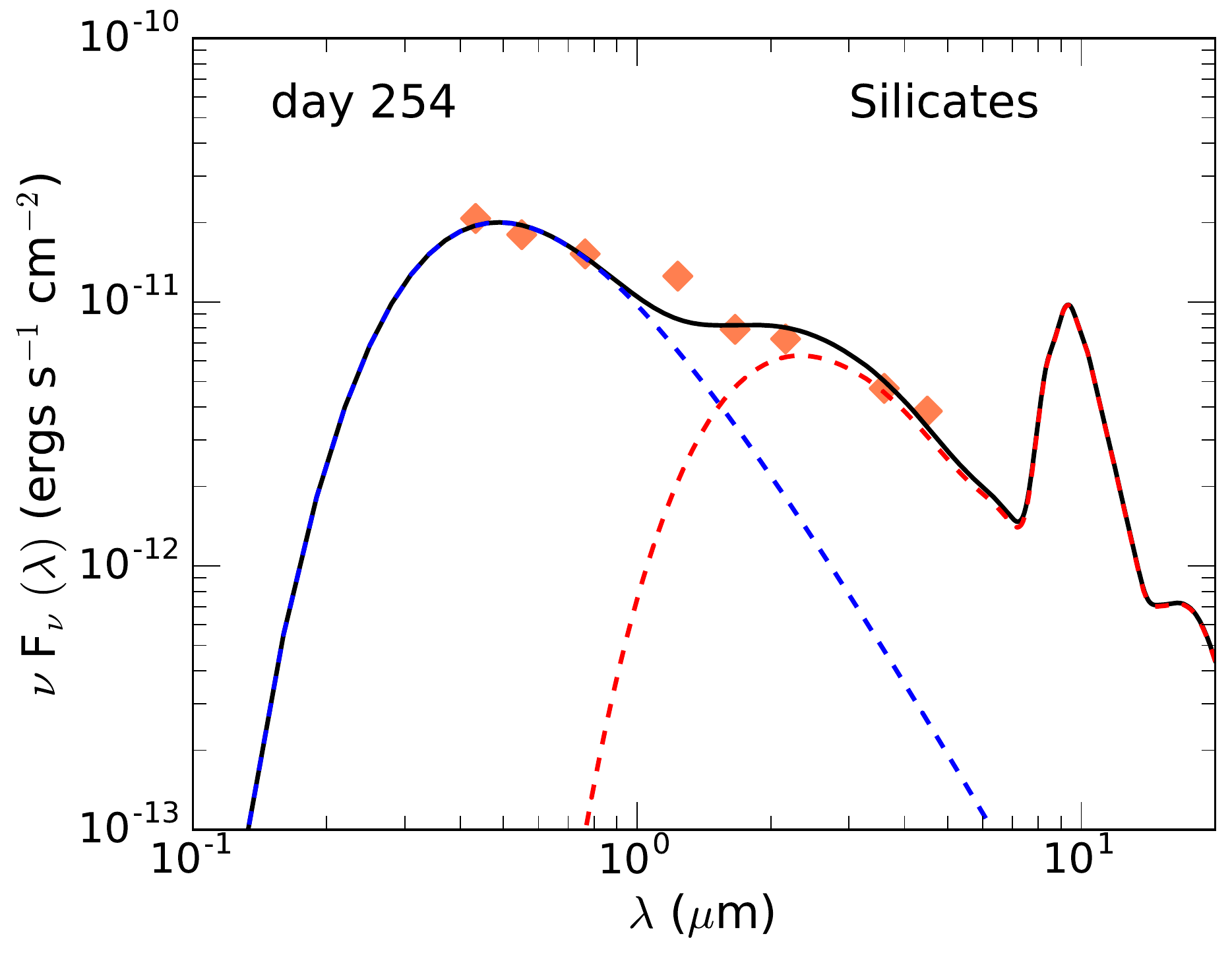}
\includegraphics[width=2.2in]{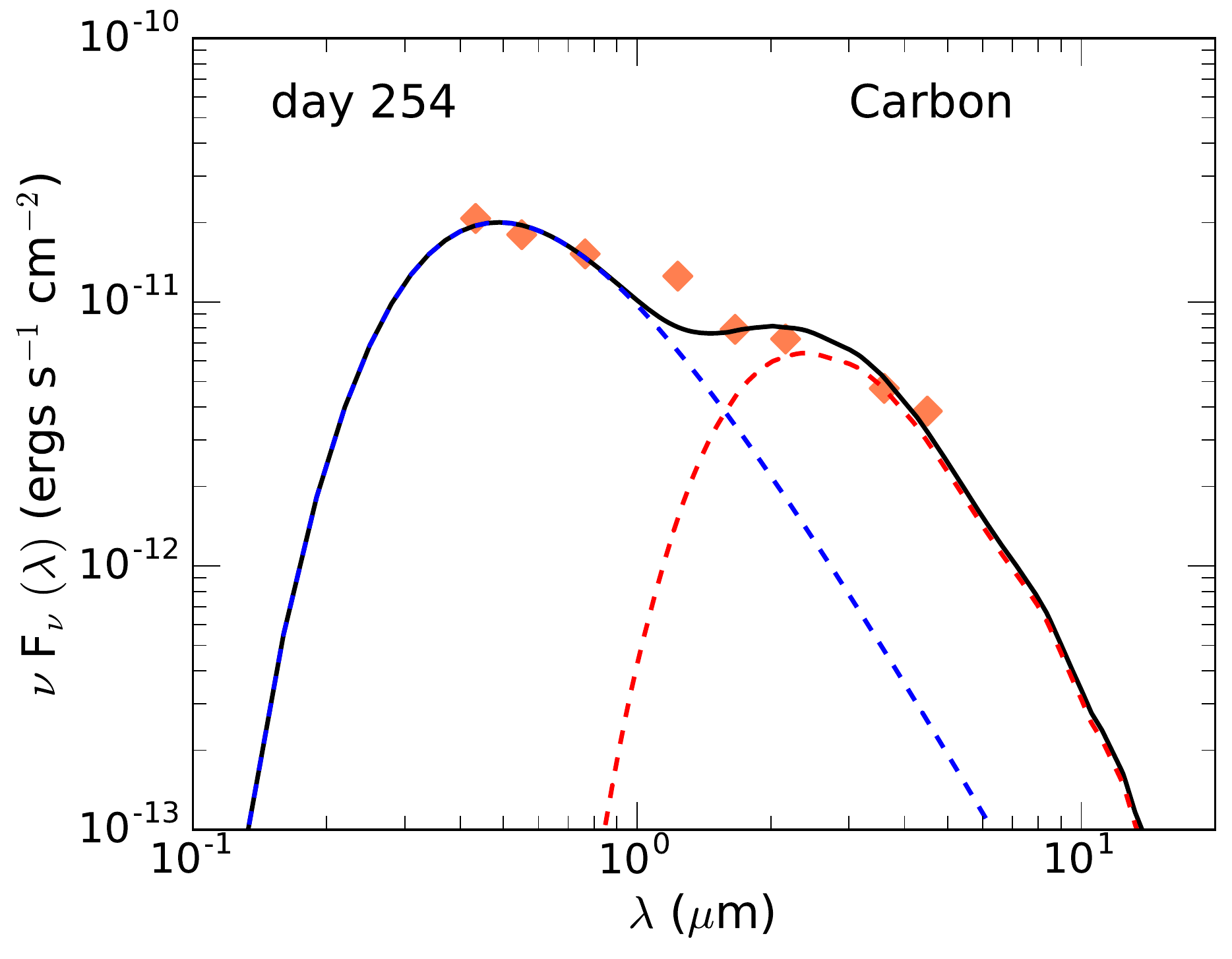}
\includegraphics[width=2.2in]{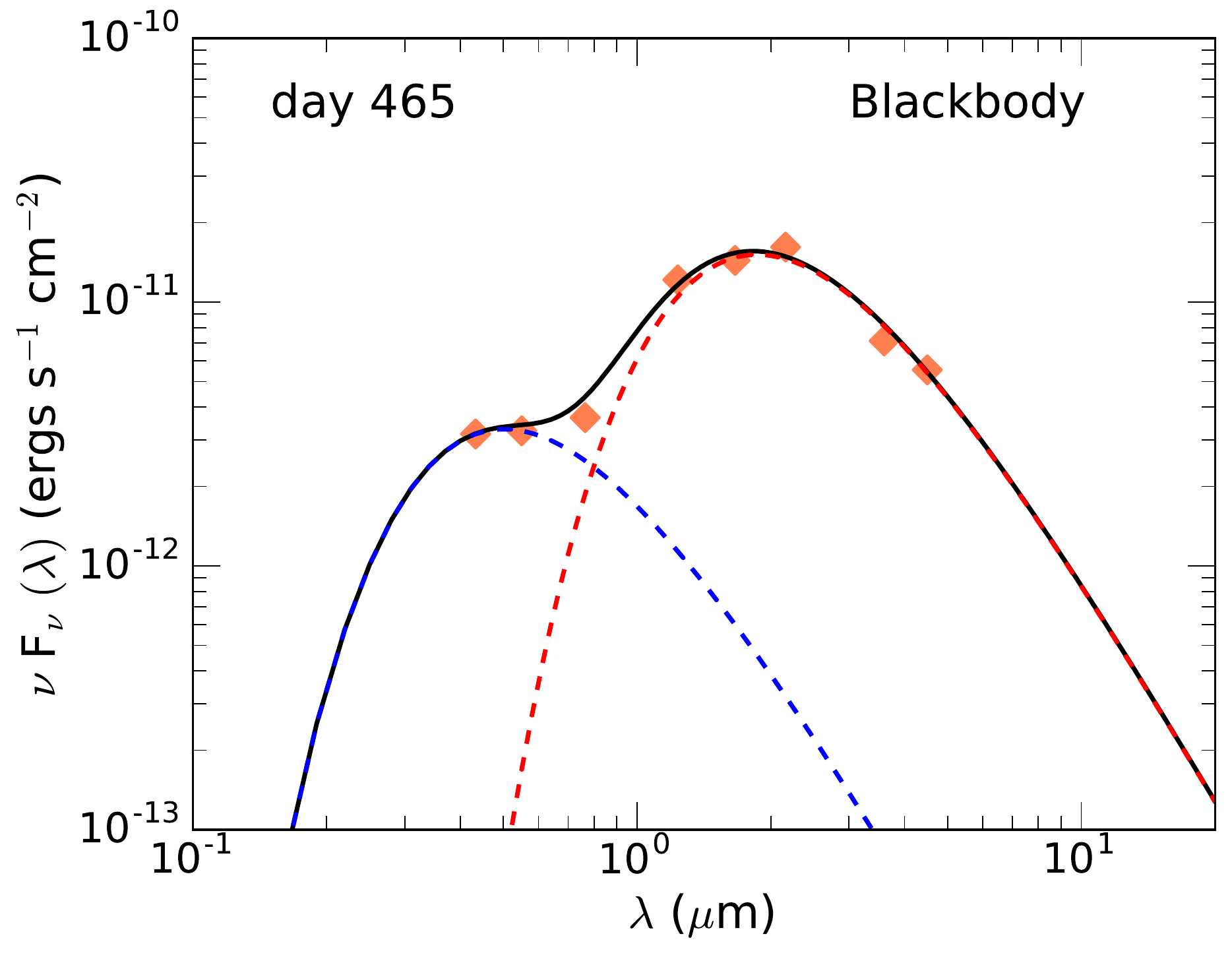}
\includegraphics[width=2.2in]{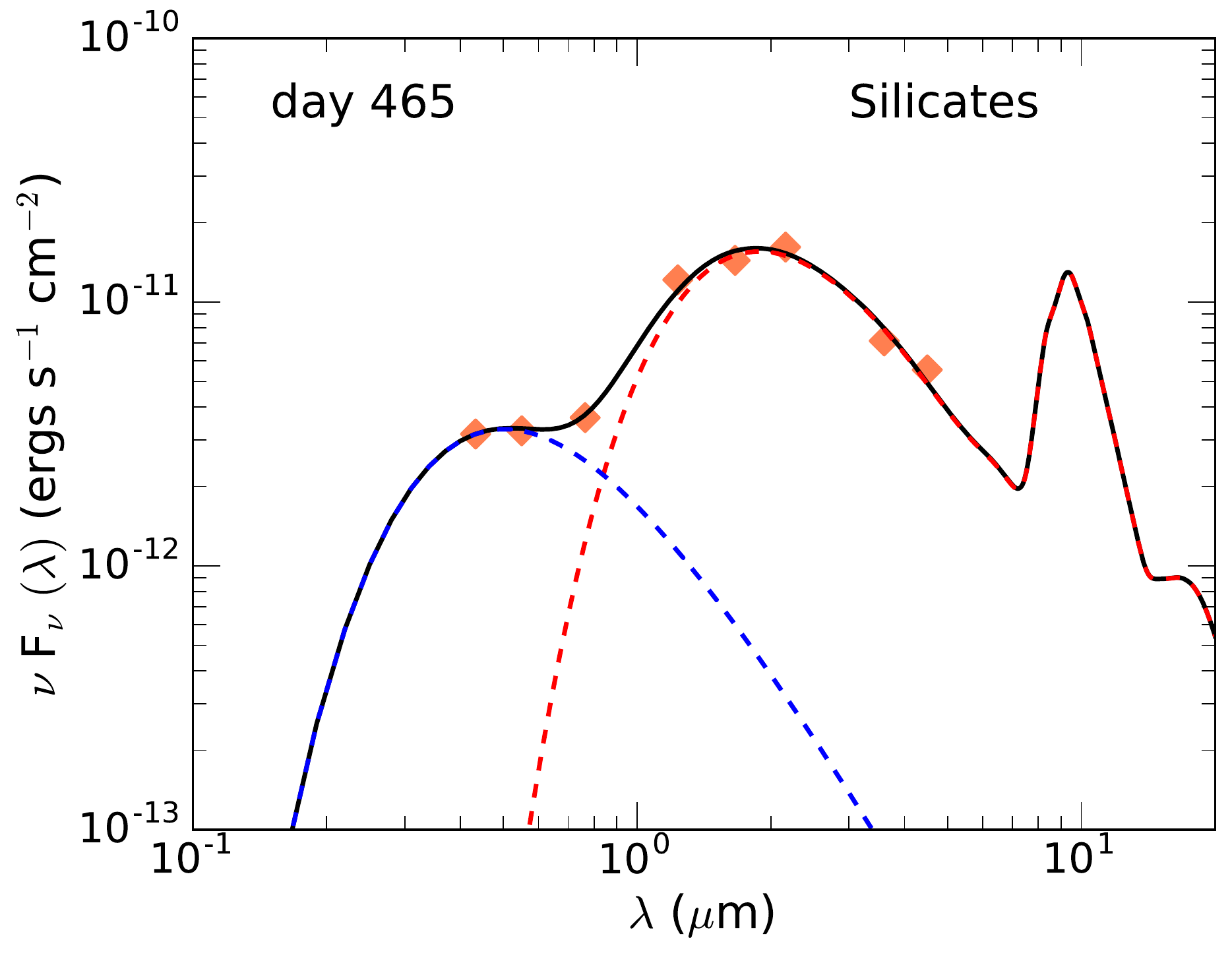}
\includegraphics[width=2.2in]{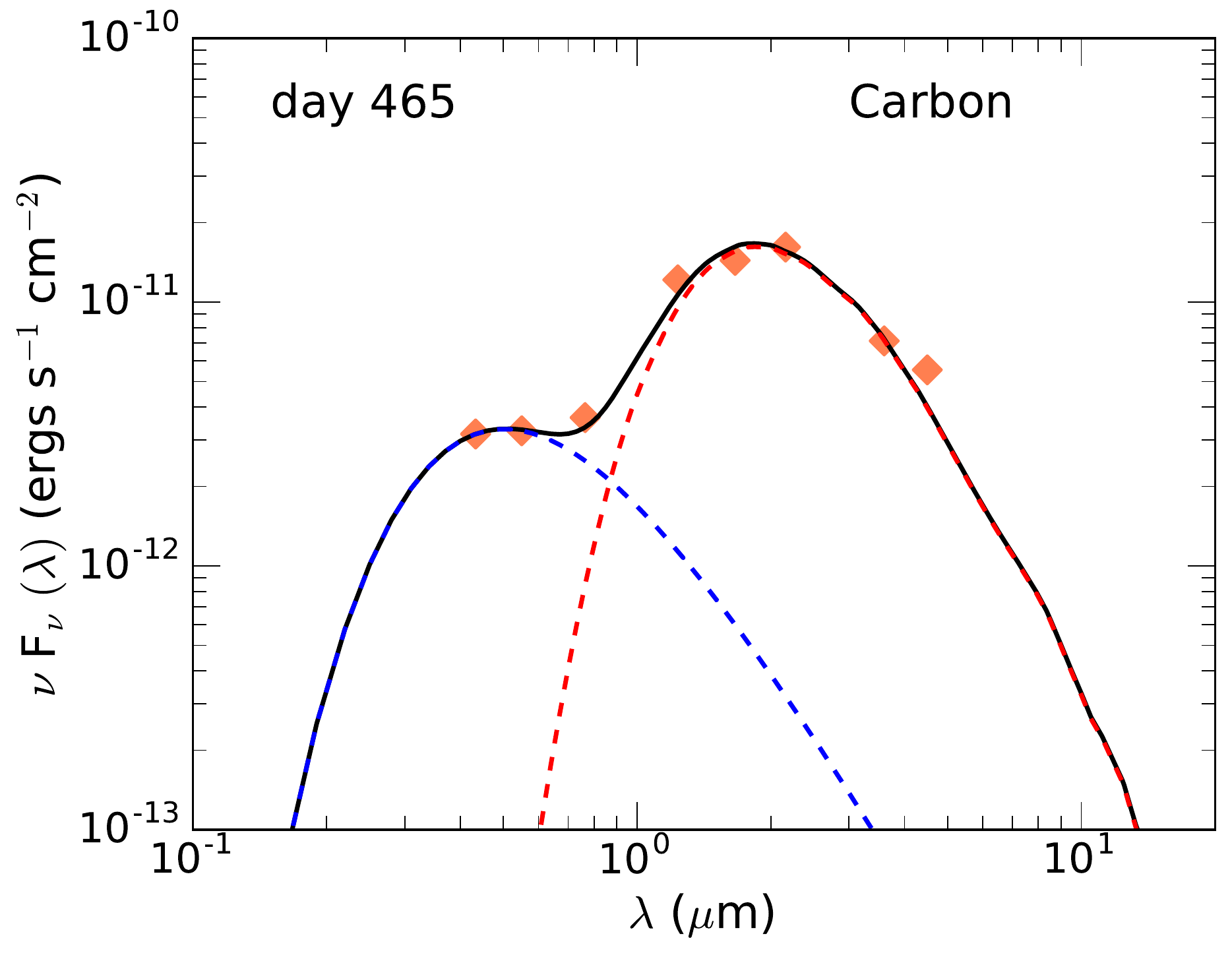}
\includegraphics[width=2.2in]{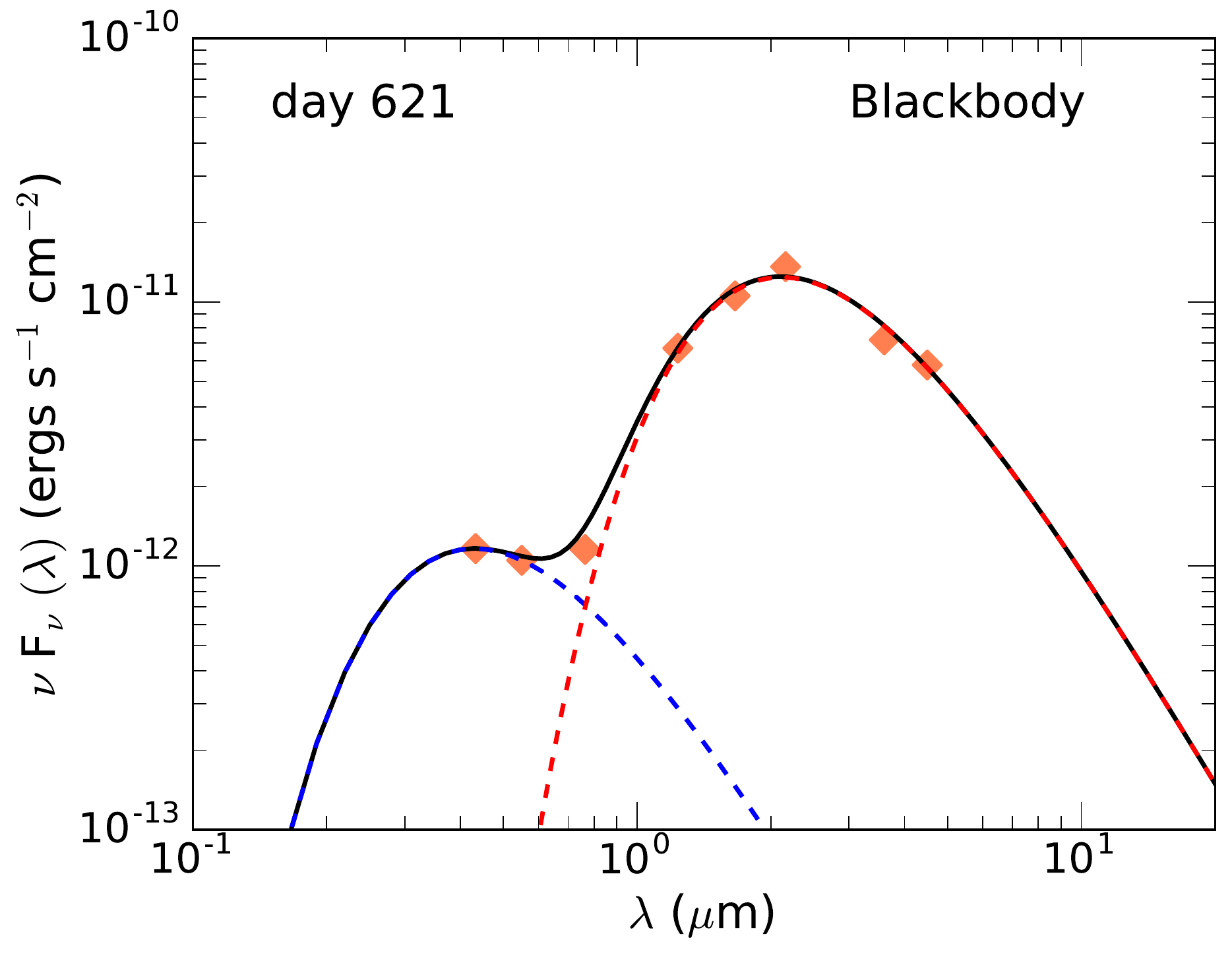}
\includegraphics[width=2.2in]{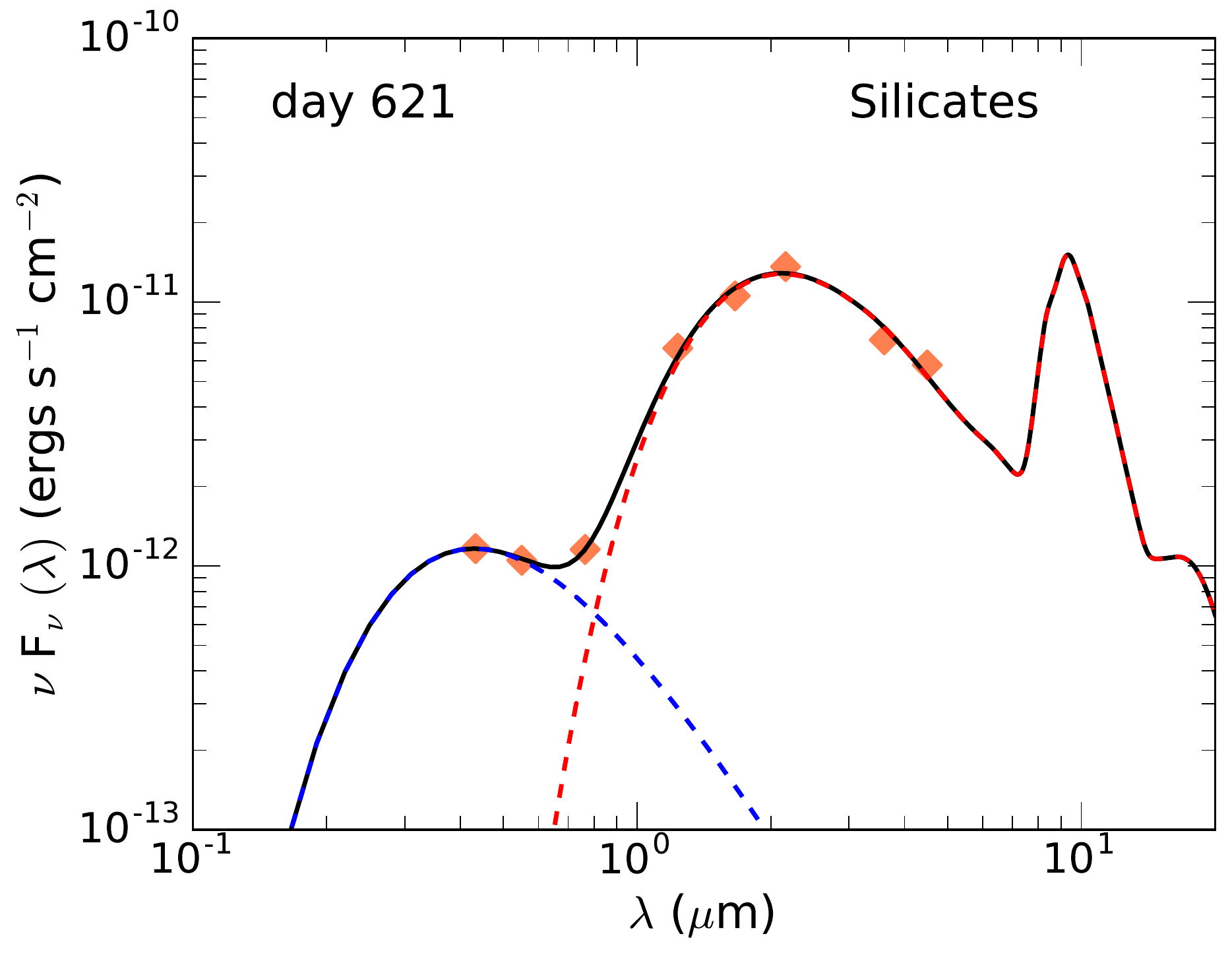}
\includegraphics[width=2.2in]{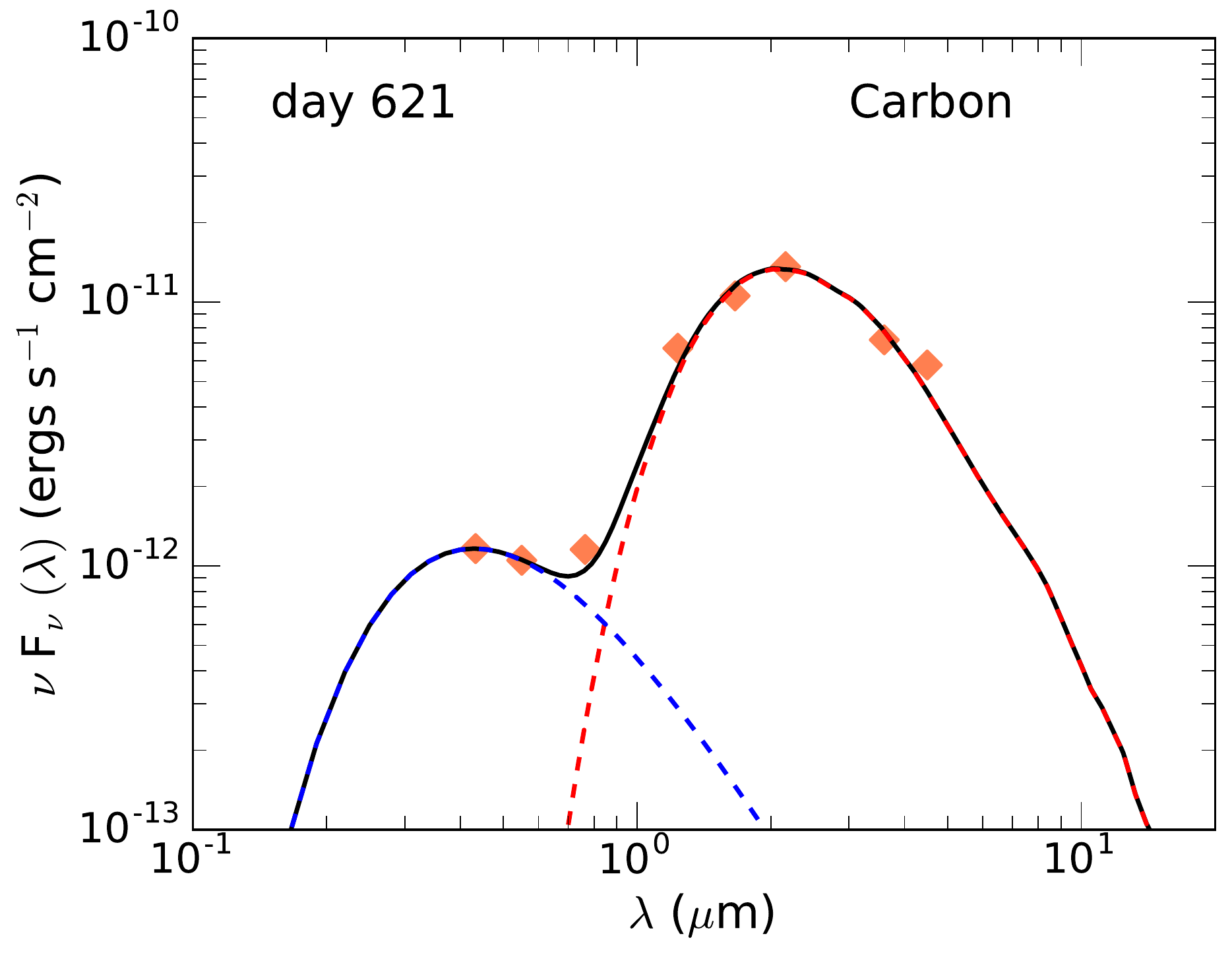}
\includegraphics[width=2.2in]{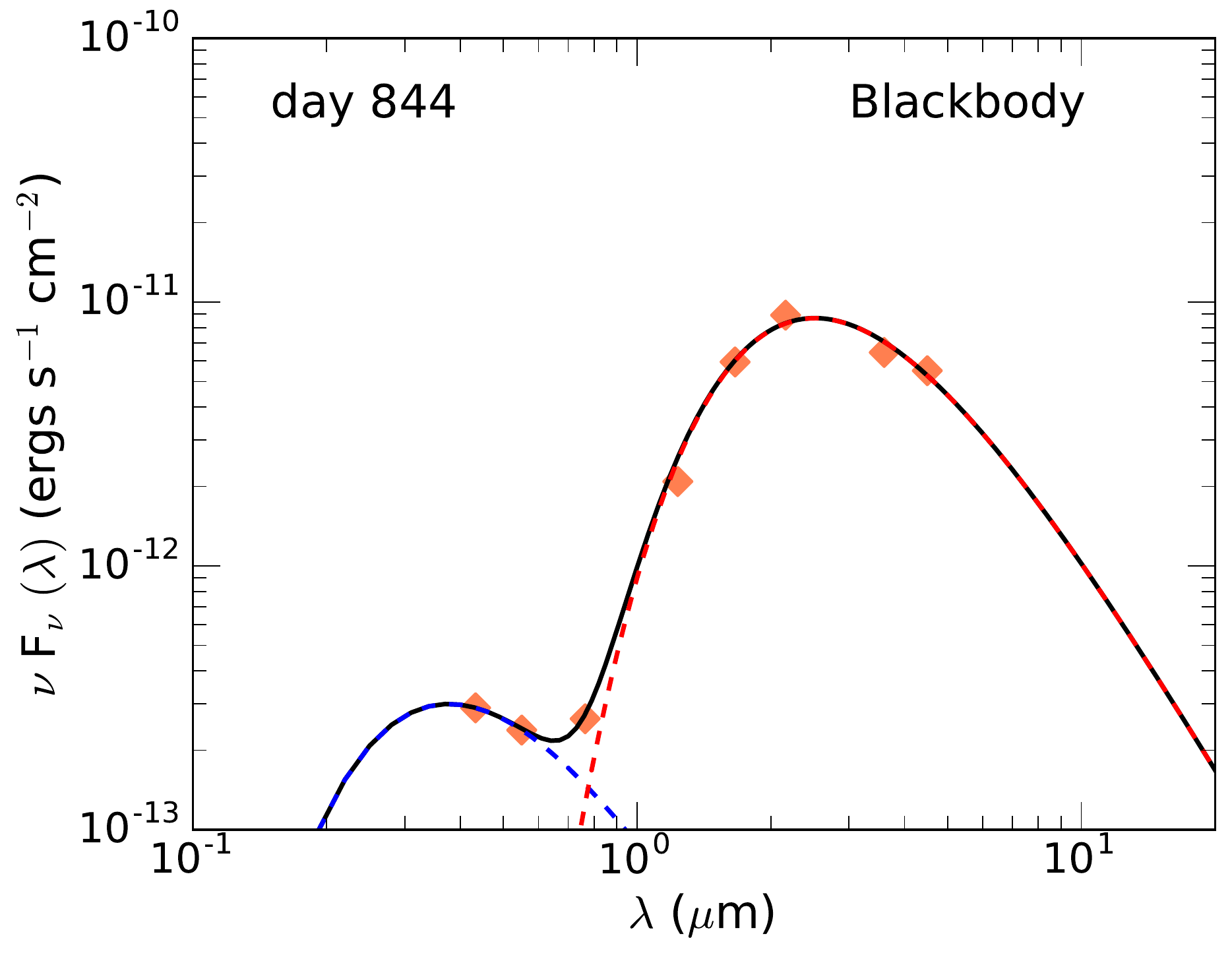}
\includegraphics[width=2.2in]{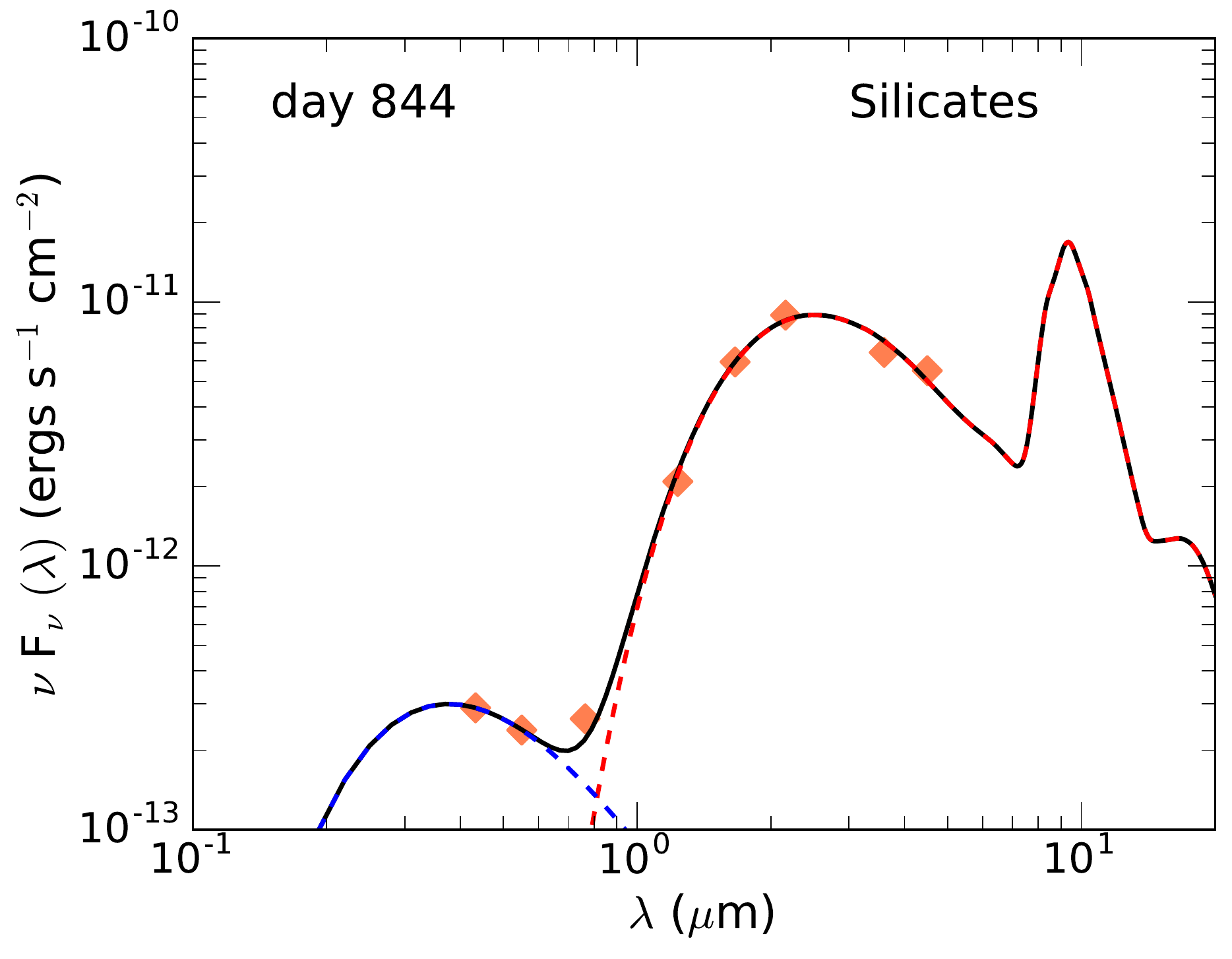}
\includegraphics[width=2.2in]{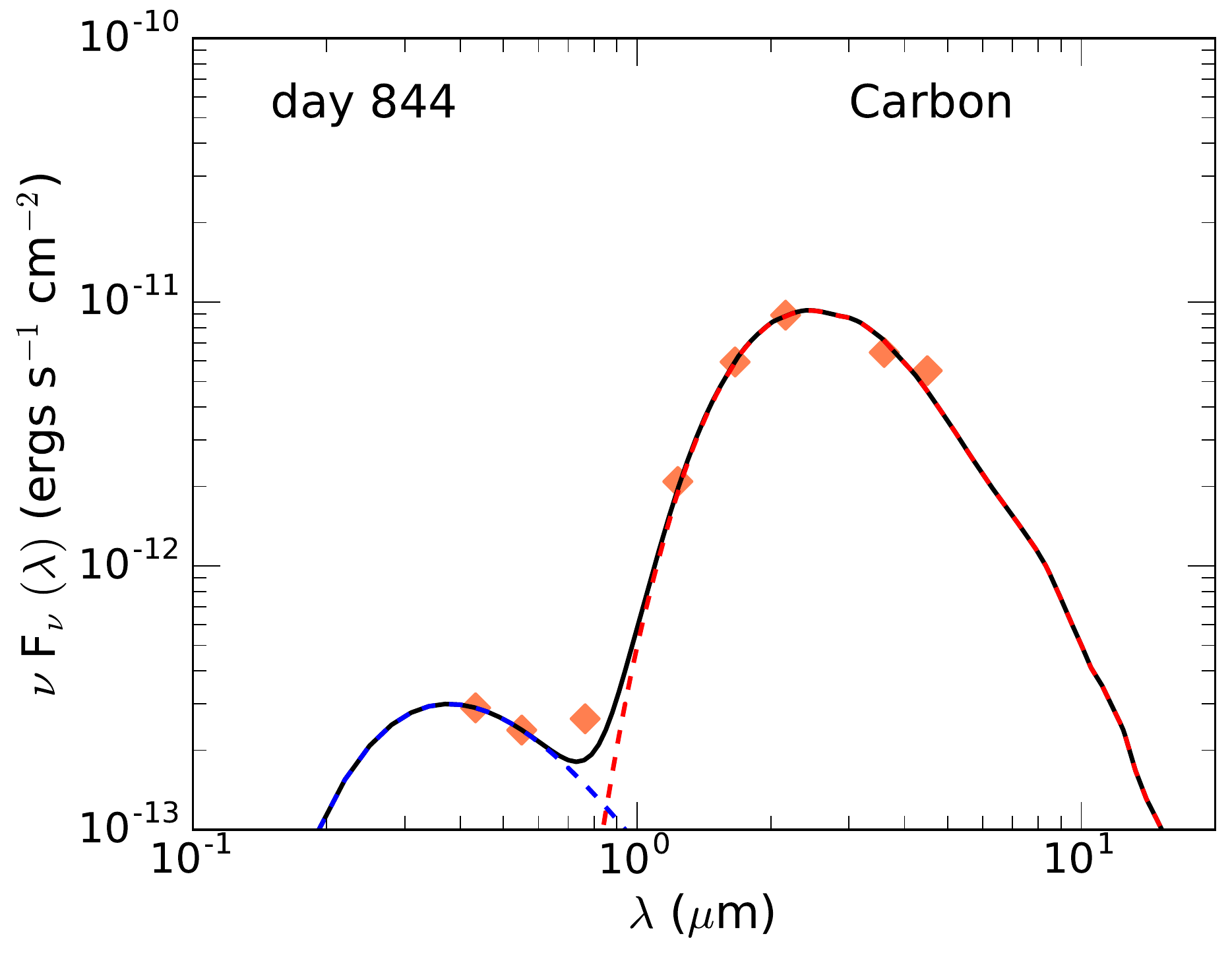}
\caption{\label{fig:fitting} \footnotesize{Fluxes in the optical ($B$, $V$, $i^{\prime}$) and the NIR ($J$, $H$, $K_s$) bands of SN~2010jl along with those in the \textit{Spitzer} 3.6 \& 4.5 \mic\ bands at day 87, 254, 465, 621 and 844  \citep{fra14,and11,fox13} are shown in the figure. The fluxes are fitted using a two-component fit with a blackbody fitting the UVO component (in blue) and either of (a) a pure blackbody (b) astronomical silicate \citep{wei01} or (c) amorphous carbon dust \citep{zub04} fitting the IR component (in red). The error-bars on the data points are too small to be visible on the figures. The best-fit scenarios obtained from the study are presented in Table~\ref{dust_epoch}. }} 
\end{figure*}

  
\begin{figure*}
\centering
\includegraphics[width=3.3in]{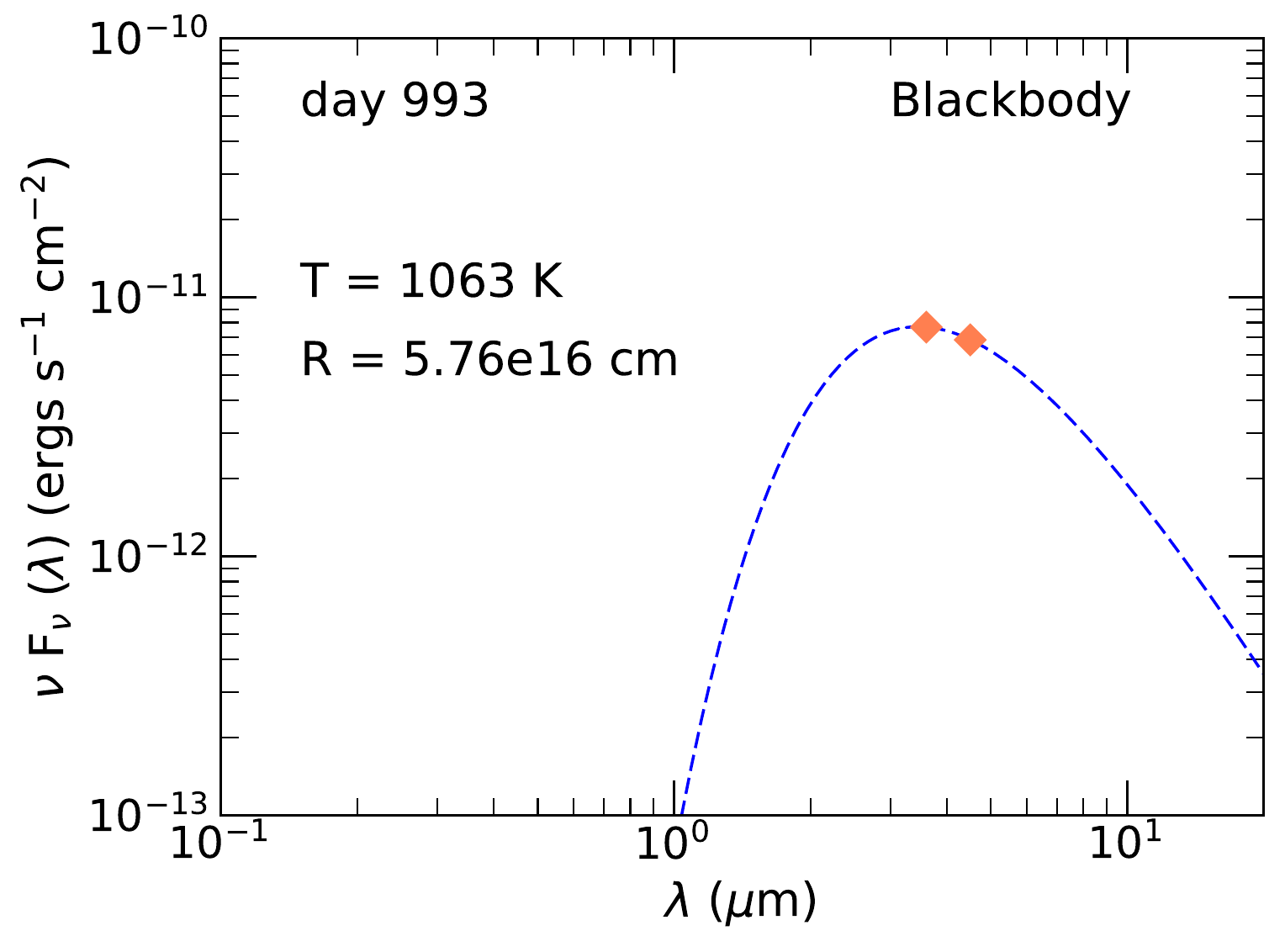}
\includegraphics[width=3.3in]{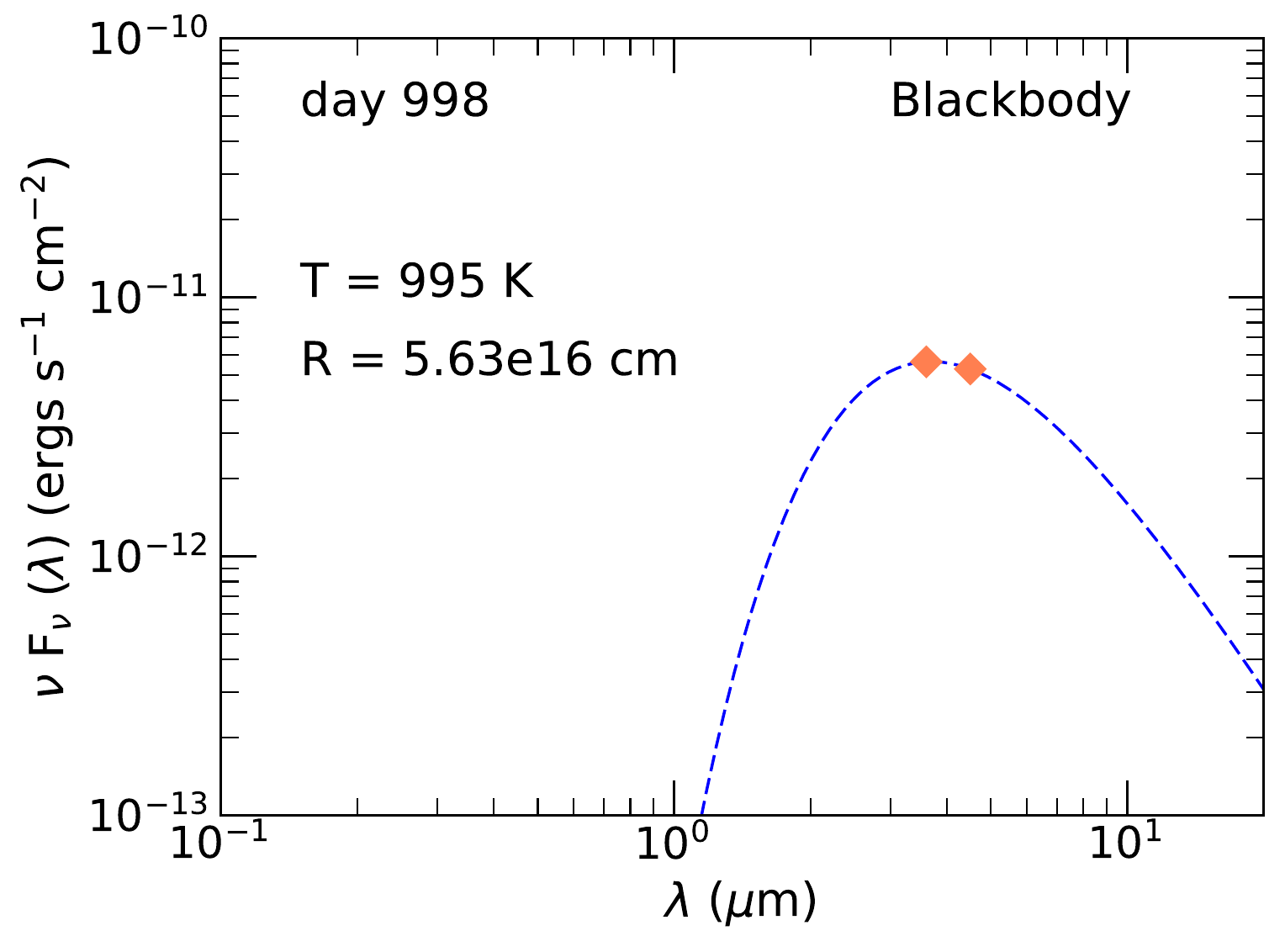}
\includegraphics[width=3.3in]{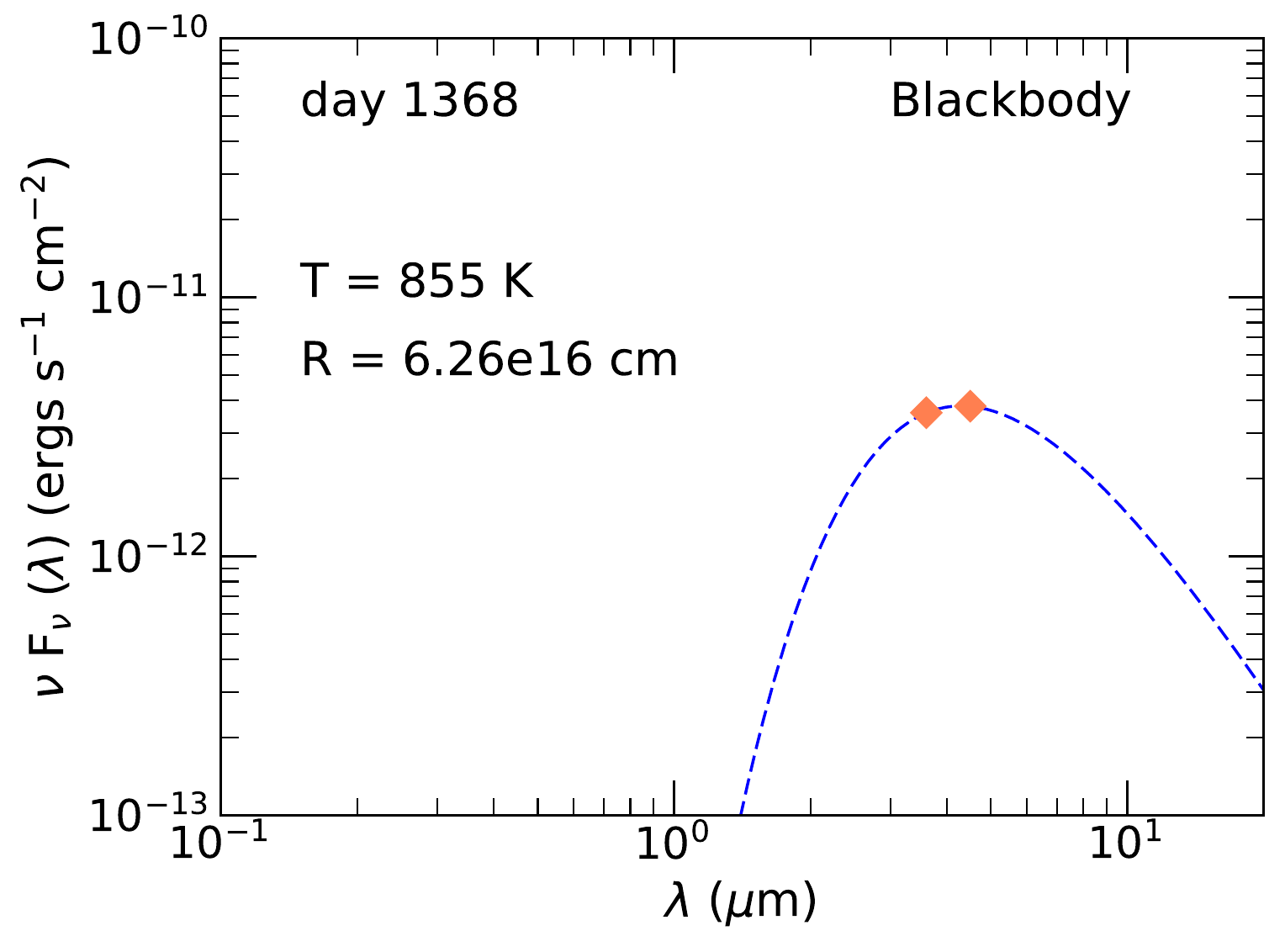}
\includegraphics[width=3.3in]{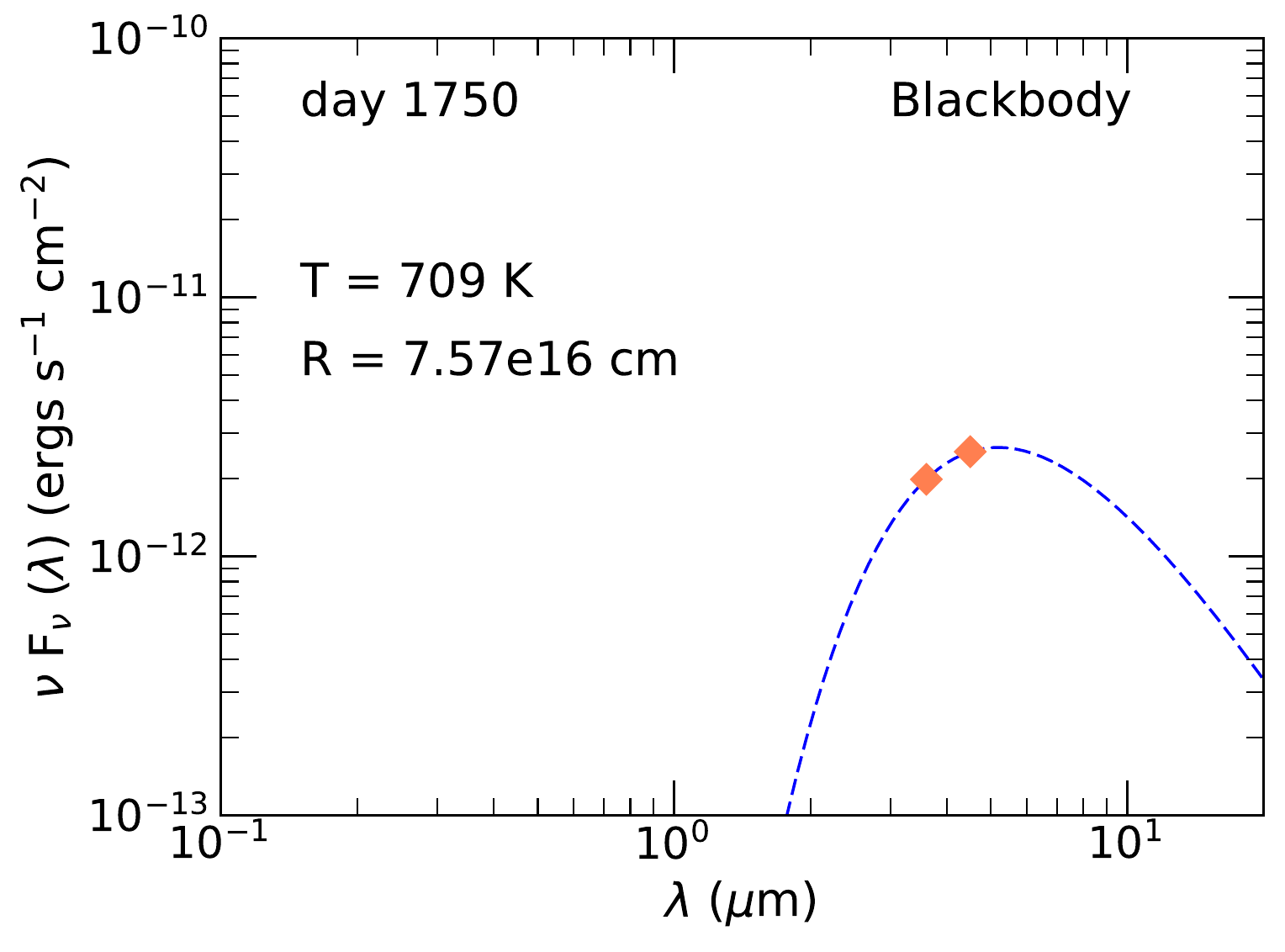}
\caption{\label{fig:lateIR} \footnotesize{A pure blackbody is used to fit the \textit{Spitzer} 3.6 and 4.5 \mic\ data at late times (\textit{Spitzer IRSA}) in order to derive the IR black body radius and the range of dust temperatures at late times. }} 
\end{figure*}
  

\begin{figure*}
\centering
\includegraphics[width=3.3in]{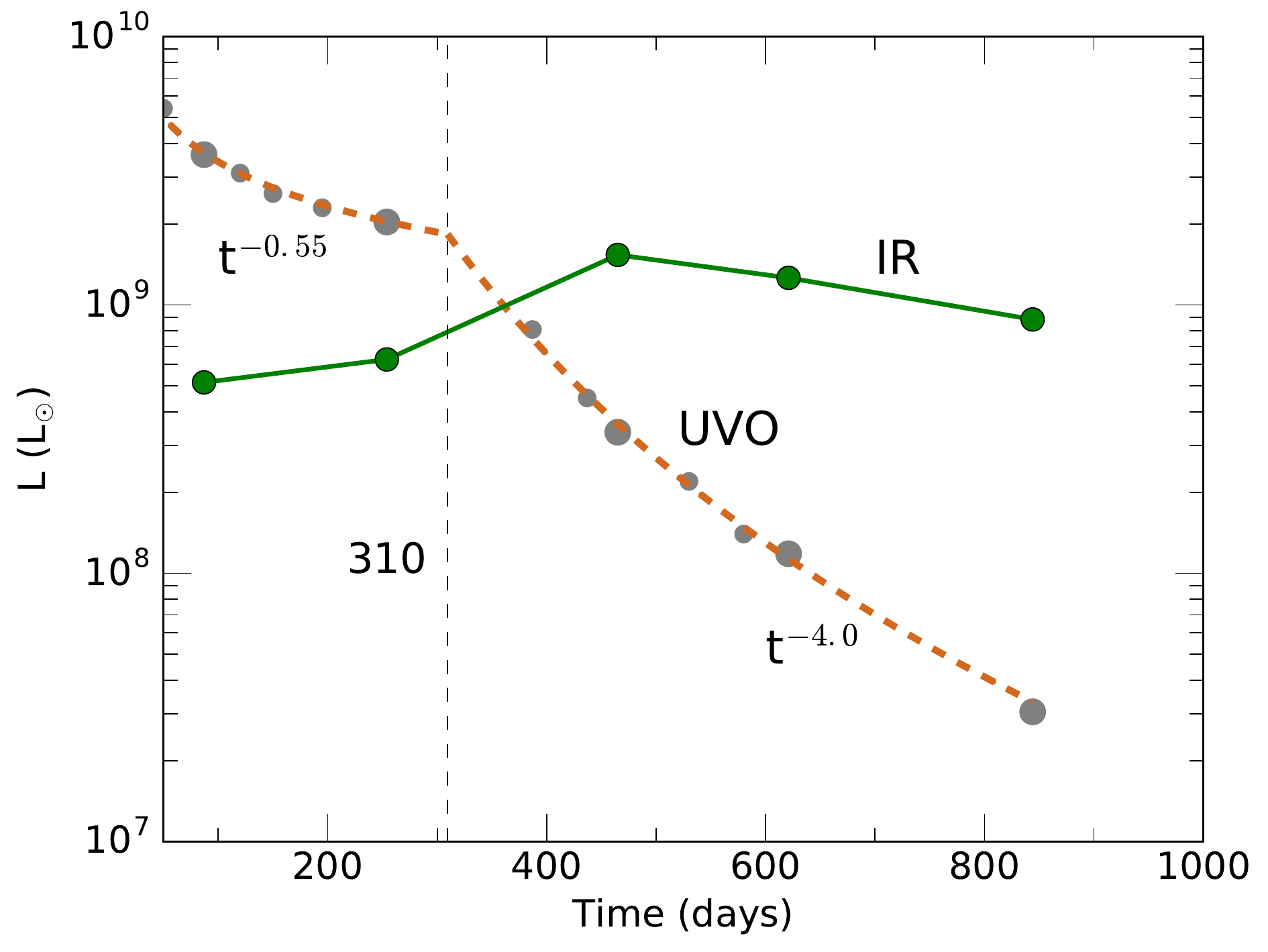}
\includegraphics[width=3.3in]{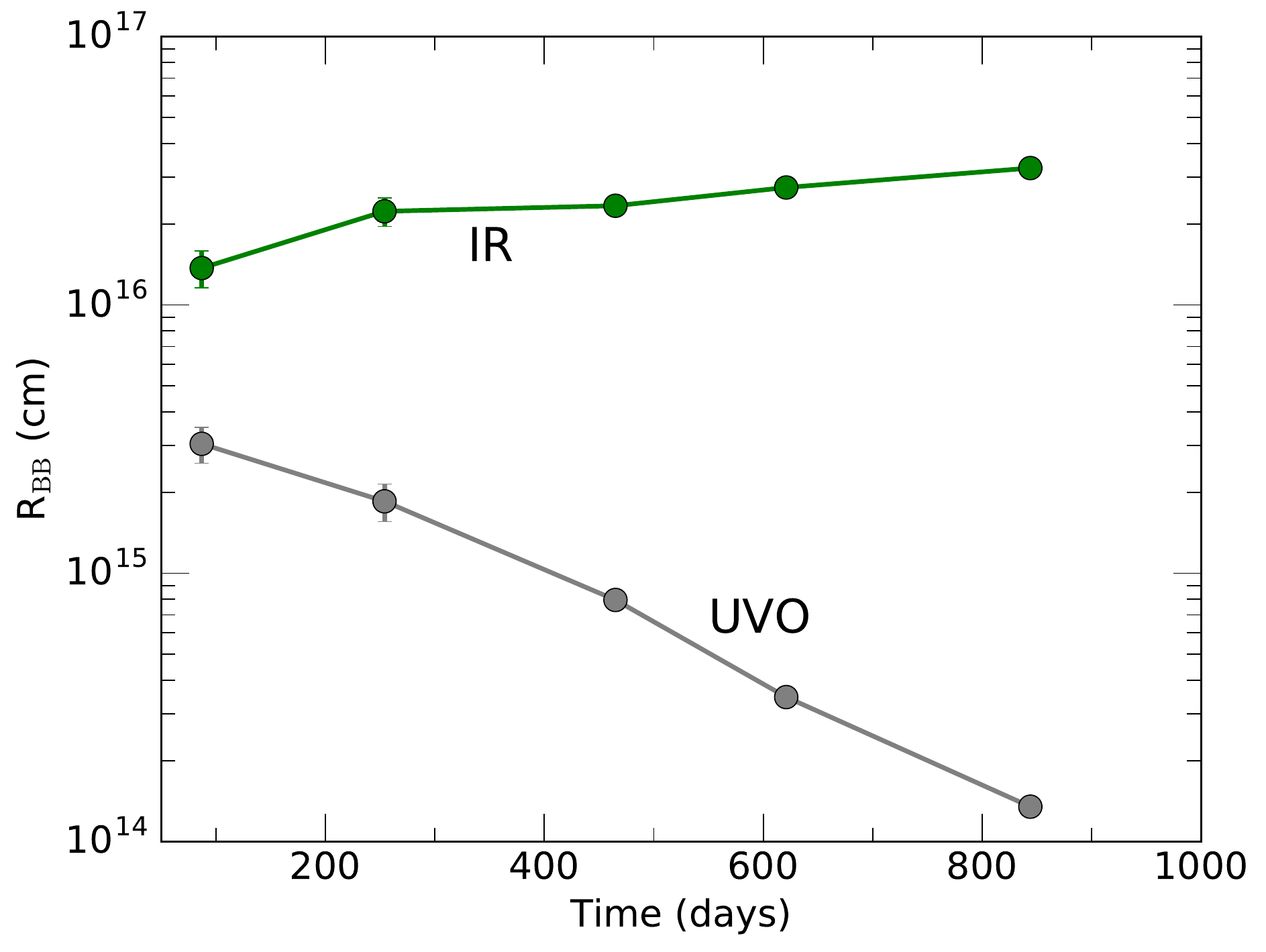}
\includegraphics[width=3.3in]{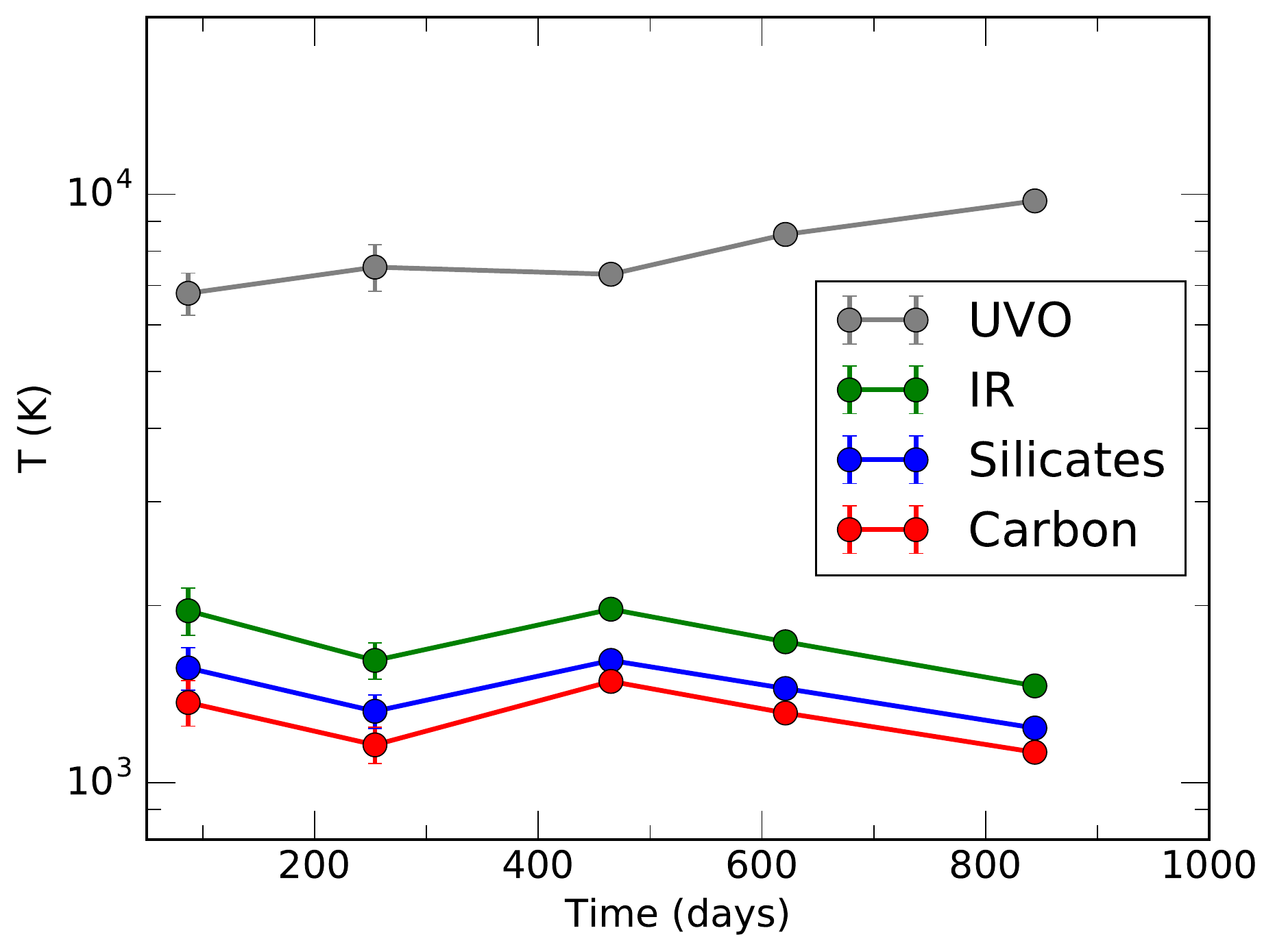}
\includegraphics[width=3.3in]{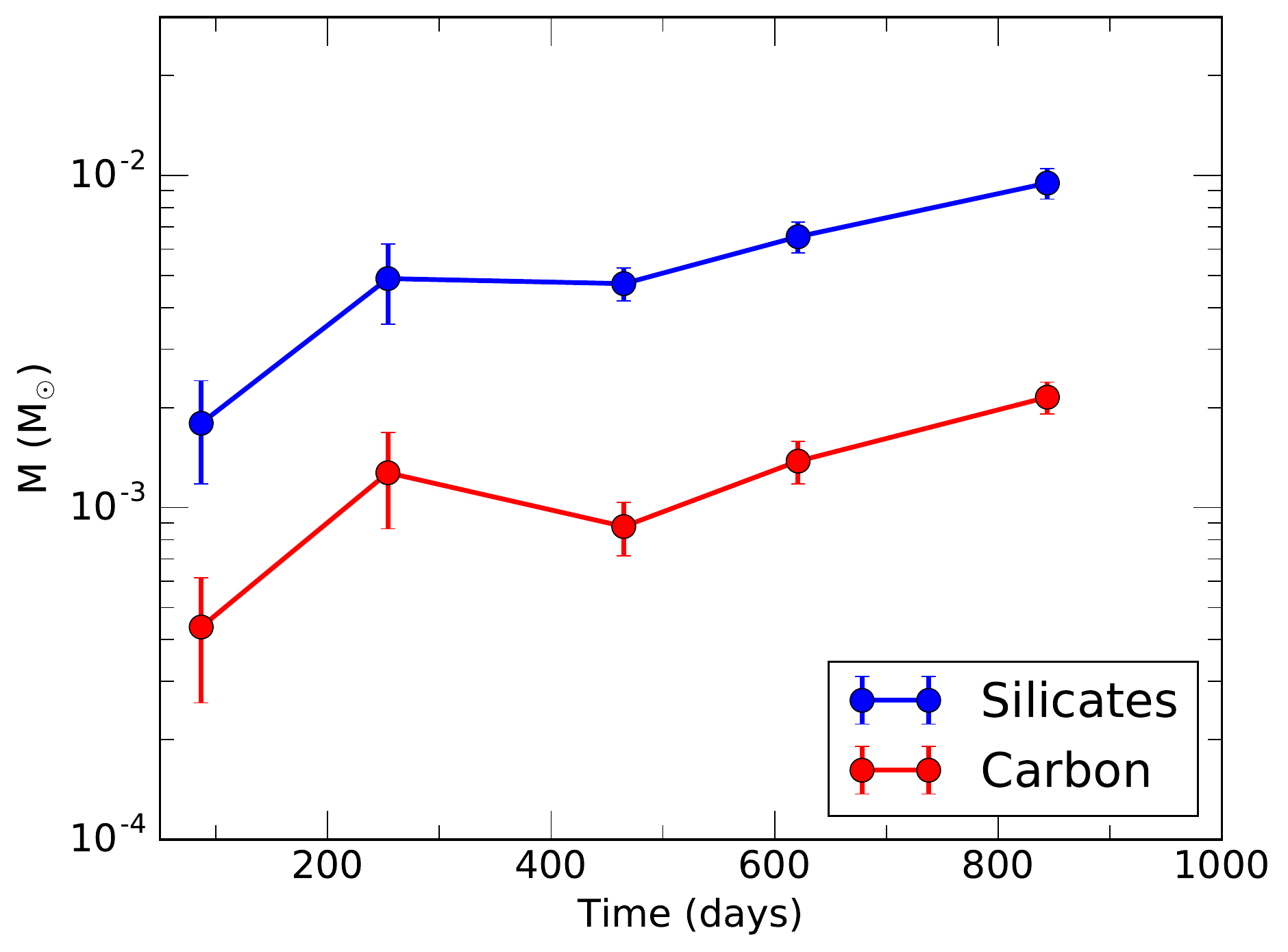}
\caption{\label{fig:fitpars} \footnotesize{The post-explosion conditions, obtained by the fitting of the optical and IR bands, are presented in the figure: (a) the luminosity of the UVO component and the IR component (b) the radius of the two blackbody components (c) the temperatures of the optical as well as the three different IR components (d) the masses of astronomical silicates and amorphous carbon. The luminosity was obtained by interpolating and fitting the the optical ($B$, $V$, $i^{\prime}$) and the NIR (J-H-K$_s$) fluxes from \cite{fra14} at several epochs, in addition to the five epochs (day 87, 254, 465, 621, 844) for which we have the $Spitzer$ data. Thereafter, luminosity, $L(t)$, was fitted using a power law of time. The exponents are found to be -0.55 \& -4.0 with a switch at $\sim$ day~310, as shown on the figure. }} 
\end{figure*}
  

%
%

\begin{table*}
\centering
\caption{Best-fit parameters to the photometric data of SN~2010jl}
\label{dust_epoch} 
\begin{tabular}{c c c c c c c c c}
\hline \hline
epoch & T$\rm _{UVO}$ & $R\rm _{UVO}$ & L$\rm _{UVO}$ & IR & T$\rm _{IR}$ & R$\rm _{IR}$ & L$\rm _{IR}$ & Dust mass   \\
(days) &  (K) &  (cm) & (\Ls) & Component & (K) & (cm) & (\Ls) & (\Ms)  \\
\hline
& &  & & BB & 1959 & 1.4 $\times$ 10$^{16}$ & 5.1 $\times$ 10$^{8}$  & -  \\
87 & 6787 & 3.0 $\times$ 10$^{15}$ & 3.6 $\times$ 10$^{9}$ & Sil & 1566 &  - && 1.8 $\times$ 10$^{-3}$  \\
& &  & & Am-C &  1368 & - && 4.4 $\times$ 10$^{-4}$  \\
\hline
& &  & & BB & 1613 & 2.2 $\times$ 10$^{16}$ & 6.3 $\times$ 10$^{8}$  & -  \\
254 & 7518 & 1.9 $\times$ 10$^{15}$ & 2.1 $\times$ 10$^{9}$ & Sil  & 1322 &  - & - & 4.9 $\times$ 10$^{-3}$  \\
& &  & & Am-C & 1159 & - & - & 1.3 $\times$ 10$^{-3}$  \\
\hline
& &  & & BB & 1971 & 2.3 $\times$ 10$^{16}$ & 1.5 $\times$ 10$^{9}$ & -  \\
465  & 7311 & 8.0 $\times$ 10$^{14}$ & 3.4 $\times$ 10$^{8}$ &  Sil & 1613 &  - & - & 4.7 $\times$ 10$^{-3}$  \\
& &  & & Am-C & 1486 & - & - & 8.8 $\times$ 10$^{-4}$  \\
\hline
& &  & & BB & 1735 & 2.7 $\times$ 10$^{16}$ & 1.2 $\times$ 10$^{9}$ & -  \\
621 & 8545 & 3.5 $\times$ 10$^{14}$ & 1.2 $\times$ 10$^{8}$ & Sil & 1445 &  - & - & 6.5 $\times$ 10$^{-3}$  \\
& &  & & Am-C & 1313 & - & - & 1.4 $\times$ 10$^{-3}$  \\
\hline
& &  & & BB & 1460 & 3.2 $\times$ 10$^{16}$ & 8.8 $\times$ 10$^{8}$ & -  \\
844 & 9742 & 1.4 $\times$ 10$^{14}$ &  3.0 $\times$ 10$^{7}$ & Sil & 1238 &  - & - & 9.5 $\times$ 10$^{-3}$  \\
& &  & & Am-C & 1126 & - & - & 2.2 $\times$ 10$^{-3}$  \\
\hline
993 & -  & - & - &  BB & 1063 & 5.8 $\times$ 10$^{16}$ & 7.9 $\times$ 10$^{8}$ & -  \\
998 & - & - & - &  BB & 995 & 5.6 $\times$ 10$^{16}$ & 5.8 $\times$ 10$^{8}$ & -  \\
1368 & -  & - & - &  BB & 855 & 6.3 $\times$ 10$^{16}$ & 3.9 $\times$ 10$^{8}$ & -  \\
1750 & - & - & - &  BB & 709 & 7.6 $\times$ 10$^{16}$ & 2.7 $\times$ 10$^{8}$ & -  \\
\hline
\end{tabular}
\end{table*}

\section{The optical and the NIR fluxes}
\label{optnir}

 \begin{figure}[t]
\centering
\includegraphics[width=3.5in]{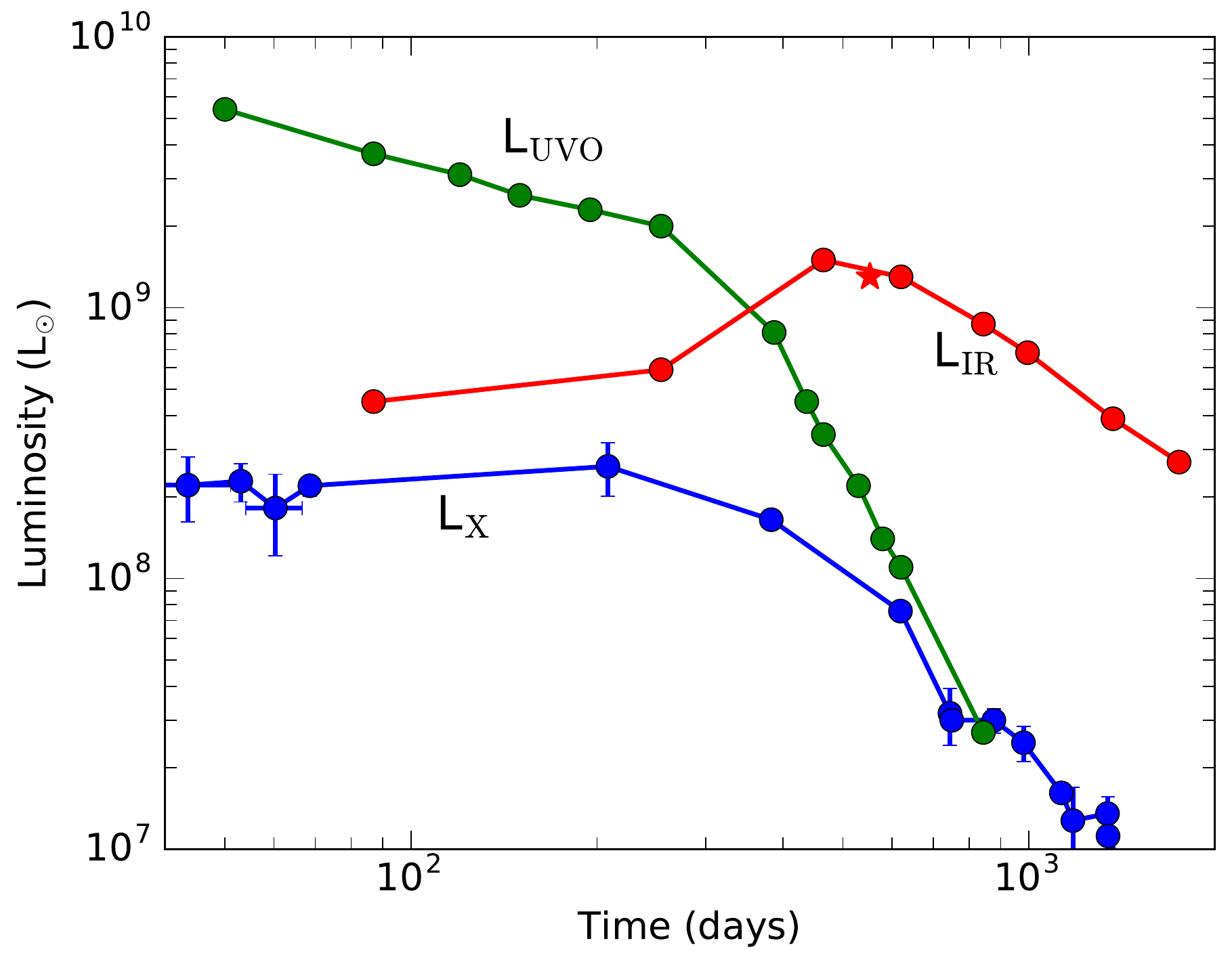}
\caption{\label{fig:allobs} \footnotesize{The luminosity of the observed UVO, X-rays and the IR as a function of time is presented, as obtained from the studies by \cite{cha15, fra14,gal14} \& \cite{mae13} (marked by an `asterix'). The IR luminosity at day 995 is an average of day 993 and 998. }} 
\end{figure}

In this section we analyze the UVO and IR observations of SN~2010jl and its implications on the SN light curve.

The optical and NIR observations of SN~2010jl at several epochs have been obtained using different facilities such as the \textit{Hubble Space Telescope} (HST), the 2.5m Nordic Optical Telescope (NOT), the KeplerCam instrument at F. L. Whipple Observatory (FLWO) in Arizona and the \textit{Spitzer Space Telescope} \citep{fra14, and11, fox11, fox13}.

Using the tabulated magnitudes provided by \cite{fra14}, we have calculated the fluxes in the optical bands $u^{\prime}$, $B$, $V$, $r^{\prime}$, $i^{\prime}$ through the Vega flux zeropoint conversions. The NIR fluxes for $J$, $H$, $K_s$ bands were calculated using the 2MASS (Two Micron All Sky Survey) standard system. The tabulated optical and NIR fluxes were interpolated on an uniform grid between the earliest and the latest epochs of observations. 

Within the timescale of UVO observations, the \textit{Spitzer} 3.6 and 4.5 \mic\ fluxes of SN~2010jl are available at days 87, 254, 465, 621 and 844 \citep{fra14, and11, fox13}. We fit the SED continuum at these five epochs. The fluxes at the $r^{\prime}$ band are likely to be dominated by the H$\alpha$ line intensities \citep{fra14}, therefore we chose to ignore this band for the purpose of continuum fitting.

The spectra were fit using a two-component fit model with a blackbody as the UVO component and either of (a) a blackbody, (b) optically thin astronomical silicate dust, or (c) optically thin pure amorphous carbon as the IR component (Figure \ref{fig:fitting}). For IR regimes of the electromagnetic spectrum, the wavelengths are much larger compared to the average grain radii ($\lambda \rm _{IR} >> a$). Within this Rayleigh limits, the derived dust masses are independent of the grain sizes. For simplicity, a single grain radius of 0.01 $\mu$m and a single temperature were assumed. 

The optical constants for the dust grains are derived from \cite{wei01} and \cite{zub04} respectively for silicates and carbon dust. The best-fit scenarios for all the fifteen cases (5 epochs, 3 cases) are shown in Figure \ref{fig:fitting}. The parameters obtained by the best-fit cases are summarized in Table \ref{dust_epoch}. 


IR fluxes of SN~2010jl for late times (day 993, 998, 1368 and 1750) were obtained from the {\it Spitzer} Heritage Archive\footnote{\url{http://irsa.ipac.caltech.edu/applications/Spitzer/SHA/}} database. There are no UVO and the near-IR fluxes recorded for those epochs. We fit the two Spitzer points at each of these late epochs using a pure blackbody and estimate the temperature, blackbody radius and the luminosity of the IR emission. The best-fit scenarios are shown in Figure \ref{fig:lateIR} and listed in Table \ref{dust_epoch}. 


The UVO component of the spectra represents the emission from the shocked cooling gas, also referred to as the photosphere. The effective temperature of this region is found to be between 6000 and 8000~K. The photospheric luminosity is found to decline with the passage of post-explosion time. This is likely to be either caused by a systematic decline in CSM density towards outer radii, a decline in shock velocity over time, or the combination of both.

 The blackbody radius ($R\rm _{UVO}$) of the UVO component, derived from the fit, represents the minimum radius of the photosphere. The $R\rm _{UVO}$ is found to recede as a function of time, as shown in Figure \ref{fig:fitpars} (top-right) and in Table \ref{dust_epoch}. In contrast, the prime source of the UVO emission is understood to be the shocked gas coupled to the outwards moving shock-front, and hence $R\rm _{sh}$ (the position of the shock) should be increasing with time. This ambiguity is caused essentially due to the shell becoming increasingly thin to optical radiation. In other words, the dense shell acts like a diluted blackbody with a dilution-coefficient $f\rm_d$ (= $R\mathrm{_{sh}}$/$R\mathrm{_{UVO}}$) which is increasing as a function of time. 

We also calculate the UVO luminosity in uniform temporal bins in order to reconstruct the UV-optical light curve. The evolution of L$\rm _{UVO}$ was thereafter fitted using a power law, $t^{-\beta}$, with $\beta$ = 0.55 before day 310 $\beta$ = 4.0 thereafter. The best-fit scenario is presented in Figure \ref{fig:fitting} (top-left).  The nature of the luminosity, $L(t)$, derived in this study compares well with the findings by \cite{fra14, jen16} \& \cite{gal14}. The rapid decline in luminosity at later times can be attributed to a change density profile at the shock-front.
A similar drop in luminosity is also reflected in the evolution of the X-rays along the line of sight \citep{cha15}.  

\subsection{Possible source of the IR emission}

In this study, good fits to the IR spectra of SN~2010jl, shown in Figure \ref{fig:fitting}, was achieved using the contribution from any one of the three sources (a blackbody, astronomical silicates, amorphous carbon) as the IR component. Therefore, the best-fit models cannot provide any strong bias towards either of the three sources to be the most likely one. 


A blackbody fit is appropriate when the dusty shell is optically thick. Additionally, the blackbody radius ($R\rm _{IR}$) represents the minimum possible radius of the dusty shell from where the IR emission originates. 

The NIR fluxes can be fit well using the absorption coefficients of optically thin astronomical silicates.  
The upper limit on the 9.7 \mic\ flux at late times was obtained from the \textit{SOFIA} observations \citep{wil15}. It indicates the absence of silicate feature. Therefore, the presence of silicates is doubtful, unless the grains are large (a $\sim$ 5 \mic) or the dusty shell is optically thick.

With only photometric data in the 2.2-4.6 $\mu$m region, many featureless dust species can fit the spectrum. We choose carbon as a likely dust type owing to its relatively large abundance in space compared to other metals. Moreover, the studies dealing with the IR spectra from the pre-explosion era also hint at the presence of carbon dust in SN~2010jl \citep{dwe17}. However, the CSM of SN~2010jl is reported to have a N-rich \citep{fra14} C-depleted environment (N/C = 25 $\pm$ 15, N/O = 0.85 $\pm$ 0.15). In such environments, O-rich dust species are known to be the primary dust components and all the carbon mostly gets locked up in CO molecules \citep{sar13}. To support this argument, the observation of N-rich dusty shell around massive stars, such as the Homunculus nebula surrounding Eta Carinae, does not show any evidence of C-rich dust species \citep{mor17}. Hence, the presence of amorphous carbon dust in SN~2010jl also becomes uncertain.

Apart from identifying the types of dust, the best-fit scenarios also provide the boundary conditions to the shell morphology. Figure \ref{fig:fitpars} presents the total IR luminosity, dust blackbody radii, temperatures and masses as a function of time. The $L\rm _{IR}$ is found to increase steadily until $\sim$ day 500, followed by a slow decline phase. Importantly, around $\sim$ day 350, it is found to exceed the declining $L\rm _{UVO}$. The dust grains are found to remains fairly at high temperature ($T$ $>$ 1000 K) at all times which provides additional constraints on the evolution of the shock radii. The dust masses grow steadily over time, with a small decrease between day 250 and 460 for both silicates and carbon dust.

 \begin{figure}[t]
\centering
\includegraphics[width=3.6in]{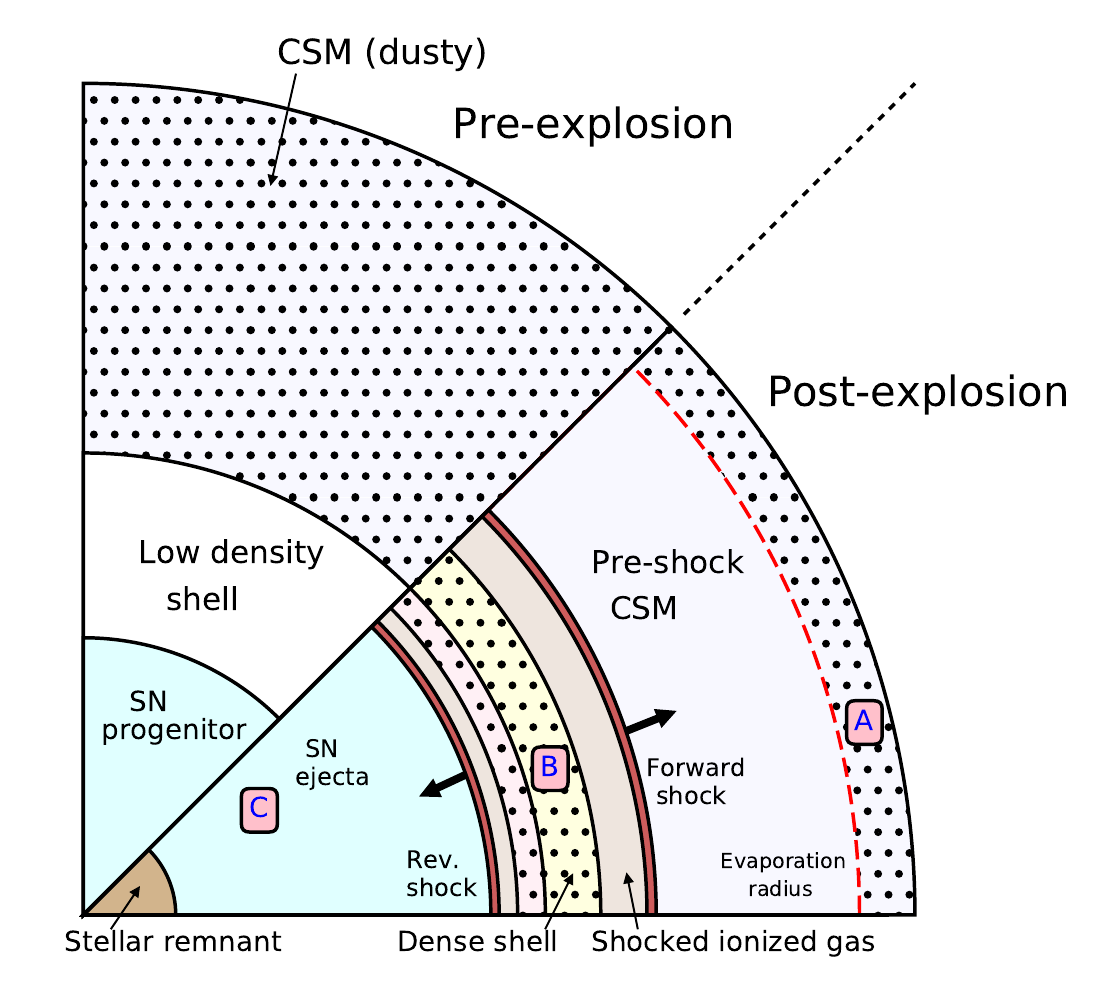}
\caption{\label{fig:shell} \footnotesize{The schematic diagram presents the pre-explosion stellar environement and the post-explosion ejcta-shell geometry of a typical type IIn SN. The possible regions where dust can reside, marked in the figure with dotted mesh, are the following:  (A) the pre-existing dust outside the evaporation radius (B) newly synthesized dust grains in the dense shell formed in the post-shock cooling gas (C) new dust formed in the SN ejecta. Therefore, if IR emission recorded in the post-explosion era, that should originate from either of these regions. }} 
\end{figure}

\section{Shell morphology}
\label{morphology}
Based on observational constraints, we formulate a schematic model of the pre-explosion star and post-explosion supernova, shown in Figure \ref{fig:shell}. For simplicity, we assume a spherically symmetric geometry. 

Prior to the explosion, the progenitor of a typical type IIn supernova constitutes of the central star, most likely in its supergiant phase, surrounded by a dense CSM \citep{chu94}. The region between the stellar photosphere and the surrounding CSM is characterized by a low density region, which might be extremely thin, making the CSM and the photosphere almost adjacent to one another, or it might also be a sizable fraction of the total CSM thickness \citep{dwa07}. The size of this region is determined by the mass-loss history of the central star, which may vary from being periodic to continuous in nature. 

Following the explosion, the SN produces a blast wave that encounters the CSM within a few days. Thereafter, the forward shock traverses outwards through the CSM. Simultaneously, due to the ejecta-CSM interaction, a reverse/reflected shock is also generated, that travels inwards through the ejecta. Owing to the high densities in the CSM, it is unclear that the backward shock is reverse or reflected in nature. We consider it as a reverse shock for the context of this study. The distinct regions in the shell-ejecta morphology along the radially outward direction are described as follows: 
\begin{enumerate}[label=(\alph*),noitemsep]
\item the stellar remnant in the form of a neutron star or black hole \item the expanding ejecta powered by the explosion energy \item the reverse/reflected shock traveling inwards through the expanding ejecta \item A thin layer of hot and ionized ejecta following the reverse shock \item the cool dense shell formed by the cooling of the forward and the reverse shock  \item a layer of adiabatically heated and compressed CSM immediately behind the forward shock \item the forward-shock propagating outwards \item the partially ionized CSM, ionized by the precursor radiation, lying ahead of the forward shock \item the region outside the evaporation radius where dust still survives. 
 \end{enumerate}
 
The ejecta and the CSM are taken to be separated by the contact discontinuity through out the span of the evolution.  
 
The SN forward shock is capable of heating and ionizing the CSM that produces a strong flux of ionizing radiation. Owing to the high densities, the post-shock gas forms a thin dense layer of rapidly cooling gas which emits efficiently in the UV-optical regimes of the electromagnetic spectra. The total luminosity from this thin layer of hot post-shock gas overall dominates the SN-light curve. The total luminosity of the ejecta which is powered by radioactive decay of \Ni, \Co\ and \Ti\ is much smaller than that generated by the shock-CSM interaction \citep{gal14,jen16}. Therefore, the energetics of the CSM-shock interaction determines the photosphere of the SN light curve \citep{fra14}.

\section{Origin of the dust emission}
\label{origin}

Figure \ref{fig:shell} indicates the possible regions where dust can reside in the stellar system post-explosion. The IR emission can be due to the contribution from either (i) newly formed dust in the expanding ejecta,  (ii) pre-existing dust in the unshocked CSM (iii) newly formed dust in the cooling post-shock gas, or a combination of these. In this section, we shall briefly discuss the cases of (i) and (ii) in order to explain why new dust formation in the ejecta or the surviving dust in the pre-shock CSM cannot entirely account for the IR emission in SN~2010jl. 

\subsection{Dust in the ejecta}
The SN ejecta is not expected to be the main source of dust responsible for the IR emission in SN~2010jl because of the following reasons: \\
(a) Newly formed dust in the ejecta is not likely to contribute to the observed IR emission during either the early or late epoch of the evolution of the light curve.
Theoretically, dust formation is impeded by the presence of radioactivities and the cascade of hard radiation and non-thermal electrons in the ejecta. This picture is observationally confirmed in the most extensively studies SN1987A, where dust formation commenced only after day 250 of the explosion \citep{woo93, bou93, dwe15}. In contrast, the IR excess from the SN commenced from day 30 after peak emission \citep{gal14, fra14}. \\
(b) The BB radius of the IR emitting region is 1.5 $\times$ 10$^{15}$ cm, so even if dust formed somehow early on in the metal-rich ejecta, the dust could not have traversed this distance within ~30 days at the typical ejecta velocities of ~3,000 km s$^{-1}$. \\
(c) Ejecta dust could in principle contribute to that late time ($>$300 days) IR emission. However this dust would be internal to both, the forward or reverse shock.
The UVO will therefore never be able attenuated by the dust, and its emission will consequently have to be always larger than that of the NIR emission, contrary to observations. \\
Taking into account all these factors, the ejecta of Type IIn supernovae are likely to be a rather inefficient dust producer, so the majority of IR emission must be originating from the CSM.


\subsection{The pre-existing dust in the CSM}
\label{destruc}
Pre-SN imaging of the field around SN~2010jl have provided upper limits on the UVO and NIR emission from the SN \citep{fox17, dwe17}. These upper limits suggest the need for the presence of pre-existing dust in the CSM to extinguish the UVO emission from any hot progenitor star. Such dust will reprocess the emission produced by the shock breakout and by the ensuing SN shock traveling through the CSM, giving rise to an IR echo. The intensity and duration of the echo depends on the amount of pre-existing dust, and on the radius of the cavity within which the dust was vaporized by the breakout luminosity. A full discussion, and a detailed model of the echo contribution to the SN light curve will be presented in a separate publication \citep{dwe18}.

However, the echo will not contribute much to the late time (t $>$ 300 d) IR emission. The rise in the IR light curve around day 300 will require a delay time of 150 days, which corresponds to a dust shell radius of 4 $\times$ 10$^{17}$ cm. As shown in \cite{dwe18}, the mass of dust required to produce such echo exceeds the abundance of refractory elements likely to reside in the CSM at that distance. Also the dust temperatures at such large distance do not comply with the range of temperatures derived from observations.

The ejecta dust or the pre-existing dust is therefore inadequate to account for the IR luminosities in SN~2010jl. 
Newly formed dust in the CSM is therefore the dominant source of the IR emission at times later than day ~300. In the following section we present a detailed analysis of the formation of dust in the CSM.

\section{The post-shock gas dynamics}
\label{constraints}
Formation of dust grains in the shocked cooling gas requires appropriate gas phase conditions. In order to resolve the post-shock gas dynamics, in this section we focus on the constraints on the model, which are derived from the studies of the pre-supernova CSM and the evolving SN-shock. 

\subsection{CSM dynamics}

The evolution of the shock through the CSM is addressed analytically by several studies \citep{che82, chu09,ofe10, ofe14a}. In case of SN~2010jl, we adopt a power law for the evolution of shock velocity $v$($t$) \til\ $v_0 (t/t_0)^{k}$, and derive the values of k from the evolution of the hydrogen column densities. The pre-shock CSM density is considered to vary as $n \sim r^{-w}$ \citep{che82, svi12}.  


The UVO and X-ray luminosity of SN~2010jl are characterized by a broken power law with a break around day 310 (= $t_b$), as explained in Figure \ref{fig:fitpars} (top-left) \citep{fra14, jen16}. The different exponents of the light curve before and after day 310 is attributed to the variation in the density index $w$. 

The shock velocity ($v\rm_{sh}$), the radius at the shock-front ($R\rm_{sh}$), the column density and mass of the shocked gas, $N\rm_H$ and $M$ respectively, are derived as follows:



\begin{equation}
\label{vsh}
\begin{split}
v_{sh} (t) & = v_0 \Big(\frac{t}{t_0} \Big)^{k_1}, \ \ \ \ \ t \le t_b \\
& = v_b \Big(\frac{t}{t_b} \Big)^{k_2}, \ \ \ \ \ t > t_b \\
v_b & = v_0 \Big(\frac{t_b}{t_0} \Big)^{k_1} \\
\end{split}
\end{equation}

\begin{equation}
\label{rsh}
\begin{split}
R_{sh} (t) & =  R_0 + \int_{t_0}^{t} v_0 \Big(\frac{t}{t_0} \Big)^{k_1} \mathrm{d}t, \ \ \ \ \ t \le t_b  \\
& = R_b + \int_{t_b}^{t} v_b \Big(\frac{t}{t_b} \Big)^{k_2} \mathrm{d}t, \ \ \ \ \ t > t_b \\
R_b & = R_0 + \int_{t_0}^{t_b} v_0 \Big(\frac{t_b}{t_0} \Big)^{k_1} \mathrm{d}t
\end{split}
\end{equation}


\begin{equation}
\label{vsh}
\begin{split}
n (r) & = n_0 \Big(\frac{r}{R_0} \Big)^{-w_1}, \ \ \ \ \ R_{sh} \le R_b \\
& = n_b \Big(\frac{r}{R_b} \Big)^{-w_2}, \ \ \ \ \ R_{sh} > R_b \\
n_b & = n_0 \Big(\frac{R_{b}}{R_0} \Big)^{-w_1} \\
\end{split}
\end{equation}


\begin{equation}
\label{cdsh}
\begin{split}
& N_H(R_{sh}) = \int_{R_0}^{R_{sh}} n_0 \Big(\frac{r}{R_0} \Big)^{-w_1} \mathrm{d}r, \ \ \ \ \ t \le t_b \\
& = N_H(R_b) + \int_{R_0}^{R_{sh}} n_b \Big(\frac{r}{R_b} \Big)^{-w_2} \mathrm{d}r, \ \ \ \ \ t > t_b \\
\end{split}
\end{equation}

\begin{equation}
\label{cdsh_rem}
\begin{split}
& N_H (\mathrm{tot}) = N_H (R_1) \\
& N_H (\mathrm{LOS}) = N_H (R_1) - N_H(R_{sh}) \\
\end{split}
\end{equation}

\begin{equation}
\label{msh}
\begin{split}
& M(R_{sh}) = \int_{R_0}^{R_{sh}} 4 \pi r^2 \rho_0 \Big(\frac{r}{R_0} \Big)^{-w_1} \mathrm{d}r, \ \ \ \ \ t \le t_b \\
& = M(R_b) + \int_{R_0}^{R_{sh}} 4 \pi r^2  \rho_b \Big(\frac{r}{R_b} \Big)^{-w_2} \mathrm{d}r, \ \ \ \ \ t > t_b \\
\end{split}
\end{equation}



\noindent where $v_0$, $t_0$, $n_0$, $\rho_0$ and $v_b$, $t_b$, $n_b$, $\rho_b$ are the shock velocity, time, CSM number and mass density at $R\rm_0$ and at the break ($R_b$) on day 310. $N_H$(LOS) represents the H-column density along the line of sight. Using Equation \ref{rsh} and \ref{cdsh}, we can derive the time dependance of the column densities which is an observable quantity. 

\begin{equation}
\label{cdsh_derivative}
\begin{split}
& R(t) = R_i + \frac{v_i t_i}{k_1+1} \Big[\Big( \frac{t}{t_i} \Big)^{k_1+1} -1 \Big] \\
& \frac{\mathrm{d}N_H}{\mathrm{d}t} =  v(t) \times \frac{\mathrm{d}N_H}{\mathrm{d}r}   \\
& = n_i v_i \Big(\frac{t}{t_i} \Big)^{k_1} \Bigg(1 + \frac{v_i t_i}{(k_1+1) R_i} \Big[\Big( \frac{t}{t_i} \Big)^{k_1+1} -1 \Big] \Bigg)^{-w_1}
\end{split}
\end{equation}

\noindent where $i = 0$ and $b$ before and after the break respectively.

\begin{figure}[t]
\centering
\includegraphics[width=3.3in]{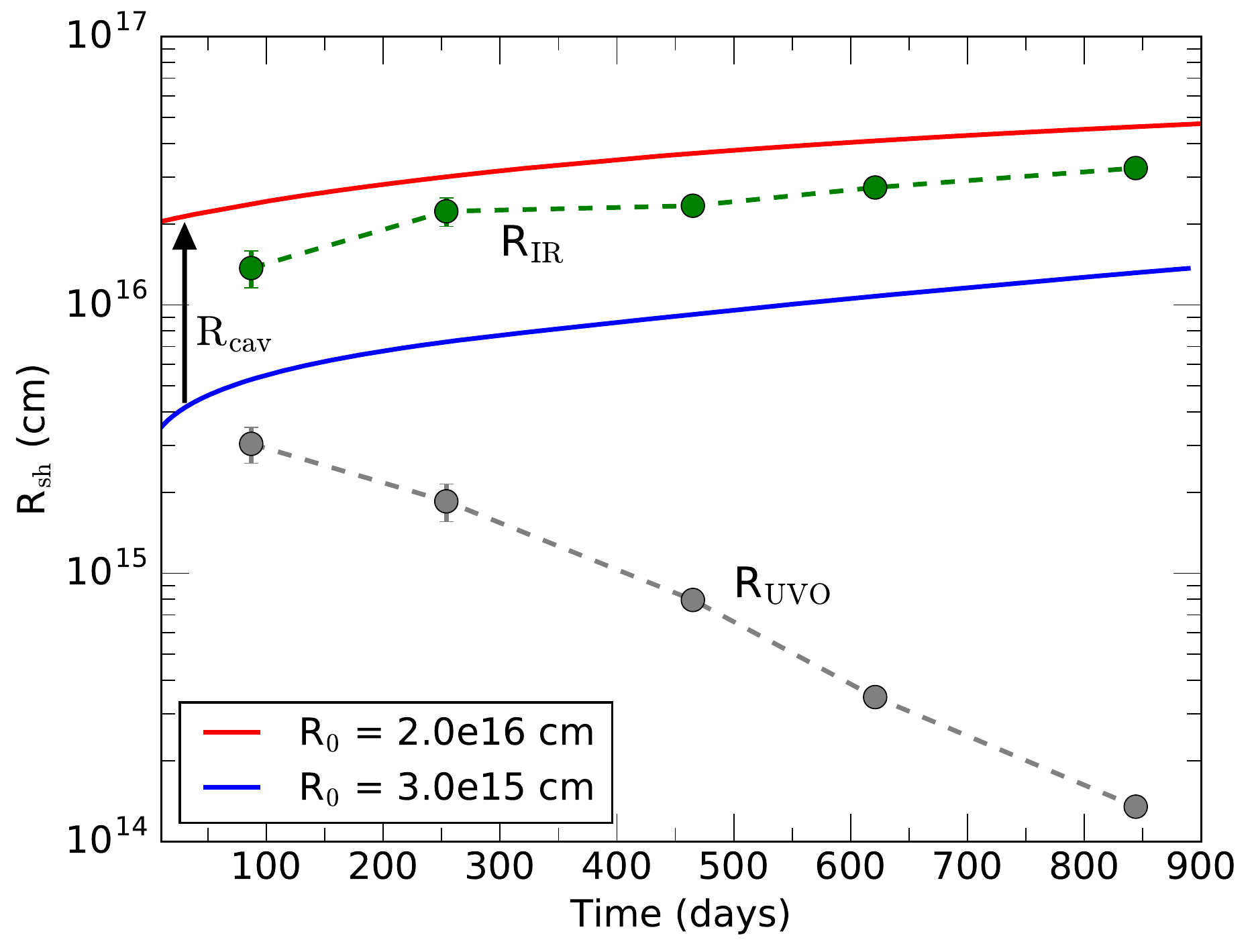}
\caption{\label{fig:withgapr} \footnotesize{The evolution of the shock radius as a function of post-explosion days is presented for two set of $R_0$. The shock radius ($R\rm _{sh}$) is compared to the evolution of $R\rm _{UVO}$ and $R\rm _{IR}$ in the figure. Assuming dust forms in the post-shock region, the shock radius should be at least at $R\rm _{IR}$, as that is the minimum radius of the dusty shell. In order to satisfy this condition, the $R_0$ is chosen to be at at least 2.0 $\times$ 10$^{16}$ cm, and a low density shell of thickness $R\rm _{cav}$ accounts for the gap between the ejecta and inner radius of the CSM. }} 
\end{figure}

\begin{figure}[t]
\centering
\includegraphics[width=3.3in]{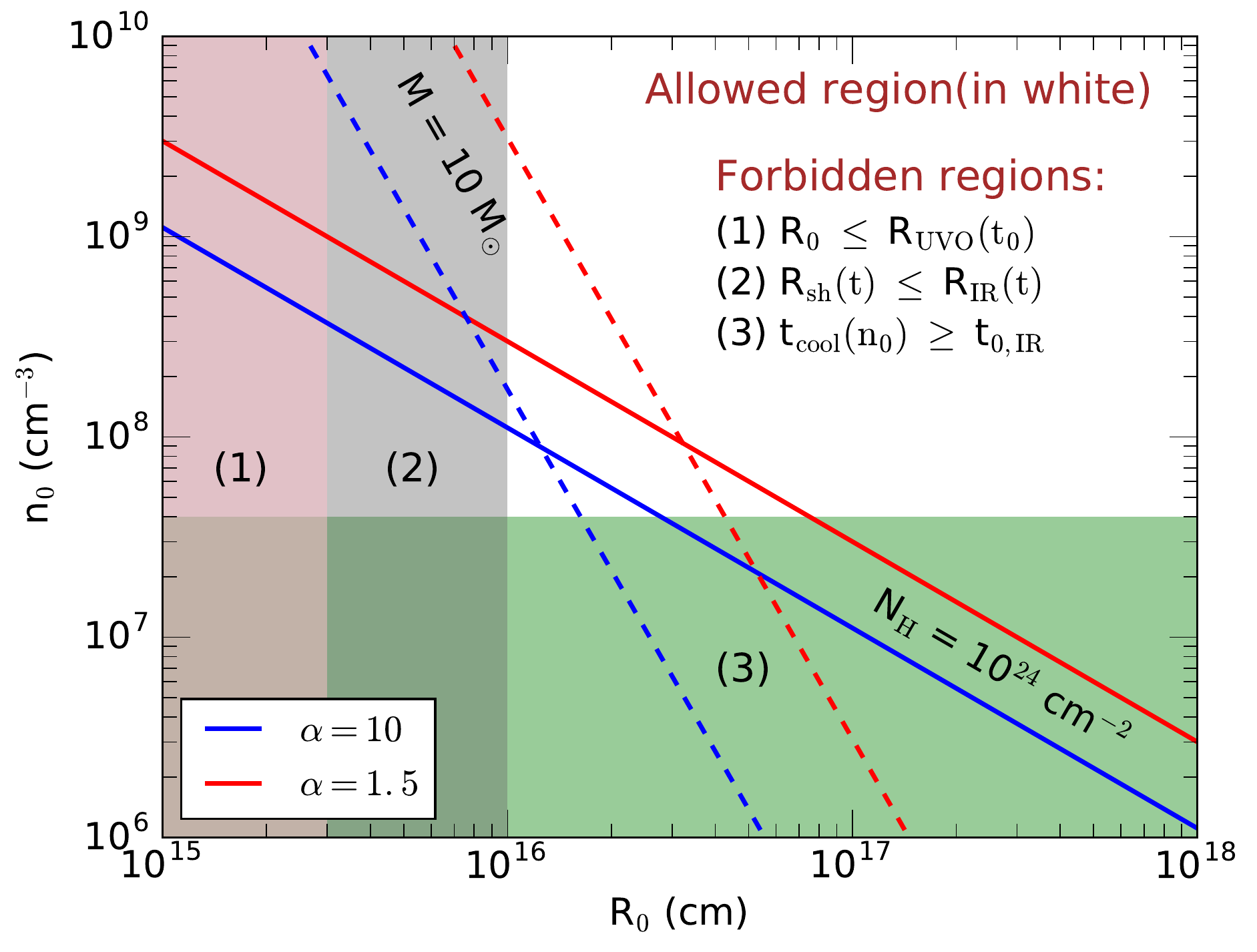}
\caption{\label{fig:boundary} \footnotesize{The figure shows the boundary conditions on the parameter space, in terms of $n_0$ \& $R_0$. The solid lines corresponds to the column density of 10$^{24}$ cm$^{-2}$ where as the dotted lines correspond to the total CSM of 10 \Ms, for two sets of $R_1$/$R_0$ ratio ($\alpha$ = 1.5, 10). The white region in the figures are allowed region, whereas the shaded regions are forbidden due to various factors as numbered on the figure. (see Section \ref{constraints}). }} 
\end{figure}
  

\subsection{Constraints from observations}

\cite{cha15} and \cite{ofe14b} study X-ray emission data from SN~2010jl using \textit{Chandra}, \textit{NuSTAR} and \textit{Swift-X-ray} telescopes. The hydrogen column densities in front of the shock along the line of sight is derived by \cite{cha15}, which provides the necessary limiting conditions on the shock parameters at a given epoch. The total H-column density of the pre-supernova CSM is estimated to be \til\ 1-2 $\times$ 10$^{24}$ cm$^{-2}$.

The radial density dependance in the first phase, before the break in the density law, was assumed as $w_1$ = 2, which resembles a freely expanding outflow. 
The lower limit on $R_0$ is the photospheric blackbody radius $R\rm _{UVO}$.

The X-ray observations provide the gas column densities along the line of sight, i.e. the pre-shock CSM at a given time. We fit the hydrogen column densities from SN~2010jl using the above equations, to determine the initial conditions such as $m, w, R_0, n_0, v_0$ etc. Important to note, there is no unique solution to the column density function, as it depends on the appropriate combination of all these parameters. In Figure 
\ref{fig:withgapr} the temporal variation of the shock radius is shown for two set of inner radii, $R_0$ = 3$\times$10$^{15}$ cm, which is the lower limit of $R_0$ given by $R\rm _{UVO}$ and $R_0$ = 2$\times$10$^{16}$ cm. 

Despite being a good fit to the observed column densities, the model with $R_0$ = 3$\times$10$^{15}$ cm cannot appropriately justify the SN light curve of SN201jl. This is because, the shock radius in this case is always less than the blackbody radius of dust ($R\rm_{sh}$$<$$R\rm_{IR}$), as clearly seen in Figure \ref{fig:withgapr}. 

The pre-existing dust has already been ruled out to be the only source of IR emission and the X-ray observations put constraints on the maximum shock velocities. Hence, starting at $R_0$ = 3$\times$10$^{15}$ cm, the shock radius at a given time will never exceed the IR blackbody radius. On the other hand, emission from the post-shock dust requires $R\rm_{sh}$ $>$ $R\rm_{IR}$, which is satisfied by models with $R_0$ $\ge$ 1.4 $\times$ 10$^{16}$ cm, that is the minimum blackbody radius of the IR emission.

We choose an inner radius of $R_0$ = 2 $\times$ 10$^{16}$ cm that provides a good fit to the observed column densities and also the condition $R\rm_{sh}$(t)$>$$R\rm_{IR}$(t) remains valid all through.    

A typical red supergiant has a photospheric temperature between 2500-3000 K and a luminosity of about 10$^6$ \Ls. Therefore, the pre-explosion star is assumed to have a photospheric radius between 10$^{14}$ - 10$^{15}$ cm. If the inner radius of the CSM ($R_0$) is larger than 10$^{16}$ cm, there must be a low density region lying between the stellar photosphere and the surrounding CSM before the explosion. 
The  possible existence of low density cavities between the progenitor star and its CSM is not unexpected. Such cavity can be created by episodal mass loss events, and strong stellar winds \citep{dwa07, cas75, smith16}.
To derive further constraints, in Figure \ref{fig:boundary} we present the the total column density ($\sim$ 1-2 $\times$ 10$^{24}$ cm$^{-2}$) and the total mass of the CSM ($\sim$ 10 \Ms) of the CSM in the parameter space of $n_0$ and $R_0$. The $n_0$-$R_0$ relation is shown for two different $R_1$, given by $\alpha$ (= $R_1$/$R_0$) taken as 1.5 and 10. This diagram helps to identify all possible combinations of $n_0$ and $R_0$ that comply with the X-ray, UV-optical and IR observations.

The choices of $n_0$ and $R_0$ were constrained by the following requirements: 
\begin{enumerate}[noitemsep]
\item The inner radius of the CSM ($R_0$) must be equal to or larger than than the observed blackbody radius of the photosphere, $R\rm_{UVO}$.
\item The shock radius ($R\rm _{sh}$$(t)$) must be larger than the blackbody radius of the IR spectra ($R\rm _{IR}$$(t)$), so that the dust formation region is internal to the propagating shock-front. 
\item The shortest cooling time ($t\rm _{cool}$($n_0$)) of a parcel of shocked gas should be smaller than the earliest epoch of dust formation ($t\rm _{0,IR}$). This is because, the shocked gas is required to cool down below condensible temperatures before the onset of dust formation. 
\end{enumerate}

\begin{figure*}
\centering
\includegraphics[width=3.3in]{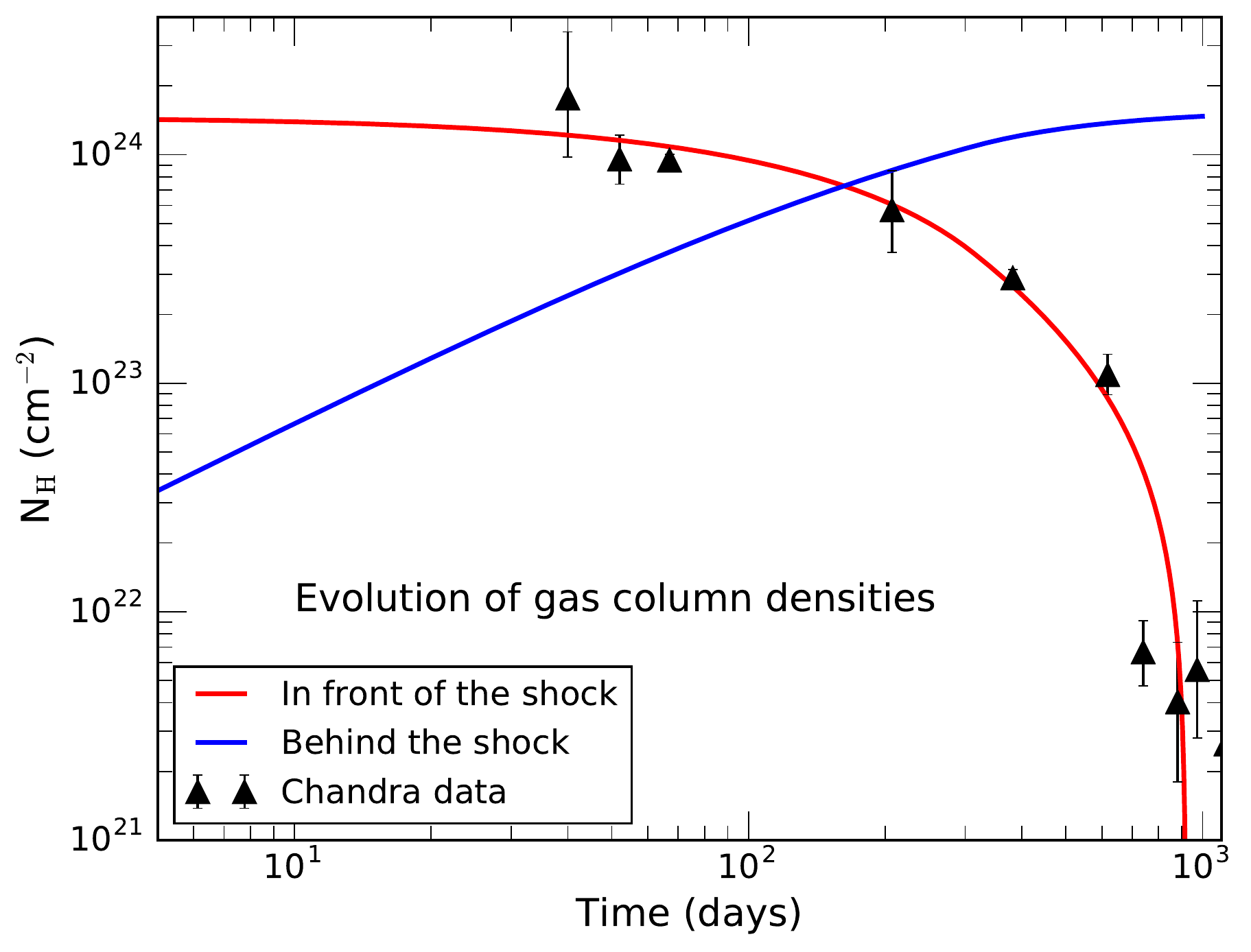}
\includegraphics[width=3.3in]{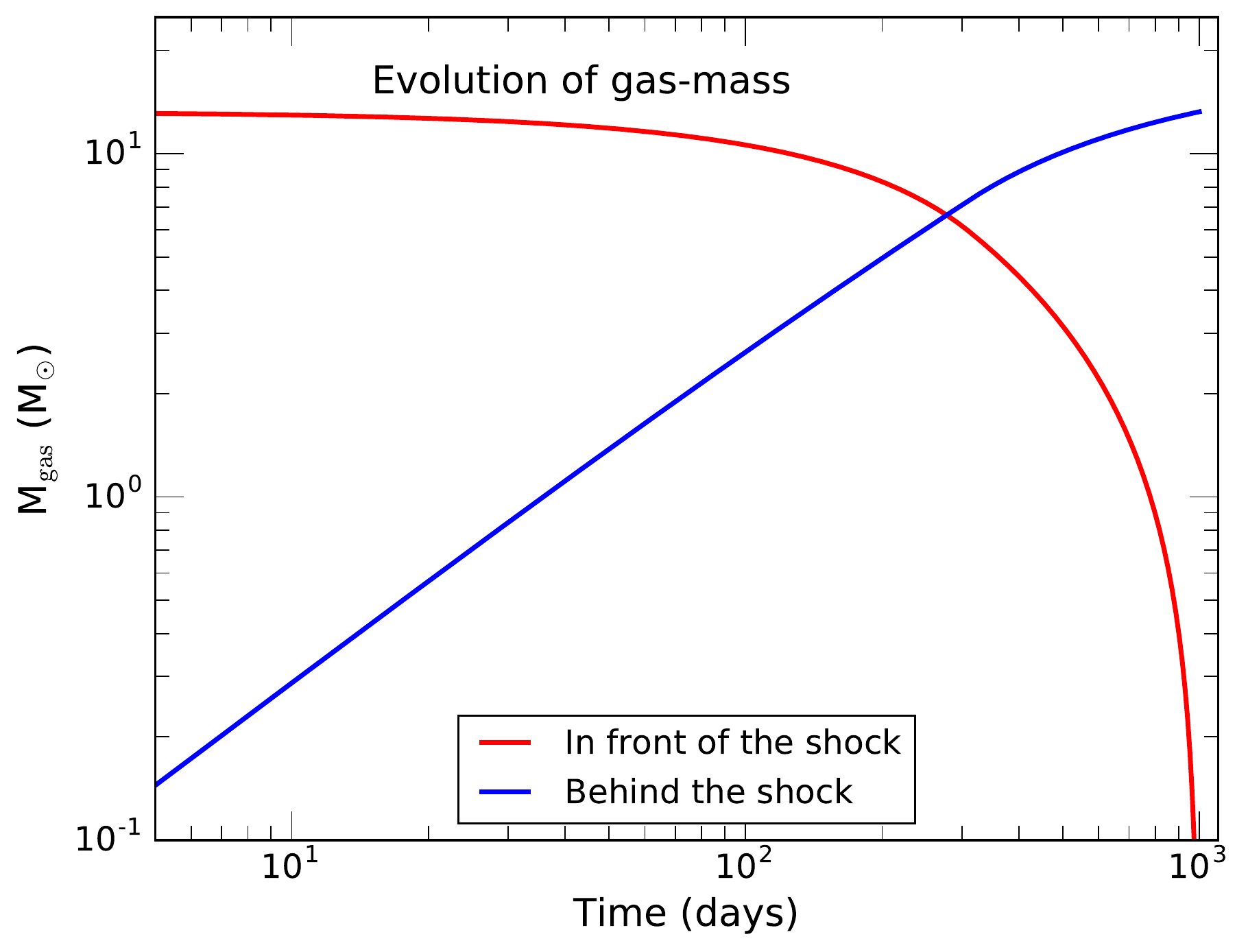}
\includegraphics[width=3.3in]{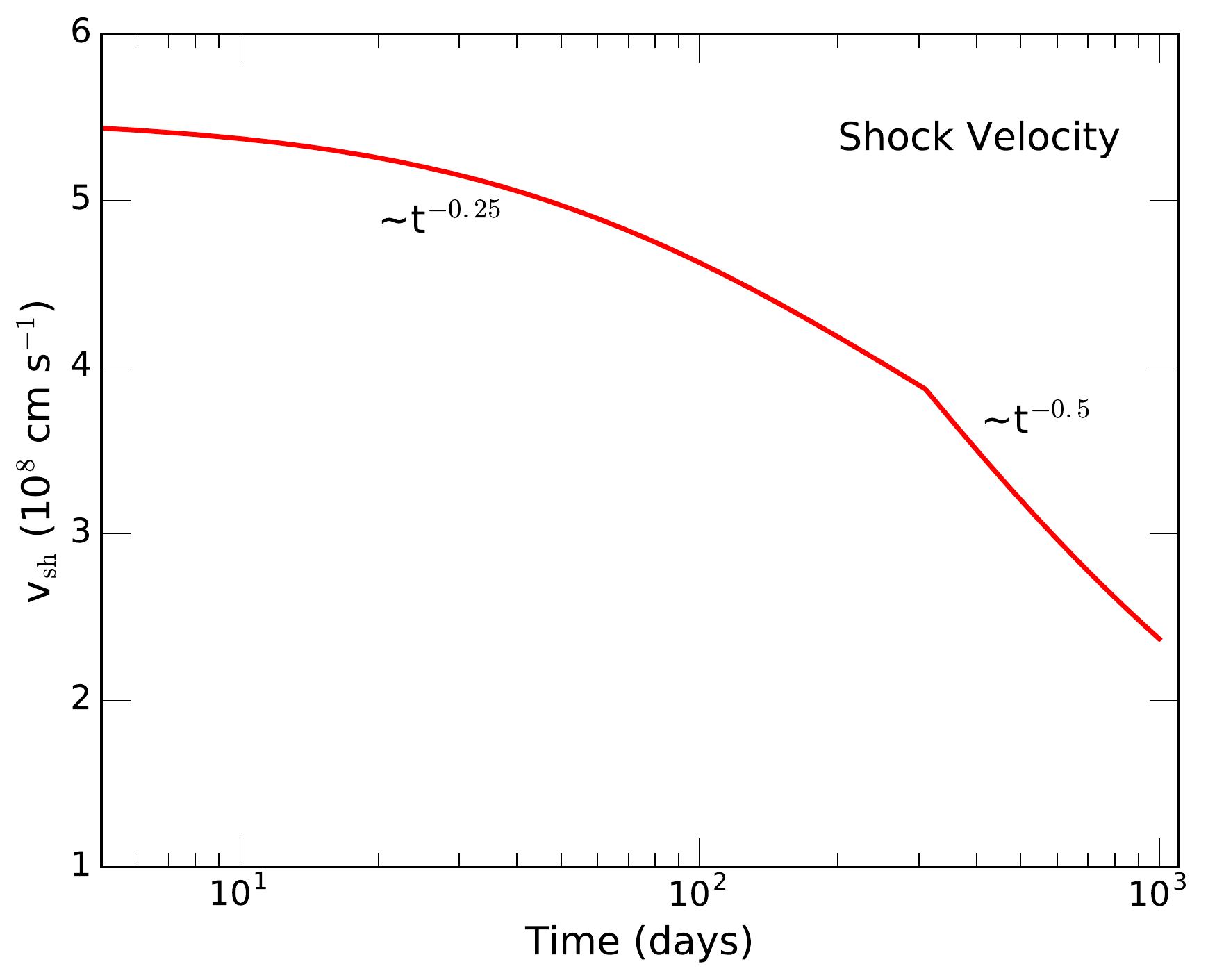}
\includegraphics[width=3.3in]{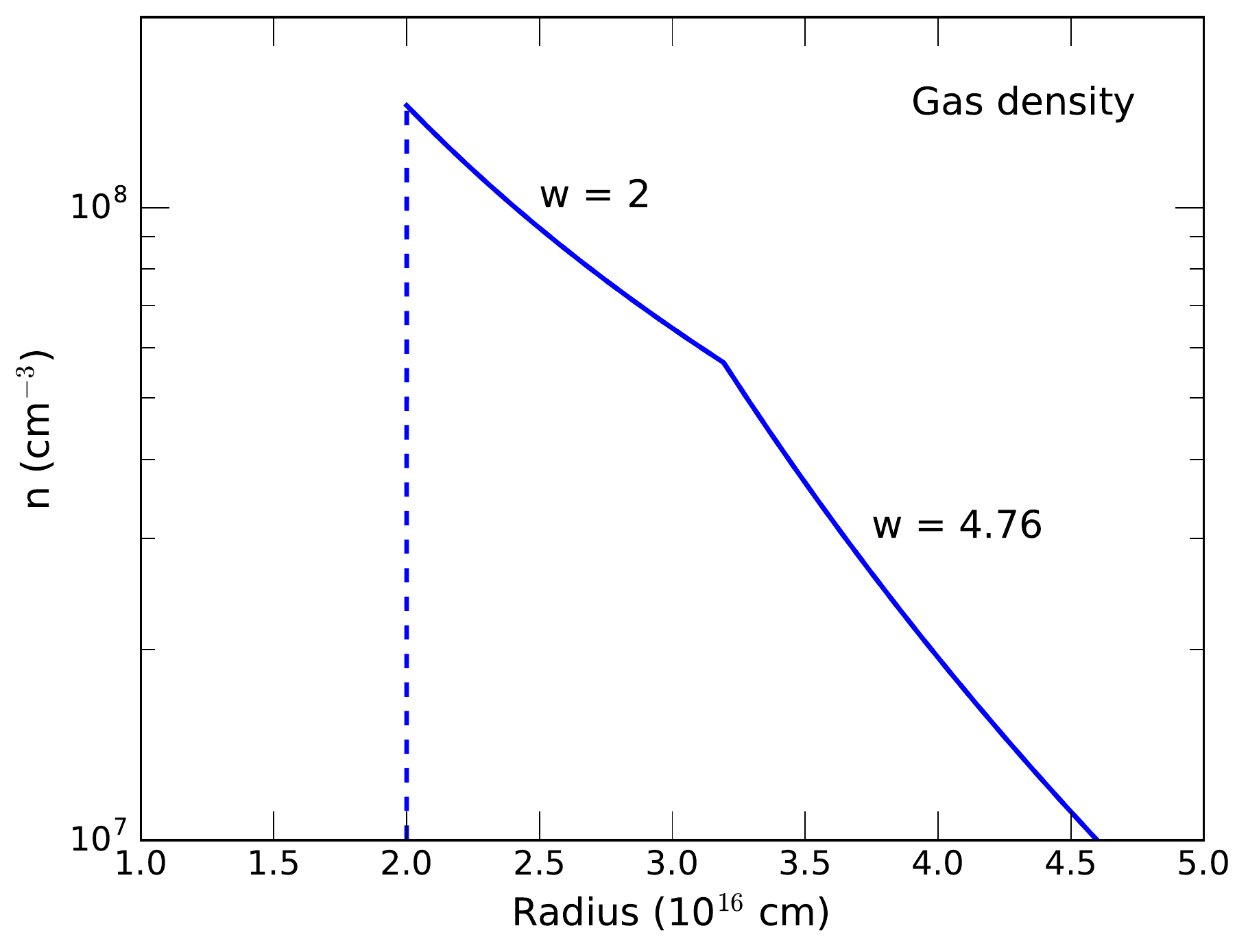}
\caption{\label{fig:case2} \footnotesize{The pre-shock CSM and the evolution of the post-shock gas is characterized by the following parameters, with reference to Table \ref{initials2}: (a) the evolution of hydrogen column density, calculated using the Equations \ref{rsh} and \ref{cdsh} (b) the evolution of CSM mass behind and ahead of the shock-front (c) the evolution of shock velocity (d) the CSM density profile as a function of radius, before the encounter with the shock. }} 
\end{figure*}
  

Based on these allowed range on $n_0$ and $R_0$ the best fit model for the observed column densities ($N\rm_H$) has been constructed. Table \ref{initials2} summarizes the initial conditions for the model and the temporal variations of the physical quantities such as density, shock velocity and shock radius. 

The fit to the hydrogen column densities along the line of sight  is shown in Figure \ref{fig:case2} (top-left) with the swept-up column in the post-shock gas. The number density in the CSM prior to the explosion is presented in Figure \ref{fig:case2} (bottom-right). The evolution of the pre- and post-shock CSM mass (top-right) and the velocity of the forward shock (bottom-left) though the dense CSM is also shown in the figure. The total mass of CSM in this case is about 12 \Ms which complies well with the studies by \cite{smi12} and \cite{fox17}. The shock velocity steadily declines with time-exponents of -0.25 and -0.5 before and after the break at day 310. 

\section{Cooling of the post-shock gas}
\label{postshock}


\begin{table*}
\centering
\caption{The number density, shock velocity and shock radius at $t_0$ and t$_b$ estimated from the best-fit to the post-explosion column densities}
\label{initials2} 
\begin{tabular}{c c c c c}
\hline \hline
parameter & value at $t_0$ ($w_1$ = 2) & variation &  value at $t_b$ ($w_2$ = 4.76) & variation   \\
\hline
 CSM density & $n_0$ = 1.45 $\times$ 10$^8$ cm$^{-3}$ & \til\ r$^{-2}$ & $n_b$ = 5.7 $\times$ 10$^7$ cm$^{-3}$ & \til\ r$^{-4.76}$ \\
Shock velocity & v$_0$ =  5.5 $\times$ 10$^8$ cm/s & \til\ t$^{-0.25}$ & $v_b$ =  3.8 $\times$ 10$^8$ cm/s & \til\ t$^{-0.5}$ \\
Shock radius & $R_0$ = 2.0 $\times$ 10$^{16}$ cm & \til\ t$^{0.75}$ & $R_b$ = 3.2 $\times$ 10$^{16}$ cm & \til\ t$^{0.5}$ \\
 \hline
\end{tabular}
\end{table*}


\begin{figure}[t]
\centering
\includegraphics[width=3.3in]{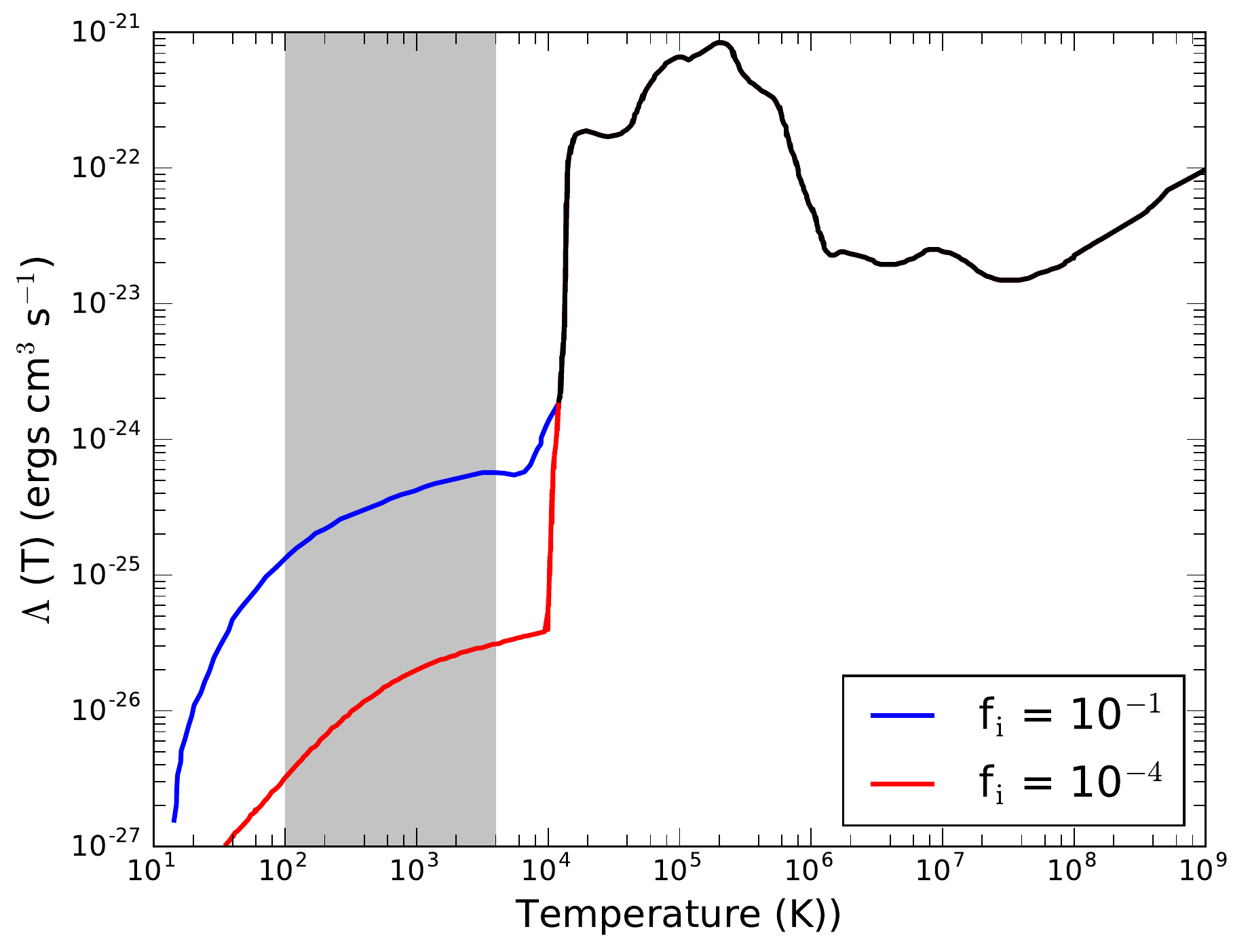}
\caption{\label{fig:coolfunc} \footnotesize{The cooling functions, $\Lambda$(T), for two different ionization fraction (f$_i$ = 10$^{-1}$, 10$^{-4}$) at low temperatures is shown in the figure, taken from \cite{dal72}. For temperatures larger than $\sim$ 10$^4$~K, the entire gas is fully ionized. The shaded region in gray represents the zone of interest, in terms of dust formation. }} 
\end{figure}



In this section we study the temporal-evolution of the shocked gas and determine the suitable conditions that facilitate dust formation in SN~2010jl. 

Various studies have analytically or numerically addressed the shock equations relevant to a parcel of gas, assuming equilibrium or non-equilibrium conditions \citep{dra93,tru99,des15}. In this case, we adopt a similar formalism to solve the energy, momentum and the continuity equations simultaneously and derive the time dependent temperature, density and velocity profiles in the post-shock shell. 

The passage of a shock with velocity of \til\ 4000 km s$^{-1}$ through a parcel of gas heats it to temperatures of \til\ 10$^8$~K. Thereafter, the gas behind the shock cools very rapidly, initially by free-free and X-ray line emission, and then through atomic line cooling, down to temperatures of $\sim 10^4$~K. The appropriate cooling functions relevant to this study are defined by \cite{dal72}, shown in Figure \ref{fig:coolfunc}. At temperatures higher than 10$^4$~K the gas is fully ionized and the cooling function reaches its maximum between 10$^5$ and 10$^6$~K. Below 10$^4$~K the cooling depends on the residual ionization of the gas (see Figure \ref{fig:coolfunc}), mainly controlled by metals such as Fe, Si and Mg. Cooling from  H$^+$, H and molecular lines are also dominant in such low temperatures. For this regime, we chose an ionization fraction of 10$^{-4}$, which corresponds to a lower cooling rate \citep{dal72}, in our calculations. The shaded region in Figure \ref{fig:coolfunc} represents a zone that is characterized by a very low cooling rate and low ionization fractions, both of which induce the formation of a warm dense shell which has minimal further cooling. The rate of cooling per unit volume of the gas is given by $\eta (T)$ = $n_e n_H \Lambda(T)$.

To study the evolution of the post-shock gas, the momentum and energy equations \citep{cox72, cox85} are solved with respect to a parcel of gas that has been shocked by the SN blastwave. The post-shock jump conditions for mass density ($\rho$), pressure ($p$), shock velocity ($v_s$) and gas temperature (T) are given by $\rho_J$ = 4$\rho_0$, $p_J$ = 3/4 $\rho_0 v_s^2$, $v_J$ = 1/4 $v_s$ (in the frame of the shock) and $T_J$ = 1.47 $\times$ 10$^{-9}$ $v_s^2$, where $v_s$ is in cm s$^{-1}$.  The density and temperature in the post-shock gas evolve as,

\begin{equation}
\label{dn}
\frac{\mathrm{d}\rho}{\mathrm{d}t} =  \frac{\rho \eta(T)}{\rho_0 v_s^2} \Bigg( \frac{5}{2}-4\frac{\rho_0}{\rho}\Bigg)^{-1}  \\
\end{equation}

\begin{equation}
\begin{split}
\label{tbyt0}
& \frac{T}{T_0} = (1+ M^2) \frac{\rho_0}{\rho} - M^2 \frac{\rho_0^2}{\rho^2},\ \ \ M^2 = \frac{\rho_0 v_s^2}{p_0} \\
& \ \ \ \ = M^2\frac{\rho_0}{\rho} \Big(1-\frac{\rho_0}{\rho} \Big) \ \ \ \ \  (\mathrm{for}, \ M_1 >> 1)  \\
\end{split}
\end{equation}
where, M is the Mach number of the medium. By integrating equation (\ref{dn}) and simultaneously using the $\rho$-$T$ relation from equation (\ref{tbyt0}), the density and temperature of a parcel of gas in the post-shock CSM is determined.  

In equation (\ref{tbyt0}), as $T/T_0$ is always positive, $\rho$ is always greater than $\rho_0$. In order to maintain continuity with the jump conditions we have $\rho \ge$ 4$\rho_0$. Therefore, the number density continues to increase over time and the drop in temperature is governed of the cooling curve (Fig. \ref{fig:coolfunc}).  

\begin{figure*}
\centering
\includegraphics[width=3.3in]{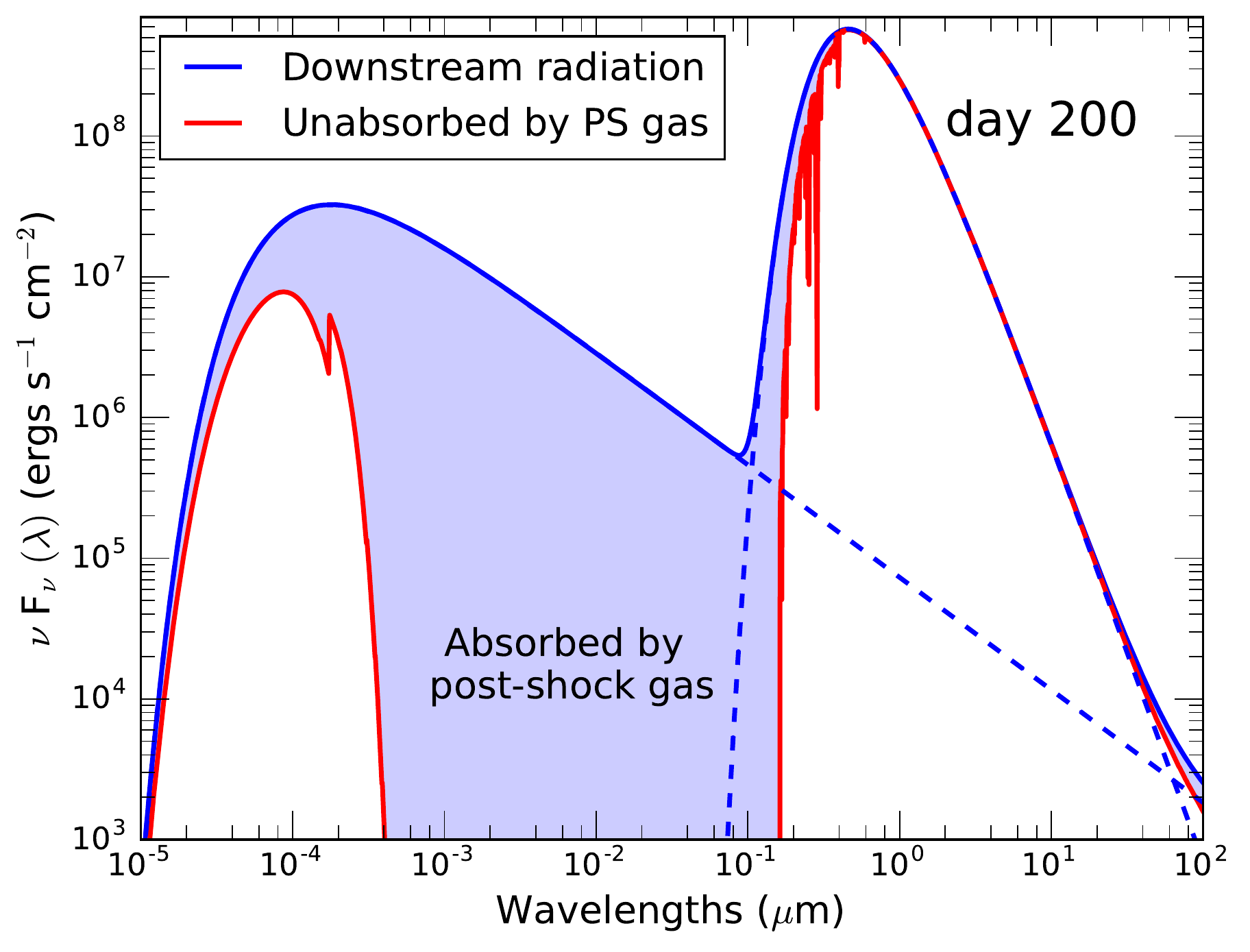}
\includegraphics[width=3.3in]{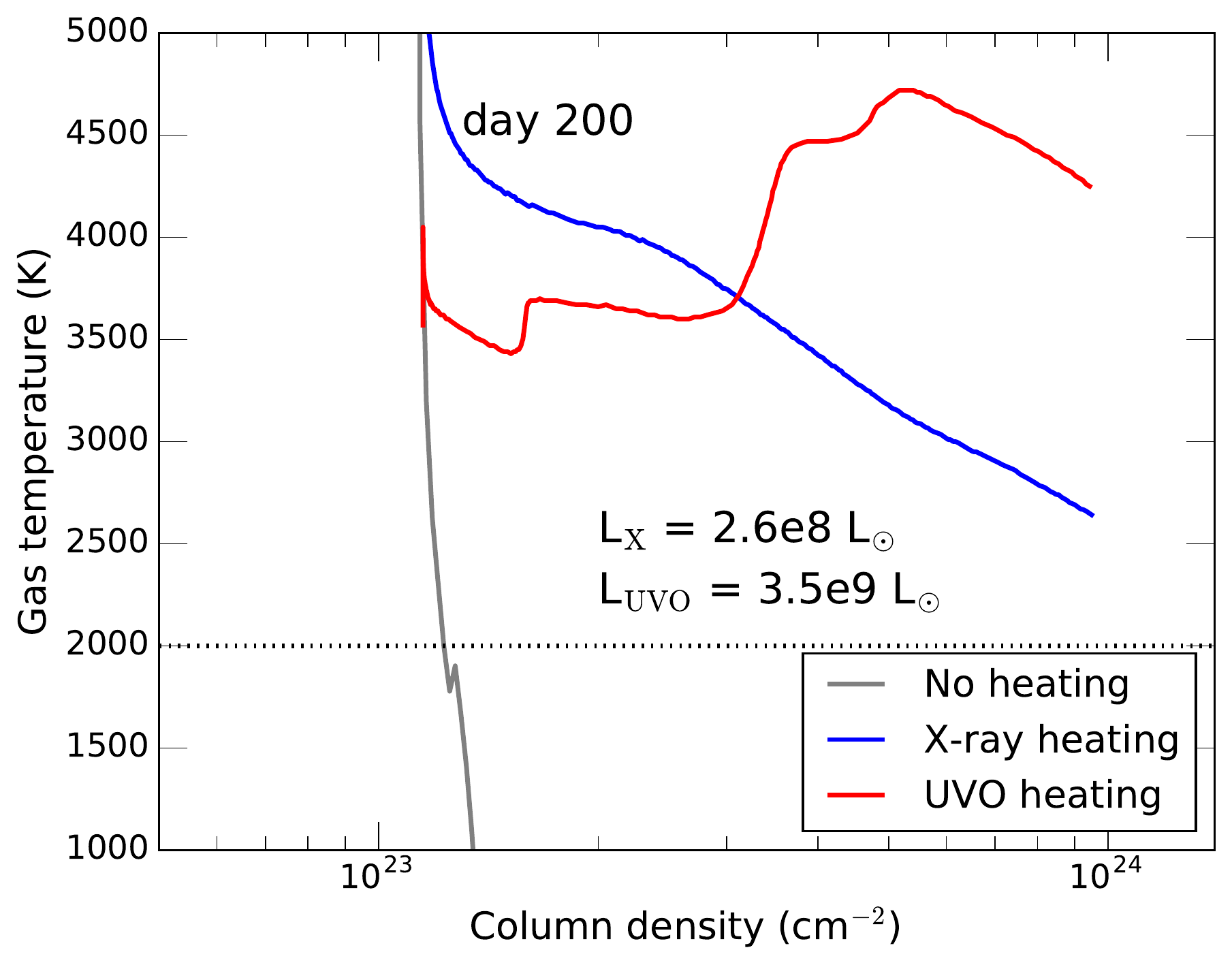}
\includegraphics[width=3.3in]{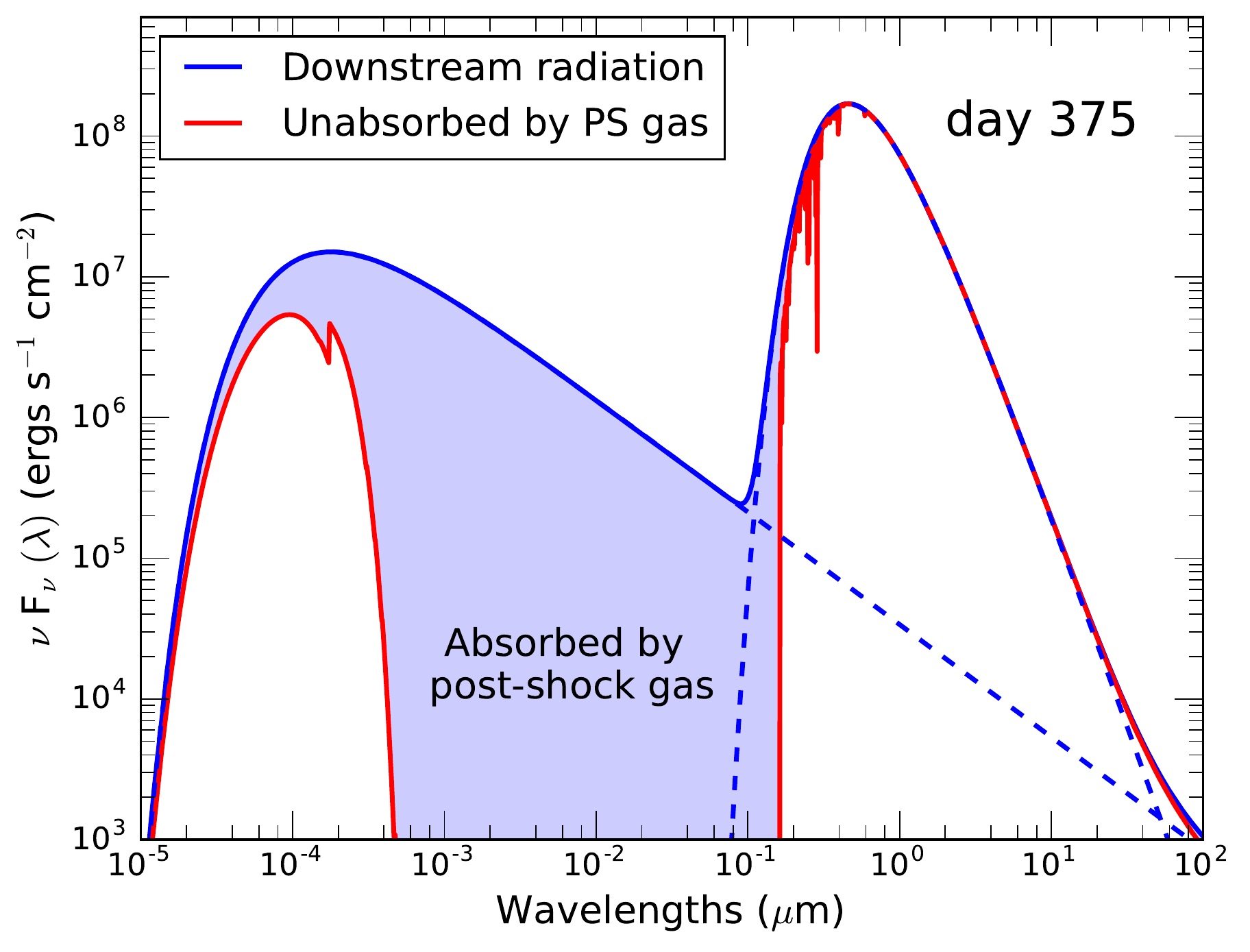}
\includegraphics[width=3.3in]{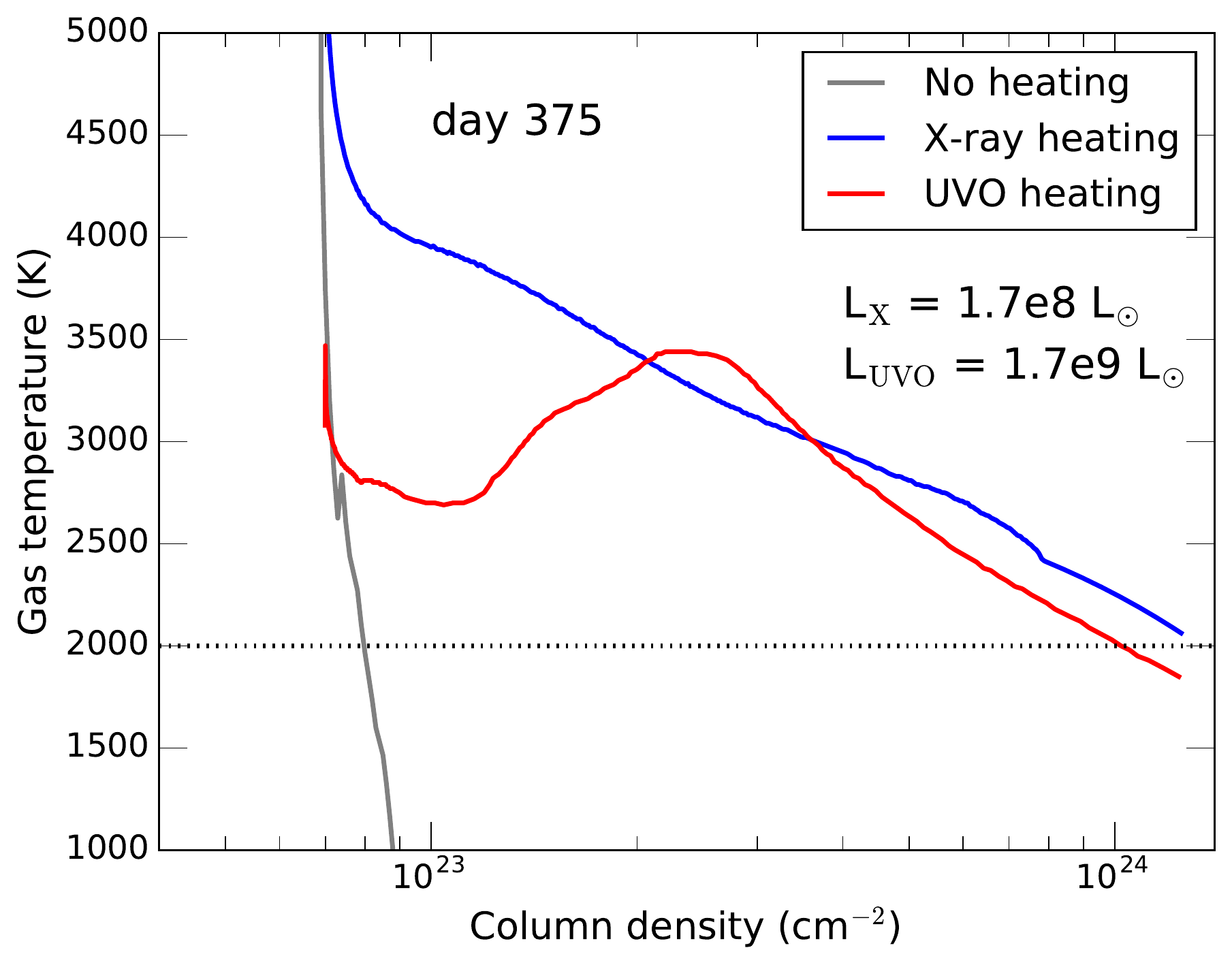}
\includegraphics[width=3.3in]{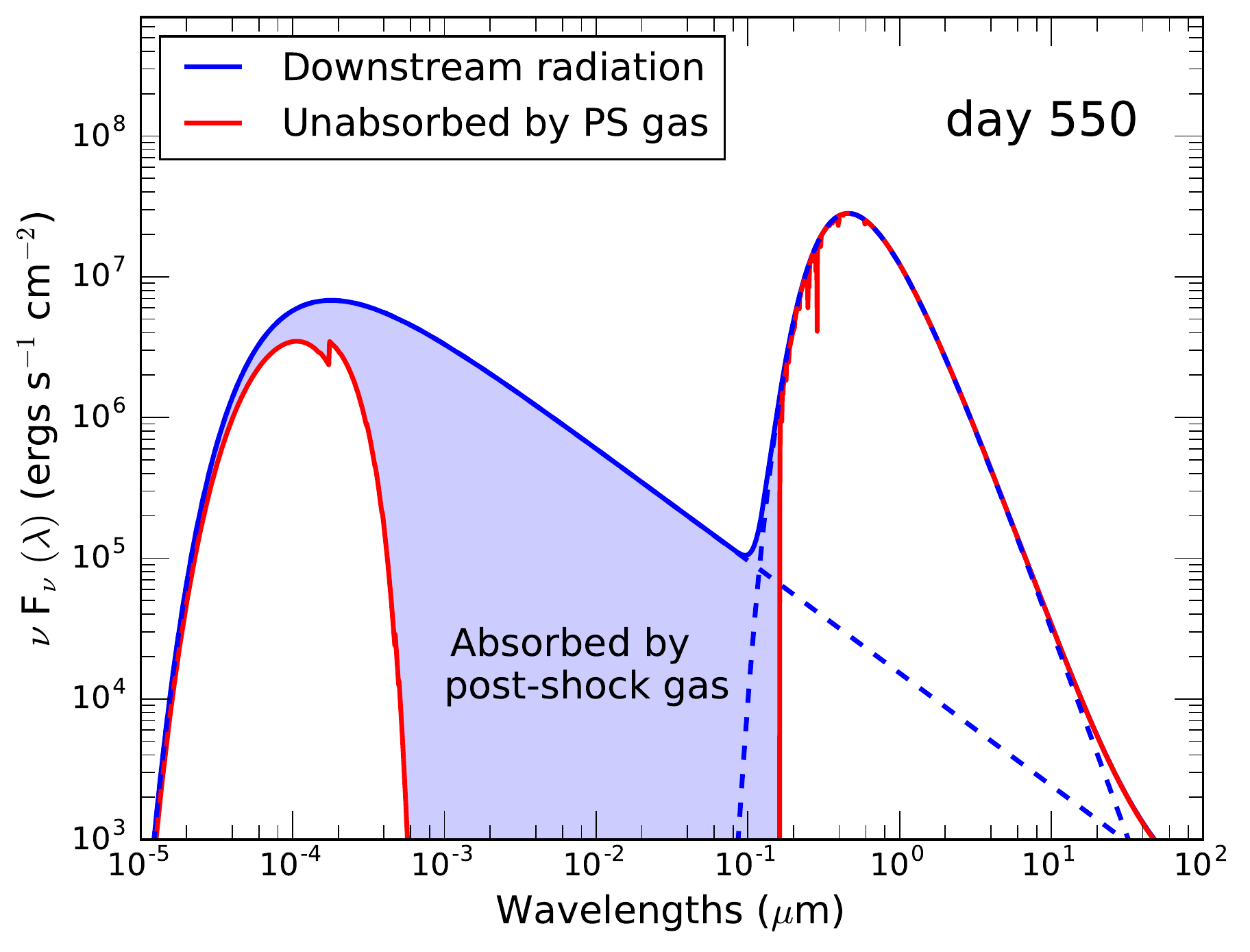}
\includegraphics[width=3.3in]{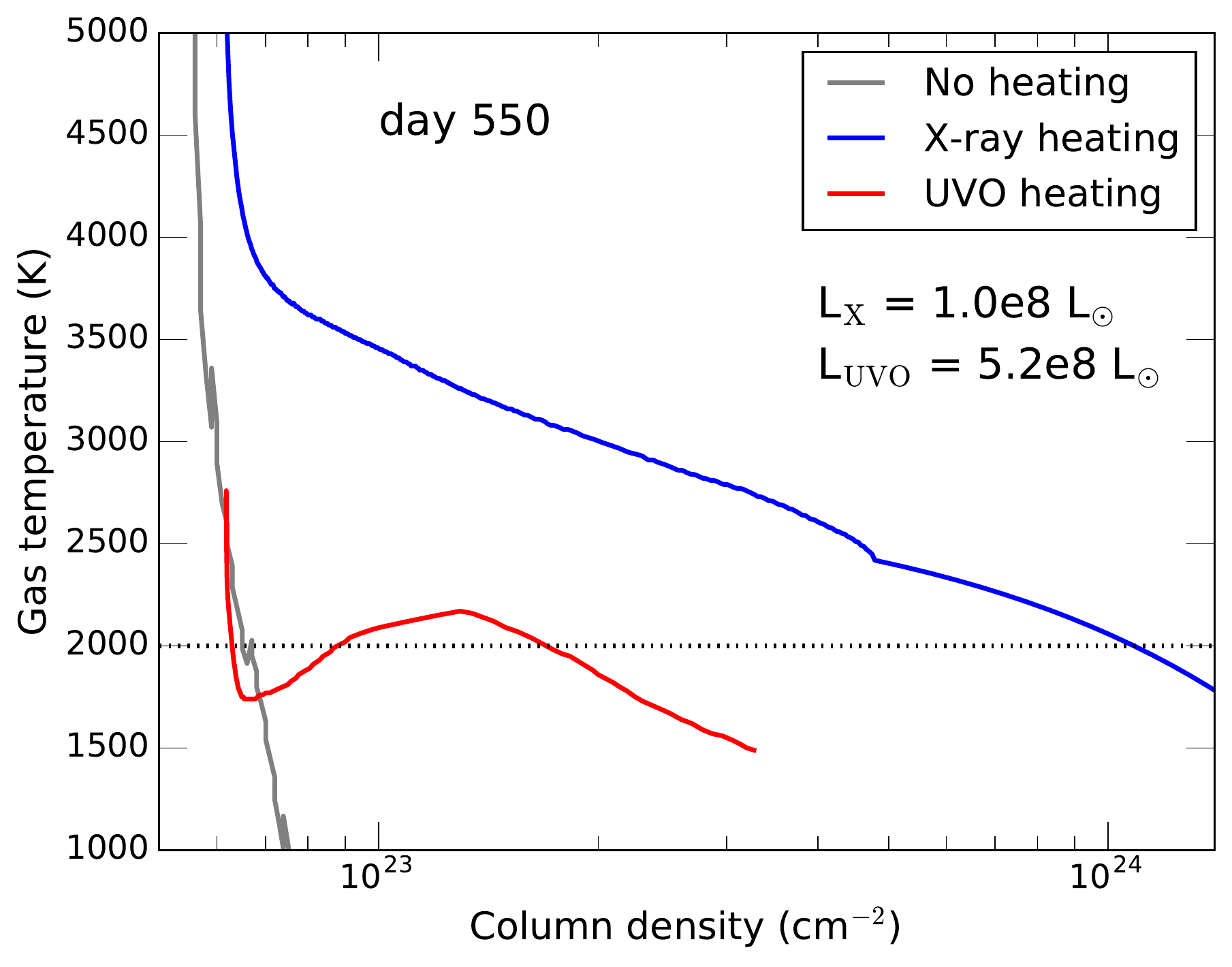}
\caption{\label{fig:uvobrem} \footnotesize{The figures in the left-panel represent the spectra of the downstream radiation in terms of luminosity per unit shock-area for three post-explosion epochs (day 200, 375, 550). The flux of downstream radiation, $F_{\nu}$($\lambda$), is calculated as $F$ = $F\rm_{obs}$($D^2/R_{sh}^2$), where $F\rm_{obs}$ is the observed flux and $D$ is the distance of the supernova from the earth. The total downstream radiation (solid blue line), generated at the shock-front, consists of the X-ray and the UVO component, shown in blue dashed lines. The spectra of the unabsorbed radiation (in red), derived using code CLOUDY, is a result of the attenuation and absorption by the column of dense dust-free post-shock (PS) gas. The radiation absorbed by the column ($N\rm _{H}$) of post-shock gas at a given time is presented as the shaded regions in the figures. The right-panel shows the resultant gas temperature, at the same epochs, assuming (a) no downstream radiation (b) downstream radiation from X-ray source, (c) downstream radiation from UV-optical source. At early times, the gas temperature is found to be controlled by the bolometric luminosity. Later, the bolometric luminosity declines fast, and the gas temperature is mainly controlled by the residual X-radiation.  }} 
\end{figure*}
  

Equations \ref{dn} and \ref{tbyt0} do not include any heating of the shocked gas by the downstream radiation from the shocked gas. In case of such high density medium, the shocked gas tends to cool rapidly within a period of few days. The shock-front, which is spatially close to the cooling gas, generates a strong flux of ionizing X-rays and UVO radiation. This radiation flows downstream and acts as a continuous heating source to the post-shock gas.
When the gas temperature is high ($>$ 10$^{4}$~K) the gas is already fully ionized and so the impact of the downstream ionizing radiation is minimal. However, at temperatures below 10$^4$~K, hydrogen starts to get partially neutral forming traces of H$_2$ molecules in the gas. Currently, there are no accurate tools available that can handle the hydrodynamics and radiative transfer through such high density semi-ionized gas to study all the relevant cooling mechanisms. We therefore calculated the post-shock temperatures using the following post-processing procedure.

The densities are sufficiently high ($>$ 10$^{12}$ cm$^{-3}$) in the post-shock cooling gas and the environment is expected to reach steady state reasonably fast. On that basis, a steady state radiation transfer model is a fair and acceptable approximation. Therefore, we use the spectral synthesis code CLOUDY \citep{fer98} to study the radiation transport through the post-shock gas at temperatures lower than 10$^{4}$~K. Even though CLOUDY is a static code, several snapshot cases were modeled in order to account for the time evolution of the post-shock gas.

The luminosity in the UVO, X-ray and the IR bands, generated by the forward shock, are shown in Figure \ref{fig:allobs}. The UVO emission is generally found to be about one order of magnitude higher than the total X-ray luminosity \citep{cha15}, and both decline steadily after day 310. Since the line intensities do not play any role in our calculations, we approximated the X-ray flux by a 10$^8$~K free-free Bremsstrahlung spectrum. Similarly, the UVO spectra can also be presented as a blackbody with appropriate temperature, as shown by \cite{fra14} and \cite{gal14}. Using this simplification, the spectra of the initial SED was constructed as an input to code CLOUDY, presented in Figure~\ref{fig:uvobrem}. 

Analytically, the energy loss term in equation (\ref{dn}) is replaced as $H$(T) =  $\eta(T)$ - $S$(T), where $S$(T) is the additional heating in units of ergs cm$^{-3}$ s$^{-1}$ accounting for the downstream radiation from the shock front. Owing to the presence of the continuous heating source, the `dn/dt' term decreases significantly, and hence n tends to remain almost unaltered at low temperatures when   $\eta(T)$ becomes comparable to  $S$(T).

The observed UVO spectra at all epochs peak around $\sim$ 0.3 $\mu$m which corresponds to a temperature of around 8000~K as also confirmed by \cite{fra14, gal14} and shown in Figure \ref{fig:uvobrem}. Therefore most the radiation in the optical band comprises of non-ionizing photons with energies lower than 13.6 ev. On the other hand, even though the luminosity of the X-rays are lower, the ionizing soft X-ray photons between wavelengths 10$^{-3}$ and 0.1 $\mu$m contribute significantly in the heating of the gas. Figure \ref{fig:uvobrem} (left-panel) shows the spectra that is generated by the CSM-shock interaction and the spectra of the downstream radiation after passage through the column of dust-free gas. The shaded region (in blue) in the figure represents the total radiation that is absorbed by the gas per unit time. So the remainder of the downstream radiation is responsible in heating the dust grains when they are formed. 

Figure \ref{fig:uvobrem} (right-panel) shows the temperatures of the post-shock gas as a function of column density, under the circumstances when (a) the downstream radiation is ignored (b) only heating by the X-rays are considered (c) only heating the UVO radiation is considered. When the downstream radiation is not taken into account, it is evident that the gas cools rapidly in absence of any heating source. However, in reality due to the heating by the downstream radiation the post-shock gas remains warm for a prolonged period. 

\begin{figure}[t]
\centering
\includegraphics[width=3.3in]{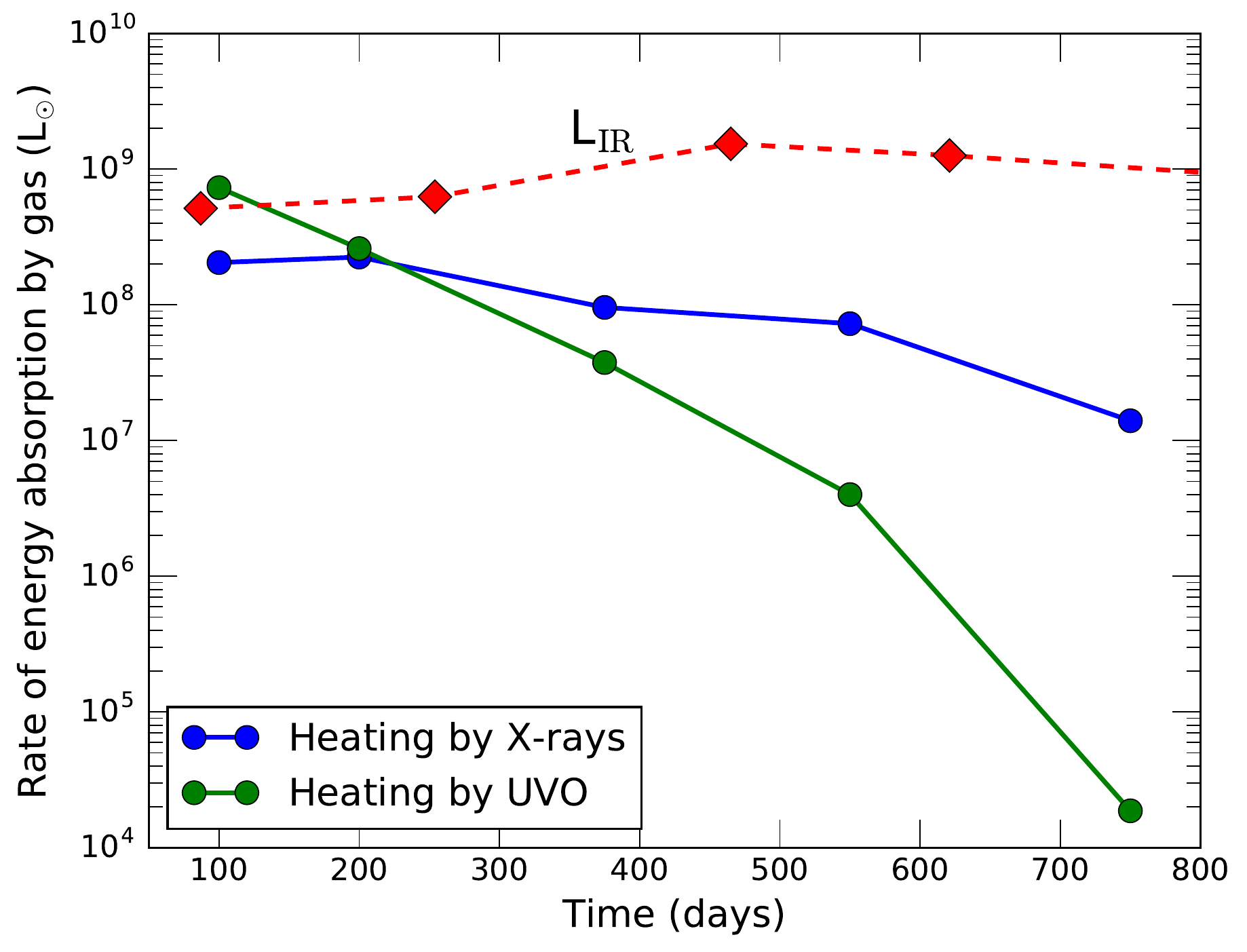}
\caption{\label{fig:enabs} \footnotesize{The rate of energy absorbed by the post-shock gas is presented, when heated by X-rays and UVO from the forward shock. Initially the heating of the gas is dominated by the UVO and later the X-rays become more significant. Further, in the figure the energy absorption rate of the gas is compared to the observed IR luminosity from the dust. $L\rm_{IR}$} is found to be larger than the energy absorbed the gas at a given time. Therefore, the exchange of heat between the gas and the dust cannot be the only source of energy that is heating the dust. } 
\end{figure}

As shown in Figure \ref{fig:uvobrem} (right-panel), at early times $\sim$ day 200, the entire column of post-shock gas remain hotter than the condensible temperatures, which is around 2000~K. Therefore formation of dust grains is unlikely in these regimes. After day 375, the luminosity of the shocked gas decreases rapidly and the post-shock gas is able to cool below 2000~K, marking the onset of dust formation. The heating of the gas is initially dominated by the UVO emission, and by the X-ray heating at later times. The rate of energy absorption by the gas due to heating by X-rays and UVO are compared in Figure \ref{fig:enabs}.

\begin{figure*}
\centering
\includegraphics[width=3.3in]{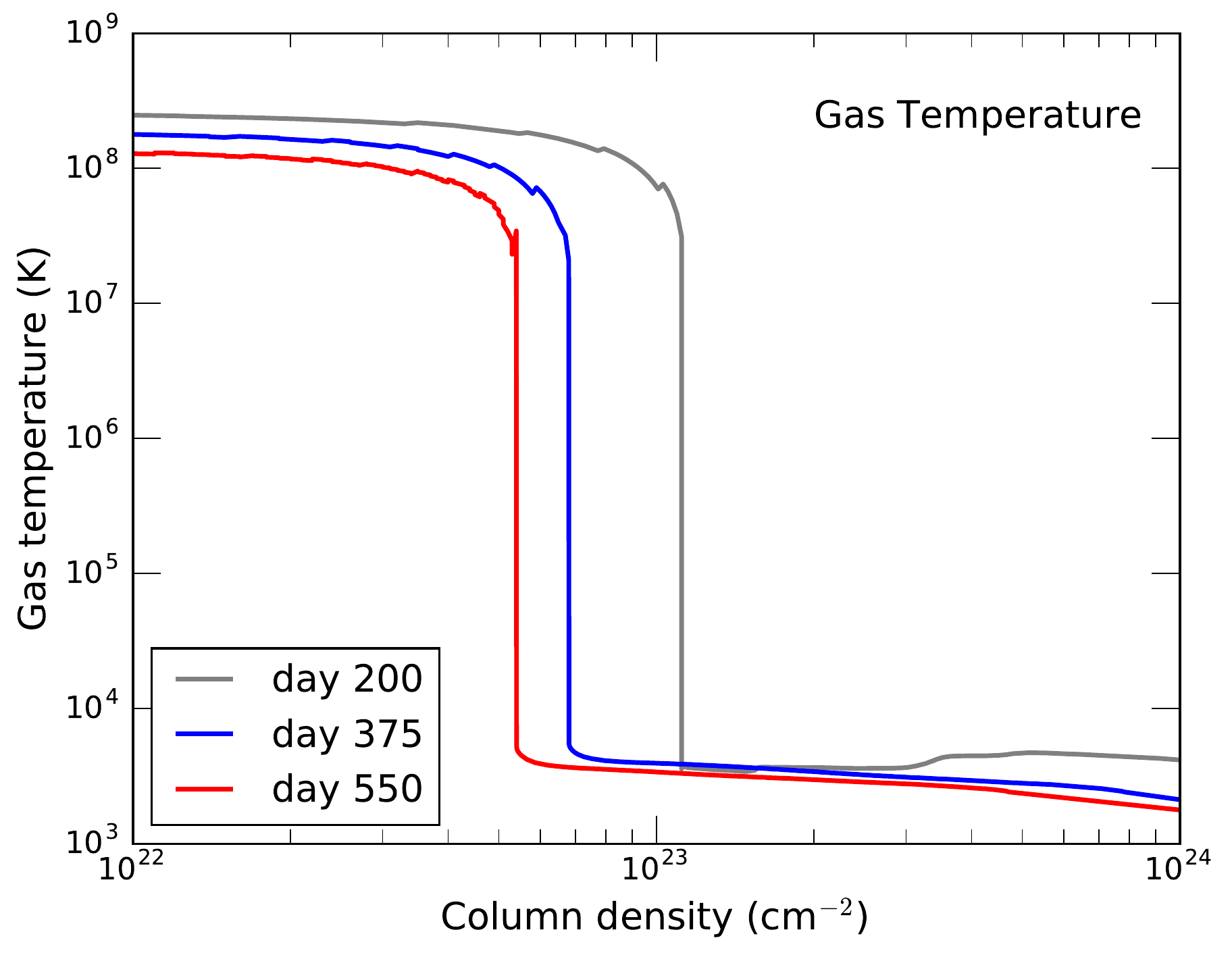}
\includegraphics[width=3.3in]{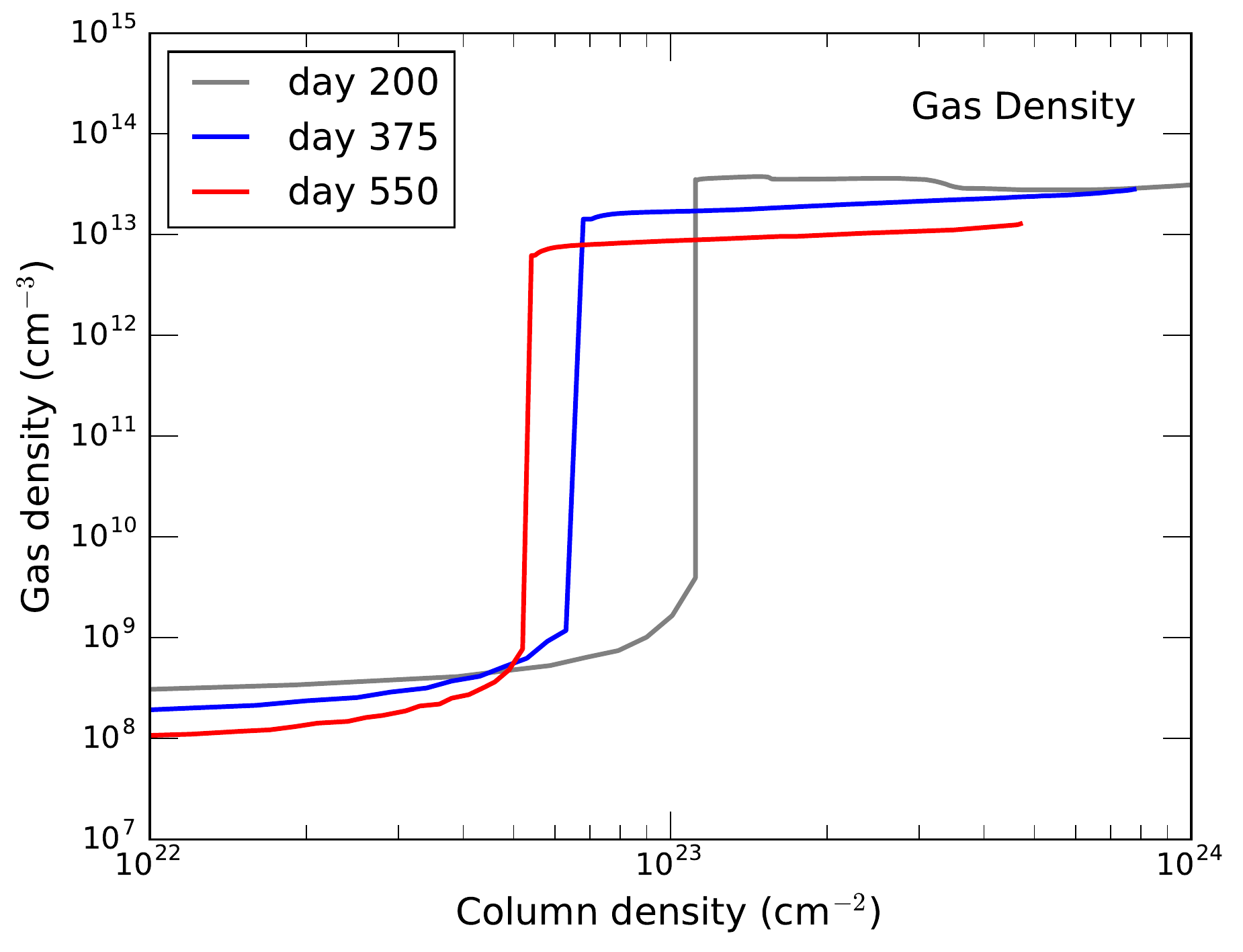}
\caption{\label{fig:tempdens} \footnotesize{The post-shock conditions of the CSM are presented in the following figures: (a) The profile of gas temperature as a function of post-shock column density for three different epochs (b) The gas density profile of the post-shock gas at three epochs. The temperature and density profiles are calculated taking into account the cooling of the shocked gas and the impact of the downstream radiation (X-rays and UVO) as a heating source.}} 
\end{figure*}

The complete temperature and density profiles of the post-shock gas in shown in Figure \ref{fig:tempdens}. At a given epoch the gas is fully ionized and temperatures are higher than 10$^6$~K up to a column density of a few times of 10$^{22}$ cm$^{-2}$ behind the SN-shock. Following that, the rest of the column in the post-shock gas are at much lower temperatures as shown in Figure \ref{fig:tempdens} (left-panel). The gas density in the post-shock gas, shown in Figure \ref{fig:tempdens} (right-panel), increases gradually from 10$^8$ cm$^{-3}$ with a steep rise at column density of $\sim$ 10$^{23}$ cm$^{-2}$ to reach a density between 10$^{13}$-10$^{14}$ cm$^{-3}$ following the $n$-$T$ relation (Figure~\ref{fig:tempdens}). 

Therefore, the post-shock gas dynamics lead to the formation of a reservoir of warm and dense gas in the post-shock shell, with a nearly steady temperatures ranging between 1000 and 3000 K. The shaded region on the cooling function in Figure \ref{fig:coolfunc} depicts the temperature zone of the warm dense shell where the heating and cooling rates tend to balance each other. Importantly, this is the region of interest for dust formation.

Calculations show that the gas temperature decreases with the amount of column density traversed by the radiation field. We define $N\rm_{min}$ as the minimum column density the radiation must traverse before the gas temperature drops down to 2,000 K. 
 $N\rm_{min}$ is proportional to the luminosity of the source. Therefore, as the luminosity of the SN-shock decreases with time, $N\rm_{min}$ also simultaneously decreases. 

\begin{figure}[t]
\centering
\includegraphics[width=3.3in]{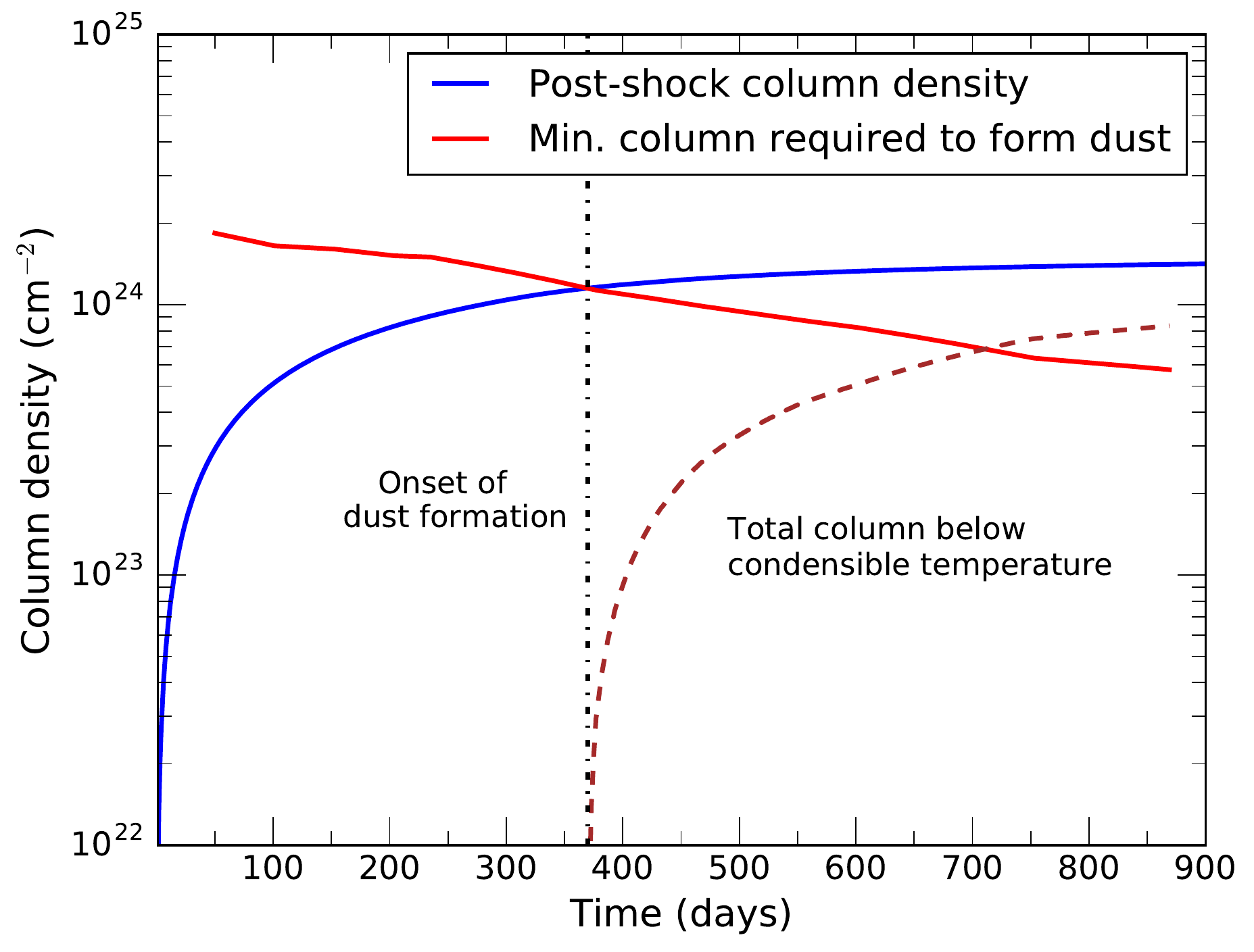}
\caption{\label{fig:cd_compare} \footnotesize{The minimum column density needed to be traversed by the shock, $N\rm_{min}$ (in red),  before the gas temperature drops below the condensation temperature is plotted versus time. Also shown is N$_H$(R$_s$) the total post-shock column density versus time (in blue). Dust can form in the post-shock gas when N$_H$(R$_s$) $>$ $N\rm_{min}$. }} 
\end{figure}

$N\rm_{min}$ was calculated using CLOUDY by studying the radiation transport through the post-shock gas at various epochs. Figure \ref{fig:cd_compare} compares $N\rm_{min}$ to the post-shock column density $N\rm_{H}$($R\rm_{sh}$) as a function of time. Importantly, the boundary condition for dust formation in the post-shock shell is therefore defined by $N\rm_{H}$($R\rm_{sh}$) $>$ $N\rm_{min}$, which occurs around $\sim$ day 380 as shown in the figure. The dashed line in Figure \ref{fig:cd_compare} represents the amount of gas in column density, that is below condensible temperature at a given time. 

\begin{figure}[t]
\centering
\includegraphics[width=3.3in]{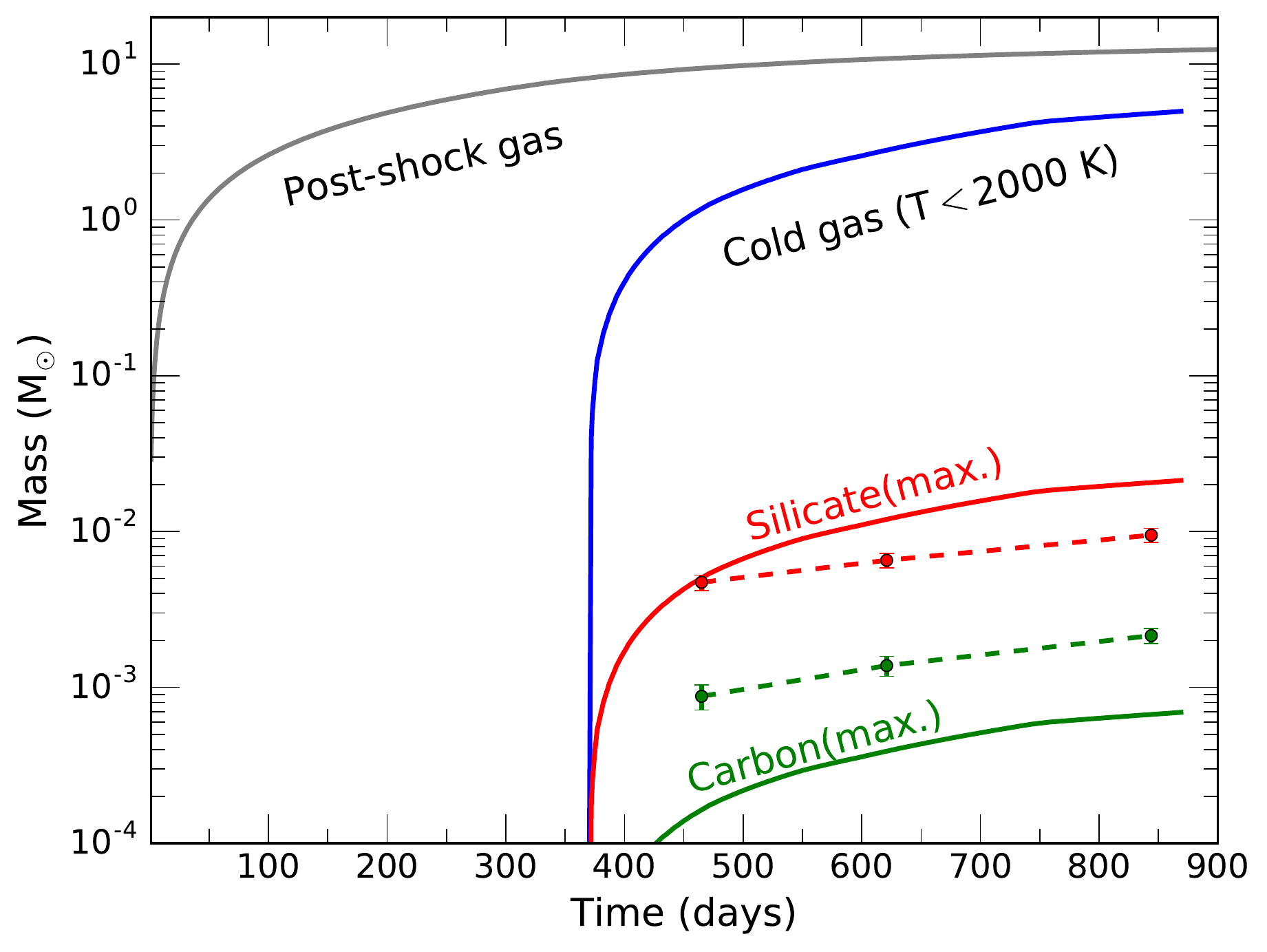}
\caption{\label{fig:coolingmass} \footnotesize{The figure shows the total mass of the post-shock gas as a function of time. Also shown is the total mass of condensible elements at temperature $T <2000 $K. From the total mass of condensible metals present in the gas, the extreme upper limits on the silicate and carbon dust masses have been derived. Also the inferred masses of silicate and carbon, obtained from fitting the IR data, are also shown in the figure as dashed lines. The CNO abundance ratios are taken from \cite{fra14}. }} 
\end{figure}

Figure \ref{fig:coolingmass} presents the total mass of the post-shock gas and the mass of the gas that is below the condensible temperatures. As evident from the figure, from day 380 onwards, a fraction of the post-shock gas becomes cold cold enough to sustain the formation of stable molecules and dust grains. Assuming solar abundance of elements in the CSM, there are sufficient atomic species that can account for the mass of silicates that is comparable to the inferred silicate mass from observations. However, owing to the carbon-poor nature of the CSM \citep{fra14}, there is not enough condensible atomic carbon that can account for the inferred carbon dust masses. This brings forward the possibility that the total dust mass can constitute of a fraction of both, silicates and carbon.

A shell of warm gas is known to cool adiabatically as it expands freely. However in this case, adiabatic cooling and expansion have minimal impact on the evolution of the post-shock gas because of the following: (a) the presence of a continuous source of energy at close proximity (b) due to the large initial radius and very small thickness of the dense shell ($\Delta R$), $\Delta R/R$ is small at all times ($\sim$ 10$^{-6}$ at day 400). So we can safely ignore further expansion of the post-shock shell due to adiabatic effects within the timescales of a few hundred days.


\section{Conditions for dust formation}
\label{dustcondtions}

In this section, we shall discuss if and how dust can form and grow chemically in such environments when the suitable gas phase conditions prevail.  

Formation of cosmic dust in such dynamic environments is a kinetic process that is controlled by simultaneous phases of nucleation and condensation \citep{sar13}. The gas phase nucleation leads to the formation of molecules and small clusters, whereas the stable gas phase clusters grow in size through coagulation and accretion \citep{sar15}. Both the phases require moderately high gas temperatures and high gas densities. The warm dense shell formed in the post-shock gas provides ideal conditions that are conducive to dust and molecule formation after day $\sim$ 380, when the average gas temperature of the shell cools down below 2000~K and the gas densities are larger than 10$^{13}$ cm$^{-3}$. 

At high temperatures, the strong flux of ionizing photons and free electrons have destructive effects on any stable molecule formed in the medium. Molecule formation in the gas phase can transpire at around 3000~K when the ionization fraction is relatively low. However newly formed dust grains with a temperature above 2000~K are most likely to sublimate instantaneously. In the context of this study, we do not implement a detail chemical kinetic scheme to study the gas phase chemistry. Instead, an empirical approach was adopted mainly focusing on the physical conditions that favor or impede dust formation in such environments and their manifestations on the SN light curve. 

\begin{table}
\centering
\caption{The timescales for different chemical processes that transpire simultaneously to form a 0.01 $\mu$m dust grain}
\label{timescales} 
\begin{tabular}{l l c}
\hline \hline
Process & Product & Timescale (s)\tablenotemark{1}  \\
\hline
 Nucleation & MgFeSiO$_4$ & 0.13 \\
 Nucleation  & C$_n$ & 17 \\
 Accretion & 0.01 $\mu$m grain & 380 \\
 Coagulation & 0.01 $\mu$m grain & 10$^5$ \\
 \hline
\end{tabular}
\tablenotetext{1}{physical conditions in the text}
\end{table}

 Table \ref{timescales} lists the timescales for important chemical processes that lead to the nucleation and condensation of silicate and carbon dust. 

A series of chemical reactions involving Si, O, H, Mg and Fe leads to the formation of the seed molecular clusters that eventually condense to form astronomical silicates, as described by \cite{gou12} and \cite{sar13}. Similarly synthesis of amorphous carbon dust proceeds through the gas phase nucleation of C chains and rings \citep{clay99, cherchneff10, sar15}. The rate of nucleation presented in Table \ref{timescales} corresponds to the slowest process in the series of reactions within the first order of approximation. The temperature and densities are taken as the ones derived in the model, shown in Figure \ref{fig:tempdens}. 



The timescale for accretion \citep{dwe11} on a dust particle given by, 
\begin{equation}
\label{accretionrate}
\tau_{ac} \sim \frac{2 \times 10^{18}}{\alpha \mathrm{_S}(T, T_d)} \ \rho_d \ \rm{ \Big(\frac{a_d}{\mu m}\Big)  \Big(\frac{n_{H_2}}{cm^{-3}} \Big)^{-1}  \Big(\frac{T}{K}} \Big)^{-0.5} 
\end{equation}
where $\rho_d$ is the density of the accreting dust particle, a$_d$ is the grain radius in microns, n$\rm_{H_2}$ is the gas density and the $\alpha \mathrm{_S}$ is the sticking coefficient. 

The time scale to grow a 0.01 $\mu$m dust grain, assuming a number density of 10$^{13}$ cm$^{-3}$ a sticking coefficient of 0.5 at 1000 K is $\sim$ 380 s.  
 
In absence of available metals to accrete, the condensation process is dominated by coagulation between small dust particles. In this density regimes, where the mean free path of particles are much greater than the grain radii, the coagulation between to particles \citep{mj05, sar15}, say $i$ and $j$, is dominated by Brownian diffusion with a rate given by, 

\begin{equation}
\label{coagrate}
k_{i,j} \sim \ \pi (a_i + a_j)^2 (\bar{v_i}^2 + \bar{v_j}^2)^{1/2} \\
\end{equation}
where $v_i$ is the thermal velocity of a particle $i$. 
For coagulation between two 10 \AA\ dust grains of silicate at 1000 K, the rate is one collision is $\sim$ 9.4 $\times$ 10$^{-10}$ s$^{-1}$, 


Assuming a dust-to-gas mass ratio of $\sim$ 10$^{-3}$, and the gas number density of $\sim$ 10$^{13}$ cm$^{-3}$, the rate of coagulation is $\sim$ 9.4 $\times$ 10$^{-3}$ s$^{-1}$. In order to form a 0.01 $\mu$m grain approximately a thousand of 10 \AA\ grains are required to coagulate between each other. Hence the timescale is $\sim$ 10$^5$ s, which is a little over one day. 

Therefore, given suitable conditions, the dust formation timescale in the post-shock gas is only about one day. As suitable conditions prevail post day 380, the post-shock dense shell in SN~2010jl becomes a site of efficient dust synthesis.

 \begin{figure}[t]
\centering
\includegraphics[width=3.5in]{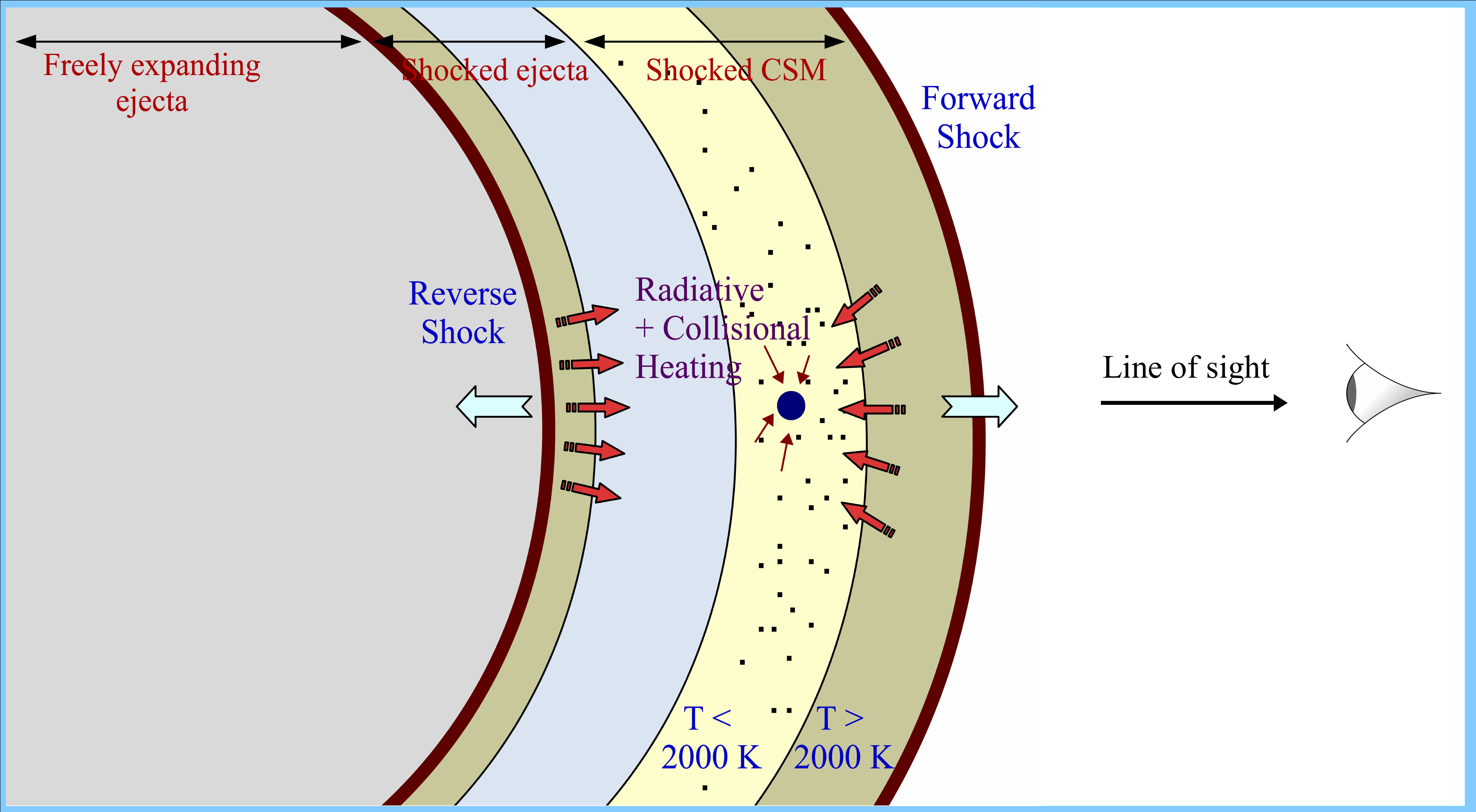}
\caption{\label{fig:heating_geo} \footnotesize{The cartoon presents the ejecta-CSM morphology at a given epoch when dust formation commences in the post-shock shell. A hypothetical grain of dust is shown to be present in the dense shell defined by gas temperatures $T_g < 2000$~K. The dust grain is subjected to radiative heating by the forward and the reverse shock. Also it is heated collisionally by the ambient gas. The arrow-pointers in red represent the radiation directed towards the dust grain. The forward shock moves towards larger radii, while the reverse shock moves inwards in the frame of the CSM. }} 
\end{figure}

\begin{figure}[t]
\centering
\includegraphics[width=3.3in]{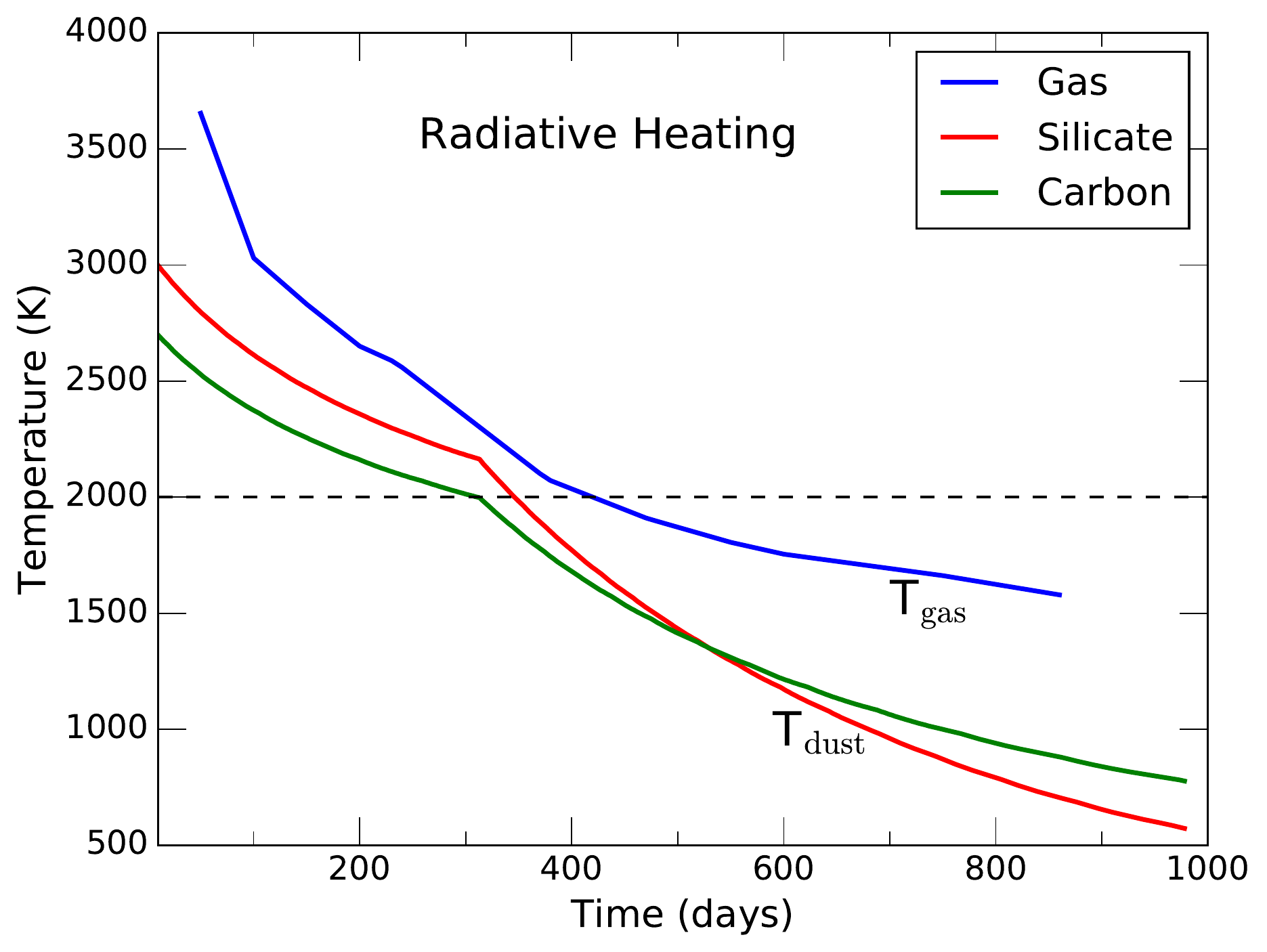}
\caption{\label{fig:dustT} \footnotesize{The figure presents the temperature of the gas and the dust in the post-shock gas, as a result of the heating by the downstream radiation only. The dotted line at T = 2000~K represents the condensible regimes. Hence, dust formation is unfeasible before day 380 because the gas, heated by the X-ray and UVO, is not cold enough and also the radiation from the forward shock evaporates any dust grain that forms in the gas. The temperatures shown in the figure does not include the impact of collisional heat exchange between the gas and the dust. }} 
\end{figure}
  

\begin{figure*}
\centering
\includegraphics[width=3.3in]{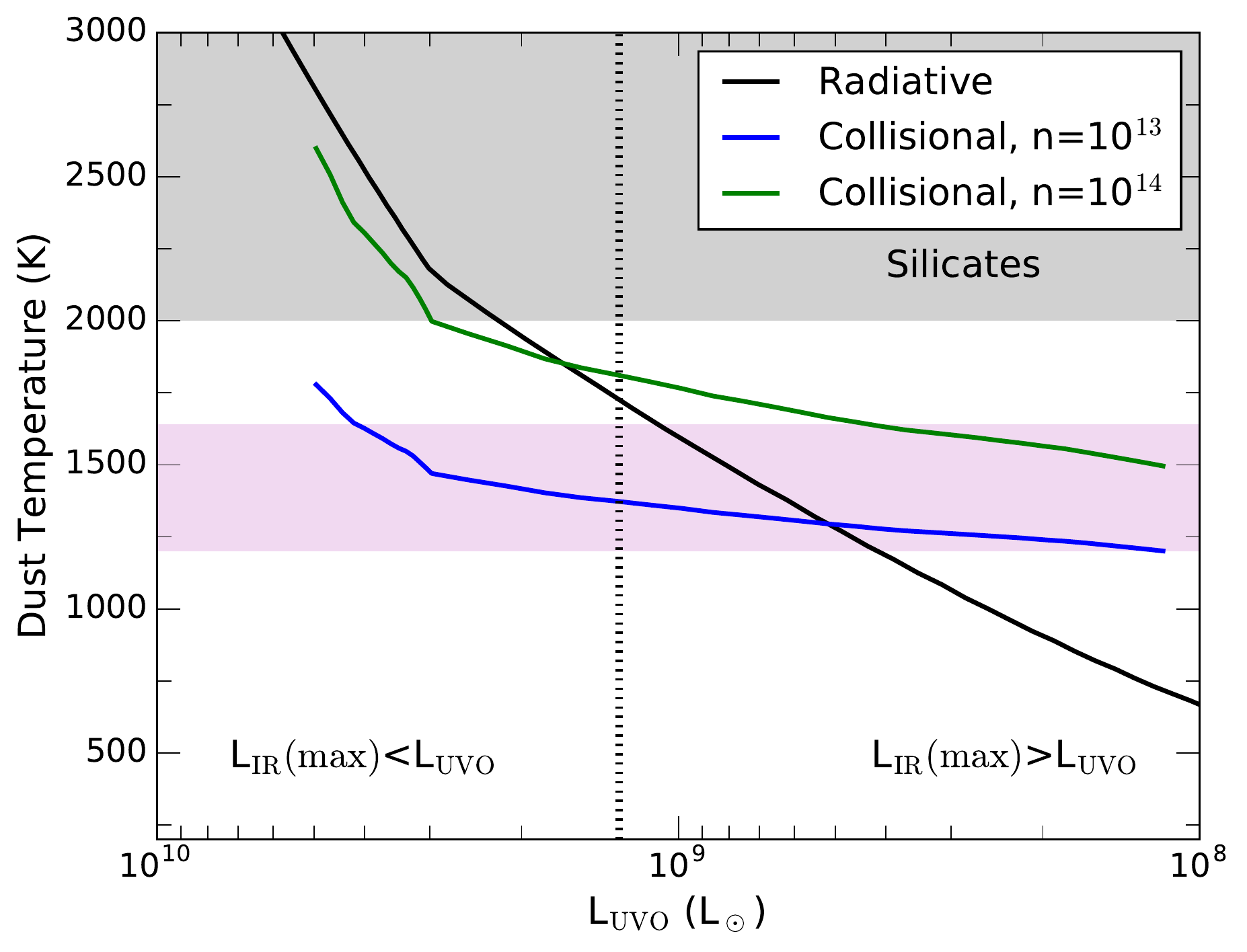}
\includegraphics[width=3.3in]{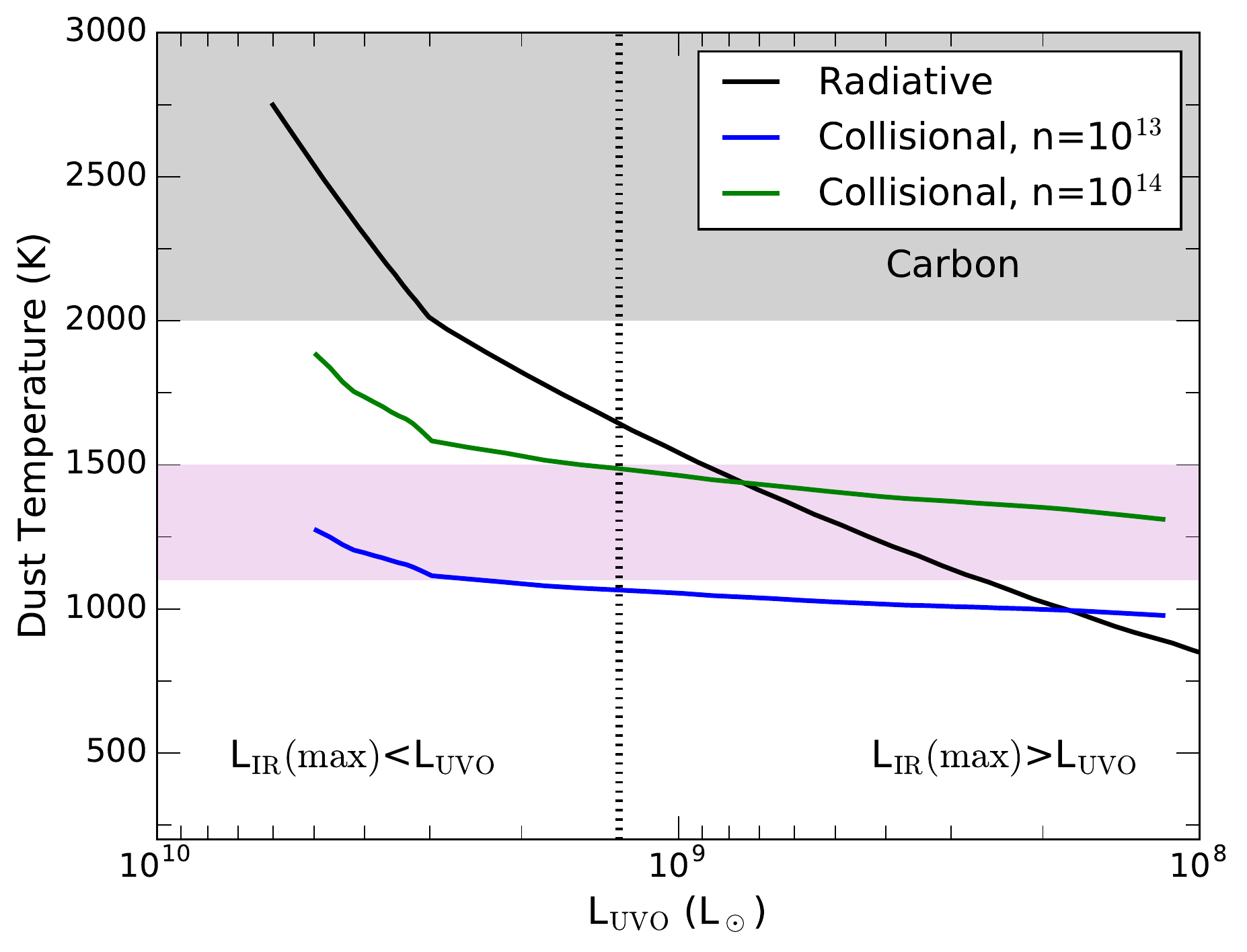}
\includegraphics[width=3.3in]{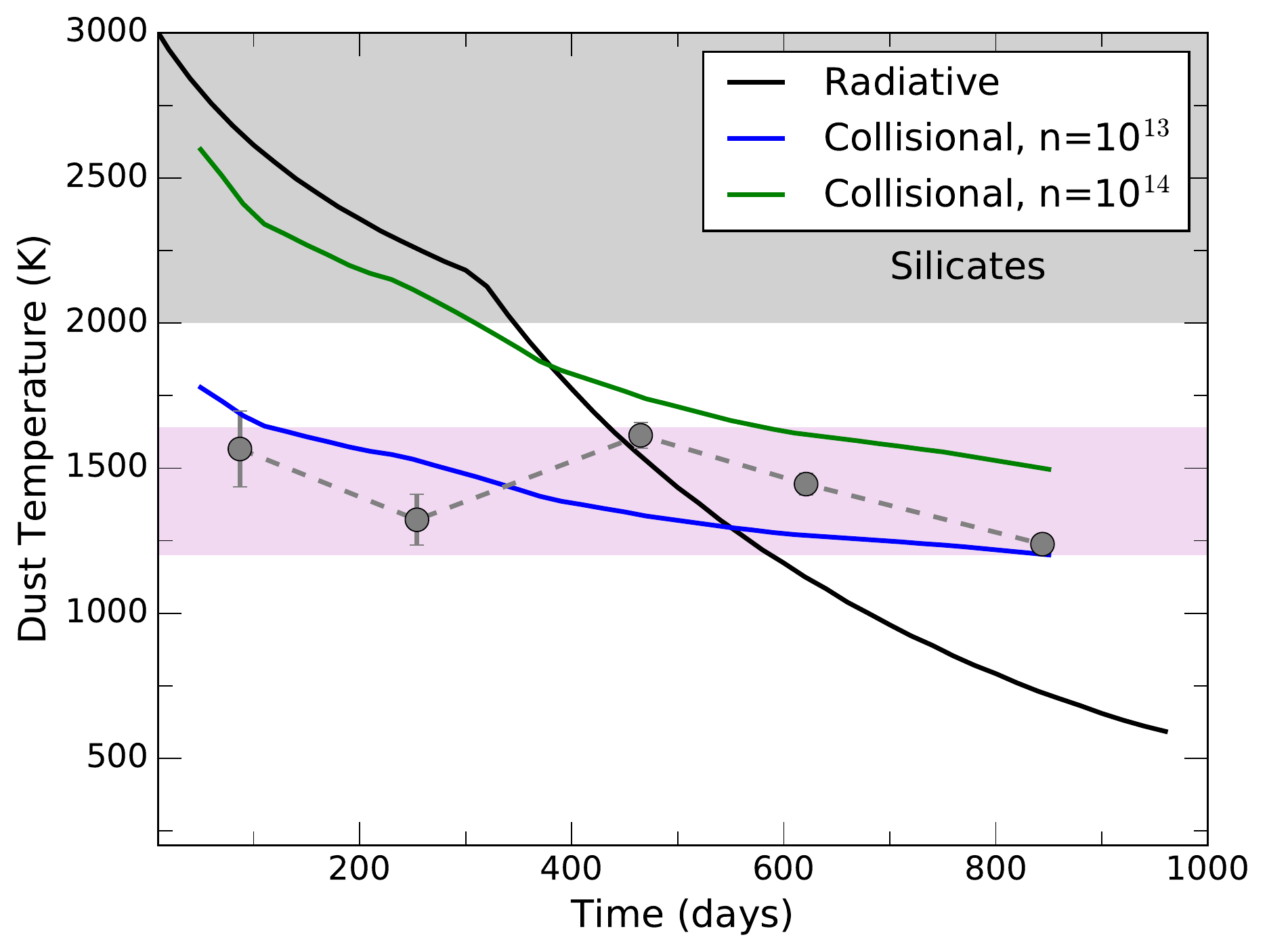}
\includegraphics[width=3.3in]{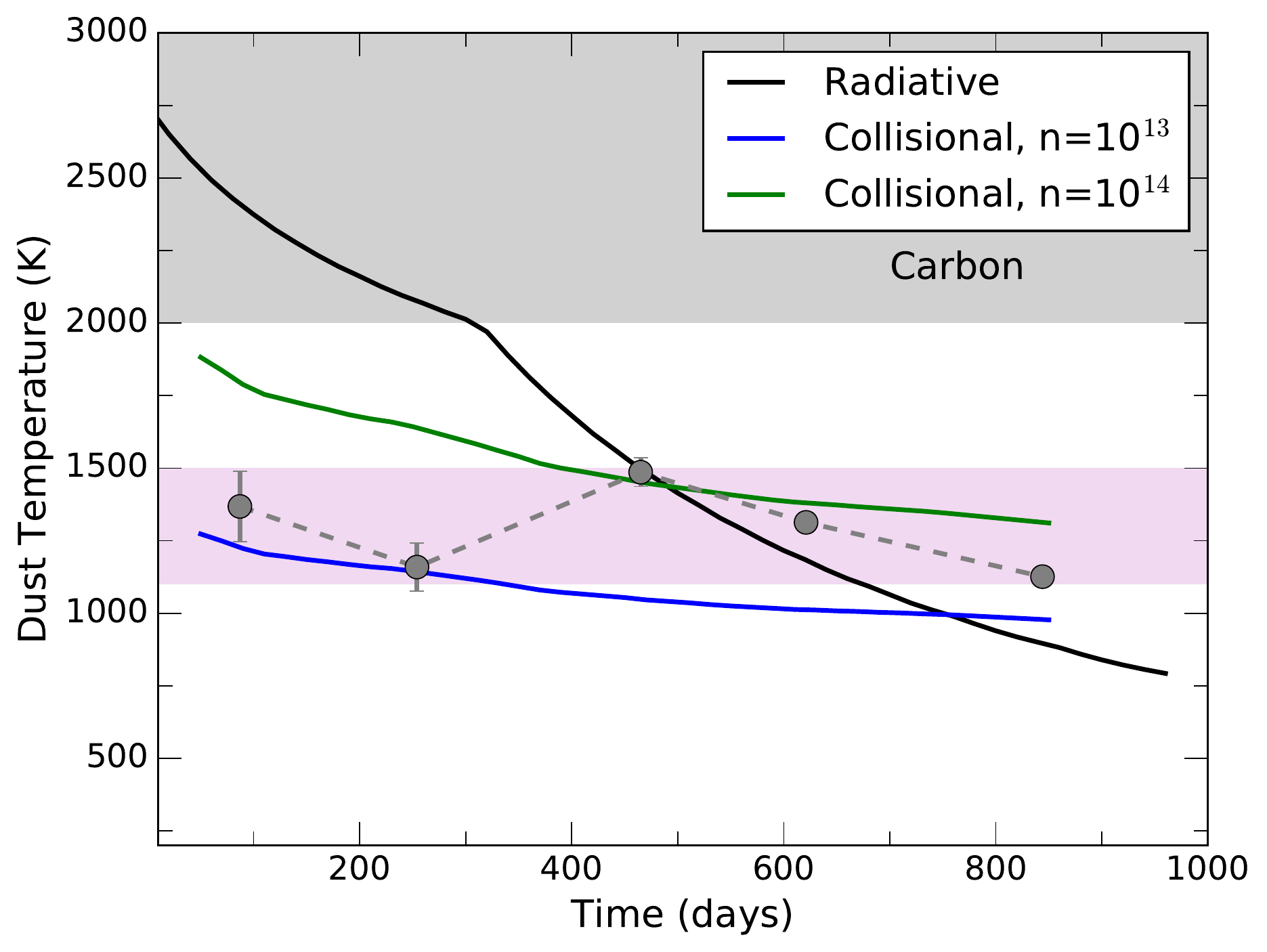}
\caption{\label{fig:combinedT} \footnotesize{The figure shows the temperatures for an astronomical silicate (Left-panel) and an amorphous carbon (Right-panel) dust grain as a function of forward shock luminosity and the corresponding post-explosion time. The impact of radiative and collisional heating on the dust temperature is calculated separately. The dust temperature at a give time $t$ due to collisional heating by the gas at $T_g$($t$) is estimated for two gas densities, 10$^{13}$ and 10$^{14}$ cm$^{-3}$. The band marked in pink represents the observed dust temperature region, whereas the band in gray shows the region above evaporation temperature. The top panel presents the dust temperatures as functions of forward-shock bolometric luminosity and the bottom panel presents the same as function of post-explosion time in days. The best-fit dust temperatures to the observed IR spectra at various epochs are also shown in the bottom panel. }} 
\end{figure*}
  

\begin{figure}[t]
\centering
\includegraphics[width=3.3in]{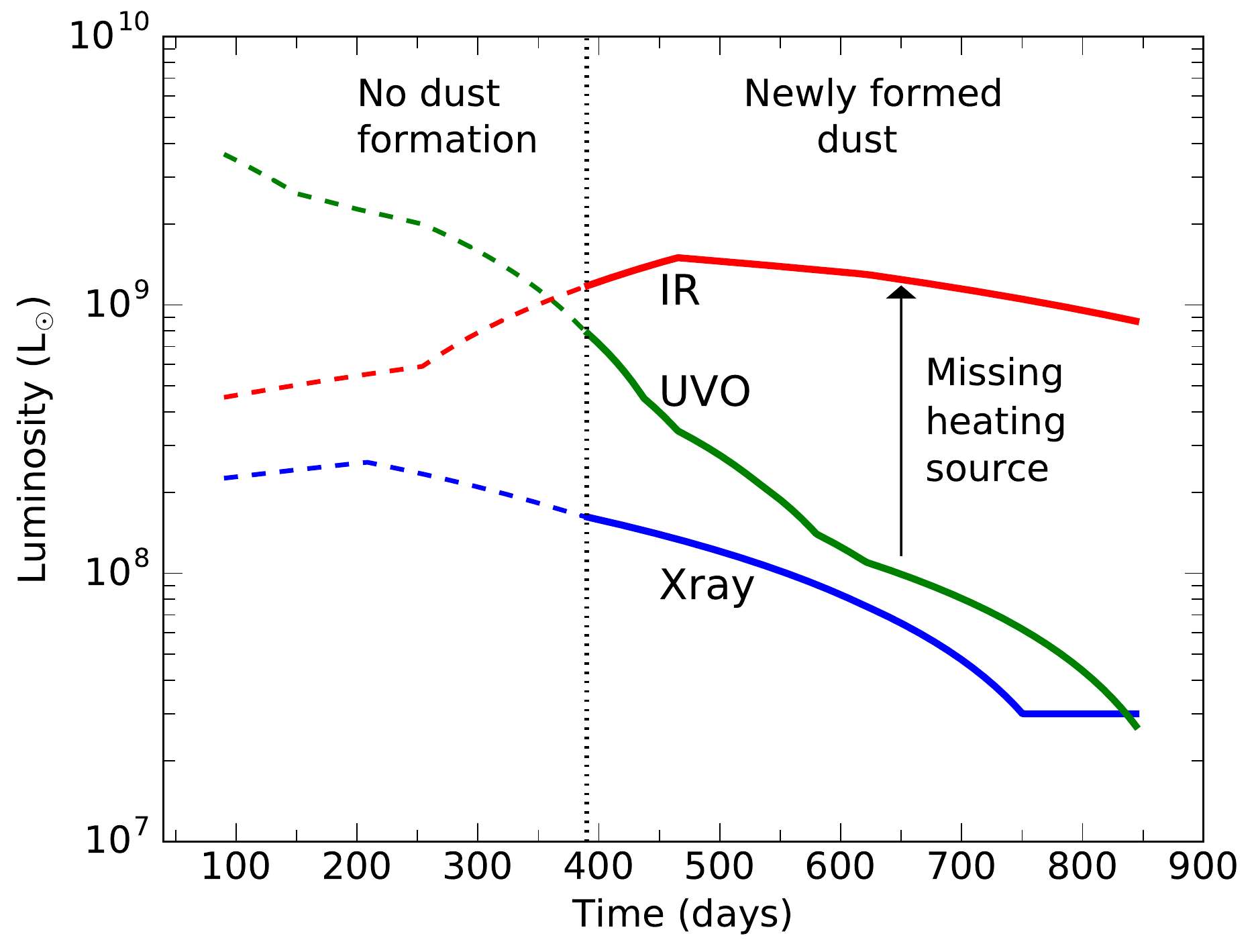}
\caption{\label{fig:dustlum} \footnotesize{The observed luminosities from SN~2010jl are presented in lights of its history of dust formation. The dotted lines before day 380 indicate the timescales where the IR emission is not from newly formed dust and rather an IR echo. Later than that, when new dust forms in the post-shock gas, the IR luminosity is always found to be larger than the X-rays and UVO from the forward shock. The upward arrow indicates the minimum luminosity of the reverse shock, which is the missing heating source, as it has no observational evidence. In all likelihood, the newly formed dust is thick to optical wavelengths and so it is efficiently scattering the incident radiation from the reverse shock that is interior to the dust forming shell. }} 
\end{figure}
  

\section{Heating of the dust grains}
\label{heating}

The preceding sections have confirmed that dust formation in the post-shock gas is feasible after $\sim$ 380 days post-explosion and the newly formed dust has sufficient time to grow in size. We shall now inspect all the possible sources of heating the dust grains, in order to account for the observed temperatures and IR luminosities.

Figure \ref{fig:heating_geo} is a schematic diagram that illustrates the heating sources of a hypothetical dust grain present in the dense shell behind the shock. 

A dust grain is radiatively heated by the ionizing photons those are traveling downstream from the SN-shock. Furthermore, the grain will also be heated by collision with the ambient warm gas in the dense shell. In addition, the dust grains are also subjected to heating by the reverse shock. The dusty shell being interior to the forward shock, the luminosity generated by the forward shock-CSM interaction will not be obscured along the line of sight by the newly formed dust.

The downstream radiation from the forward shock gets gradually attenuated while passing through the post-shock gas. The soft X-rays and UV-optical is entirely attenuated up to the Lyman-alpha continuum edge while passing though the column of post-shock gas, as shown on Figure~\ref{fig:uvobrem} (left-panel). The optical radiation with a peak at 8000~K comprises of mostly non-ionizing photons which remains unabsorbed by the gas. They act as the primary heating source to the dust.


Important to note, even though we can assume the incident spectra to resemble a blackbody, it is purely a mathematical substitution to the real physical picture. In reality, the gas around the forward shock that generates the downstream radiation is dilute (optically thin). The dilution coefficient is given by $f\rm _d$ =  $R\rm_{sh}$/$R\rm _{UVO}$, as explained in Section \ref{optnir}.

\subsection{Dust temperatures}

In this study, the temperature of a dust grain present inside the warm gas reservoir was calculated as functions of gas density, gas temperature and the flux of downstream radiation. 
The collisional heating rate of a dust grain have been addressed by \cite{dwe87} and \cite{hol79}. At high densities, the gas and dust temperatures are coupled. When $T\rm_d$ $>$ $T\rm_g$, collisions with the dust heats the gas, and vice versa. The the heating/cooling rate of a dust grain by the gas is given as,  

\begin{equation}
\frac{\mathrm{d}H}{\mathrm{d}t} = n \bar{v} f_{ed} \pi a^2 (2kT_g - 2kT_d)
\end{equation}
in terms of gas density ($n$), mean thermal velocity ($\bar{v}$), geometric cross-section ($\sigma_d = \pi a^2$) of a dust grain and the accommodation coefficient ($f\rm_{ed}$) of fractional energy deposition \citep{hol79}. 

The gas density in the shell varies between 10$^{13}$-10$^{14}$ cm$^{-3}$, whereas the average gas temperature of the reservoir gradually drop from 4000~K at early epochs to 1000~K in a couple of years. 

The first dust grains synthesized in the gas are of a few \AA\ in size. Subjected to the general interstellar radiation field, such small grains are known to undergo temperature fluctuations. However with a baseline temperature of $\sim$ 2000~K (gas temperature when the first grain condenses), the absorption of individual photons has a negligible effect on the temperature fluctuations. Furthermore, in spite of the high flux of photons, the cooling time is much smaller compared to the time between two photon absorption. So the impact of stochastic heating is negligible in this case. 

The rate of radiative heating was calculated by the equilibrium heating and cooling balance \citep{dwe87} of the grains in presence of the SN-forward shock, where the source luminosity is $L\rm _{UVO}$(t). Considering a 10~\AA\ particle as a prototype of the smallest dust grain, the equilibrium temperatures were calculated for astronomical silicates and amorphous carbon dust.
Figure \ref{fig:dustT} shows the dust temperatures for the two dust types when subjected to radiative heating by the SN-shock. The gas temperature at the back-end of the post-shock shell (region where the dust forms) is also presented as a function of time. The dashed-line at 2000~K indicates the maximum temperature temperature which can support the formation and survival of dust grains. 

Due to the high luminosity of the forward shock, the equilibrium temperature of the fiducial grain is always higher than 2000~K for both the dust types in the first $\sim$ 350 days. As the gas temperatures are also beyond condensible limits, there is no scope of dust formation or survival in this period. 

Figure~\ref{fig:combinedT} illustrates the dependence of dust temperature on the ambient gas conditions. The higher the gas density, the dust temperature is more closely bound to the gas temperature of the shell. It is evident from the figure that, for such high density mediums like the warm dense shell, the dust temperature is heavily influenced by the collisional heating by the surrounding gas. The band in pink marks the region of observed dust temperatures.  

The final temperature of the grains is a combined effect of radiative and collisional heating, and the energy exchange between the gas and dust. The figure shows that the dust temperatures fit well to the observed range of temperatures when the source luminosities are below $\sim$ 10$^9$ \Ls\ post-day 400. 

The dotted vertical line in Figure \ref{fig:combinedT} (top-panel) represents the maximum IR luminosity, which corresponds to $\sim$ day 450. Furthermore, it divides the plot into two regions defined by $L\rm _{IR}$(max) $>$  $L\rm _{UVO}$ and $L\rm _{IR}$(max) $<$  $L\rm _{UVO}$. The dust temperatures in the region given by $L\rm _{IR}$ $>$  $L\rm _{UVO}$ complies with observed band of temperatures. This supports the previous finding that dust formation in the post-shock gas is possible when the luminosity of the forward shock is below 10$^9$ \Ls\ after day 350. However, the dust being internal to the forward shock, the dust cannot obscure the UVO from the forward shock along the line of sight. Therefore, 
 the UVO luminosity cannot be the primary heating source of the dust, because in that case the IR luminosity could not have exceeded the luminosity of the heating source.

The efficiency and timespan of collisional heating of a dust grain by the warm gas in the shell is limited by the total amount of energy that is stored at a given time in the heat reservoir. Post day 300, the luminosity of the photosphere drops rapidly, hence the continuous supply of the energy to the heat-sink also drops. At the same time the IR luminosity increases steadily. So the reservoir of warm gas would simultaneously start to cool down at a much faster pace as it looses energy via IR emission from the newly formed dust.

The total energy stored in the warm gas reservoir at a given instant is approximately $\sim$ 3/2 $NkT$. Assuming a 10$^{24}$ cm$^{-2}$ column of gas, with density 10$^{13}$ cm$^{-3}$ and temperature 1500 K, the total number of particles is given by 2~$\times$~10$^{58}$ and the energy in the reservoir is therefore 4~$\times$~10$^{45}$ ergs. At an IR luminosity of 10$^9$ \Ls, the entire energy is radiated away in less than one day. 

The rate of energy absorbed by the gas, heated by the radiation from forward shock, is shown in Figure \ref{fig:enabs}, along with the IR luminosity of the dust. The $R\rm_{IR}$ is found to be larger than the combined heating rate of the gas. Therefore, the transfer of energy from the gas to the dust due to collision cannot be responsible for the large IR luminosity at late times.  


\subsection{Constraints on the reverse shock}
\label{rev_shock_const}

There UV-optical spectra from SN~2010jl corresponds to the forward shock only. As the reverse shock is internal to the dust forming region in the CSM, if the dust is optically thick it can entirely conceal the UV-optical luminosity from the reverse shock. That energy is in that case absorbed by the dust grains in the CSM and re-radiated at the IR wavelengths. 

The observations in the UV, optical or X-ray regimes do not provide any information about the properties of the reverse shock. So it can be inferred that the entire reverse shock spectra has been reprocessed by the gas and dust present in the CSM. 

Figure \ref{fig:dustlum} presents the IR luminosity in the regime when new dust forms in the dense post-shock gas. It is evident that at all times post day 380, $R\rm_{IR}$ $>$ ($R\rm_{UVO}$ + $R\rm_{Xray}$). Therefore, the minimum luminosity of the reverse shock required, in order to provided the extra heating, is given by, $R\rm_{rev}$(min) = $R\rm_{IR}$ - $R\rm_{UVO}$ - $R\rm_{Xray}$.

The hard X-rays having a low absorption cross-section is unlikely to be absorbed by the 10$^{24}$ cm$^{-2}$ column of gas in the CSM. So it provides an upper limit on the shock induced gas temperatures as it should not produce any hard X-ray photons. The following points summarizes the boundary conditions that can be inferred from this study. 
\begin{itemize}[noitemsep]
\item[--] The dust in the CSM must be thick to optical radiation. 
\item[--] The minimum $L\rm _{rev}$ is equal to $R\rm_{IR}$ - $R\rm_{UVO}$ - $R\rm_{Xray}$ at a given time.  
\item[--] The reverse shock velocity should be high enough such that the rate of mechanical energy generated is more than $L\rm _{rev}$. 
\item[--] The reverse shock velocity should be low enough such that it does not produce hard X-rays by its interaction with the ejecta. 
\end{itemize}

The IR luminosity reaches its maximum, which is $\sim$10$^9$ \Ls, around day 500. In order to generate this luminosity, the rate of kinetic energy generated by the ejecta-reverse shock interaction should be at least 10$^9$ \Ls, which can then gets transformed into heat. The minimum velocity ($v_{rs}^{\rm min}$) of the reverse shock is defined by the relation,

\begin{figure*}
\centering
\includegraphics[width=3.3in]{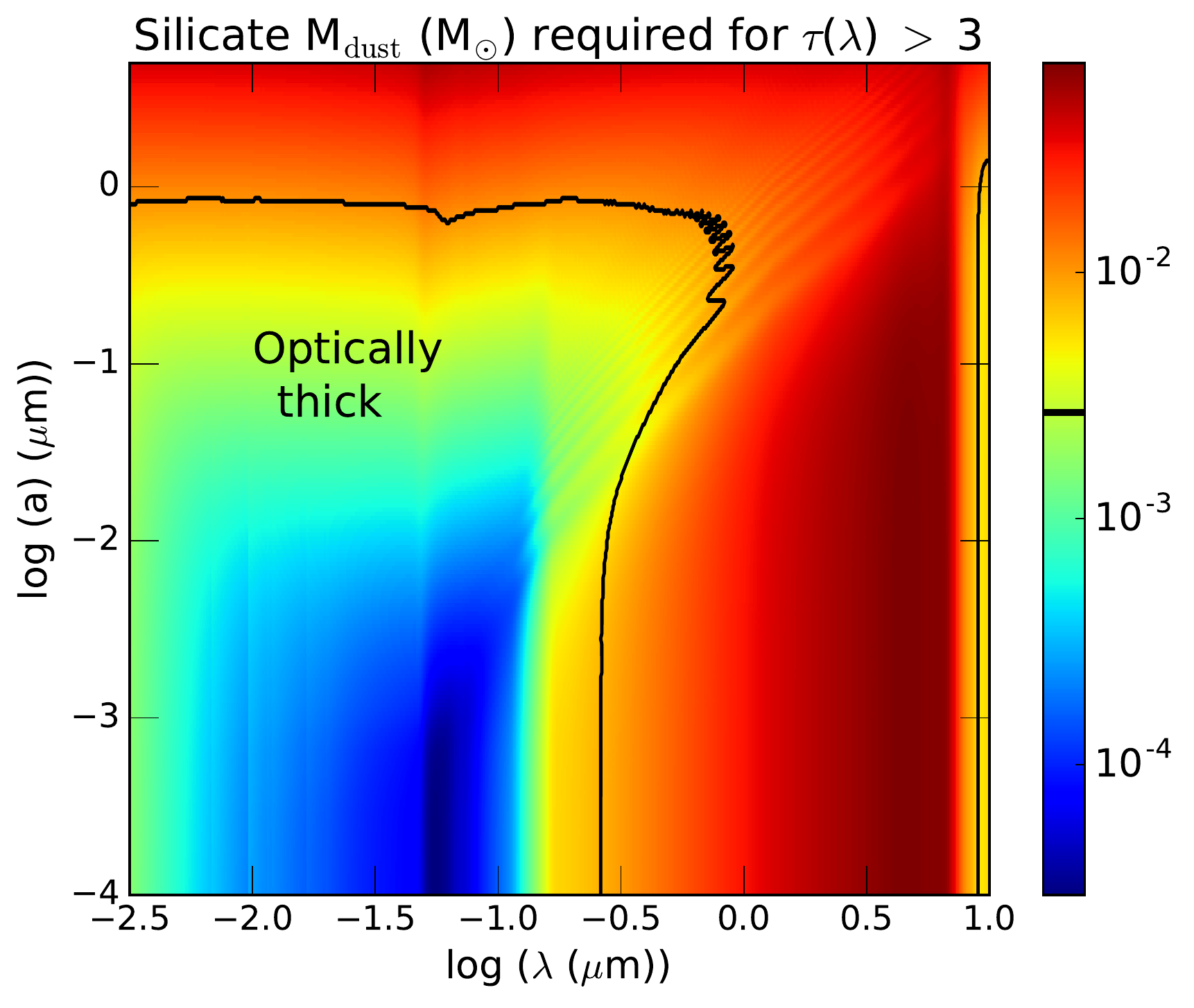}
\includegraphics[width=3.3in]{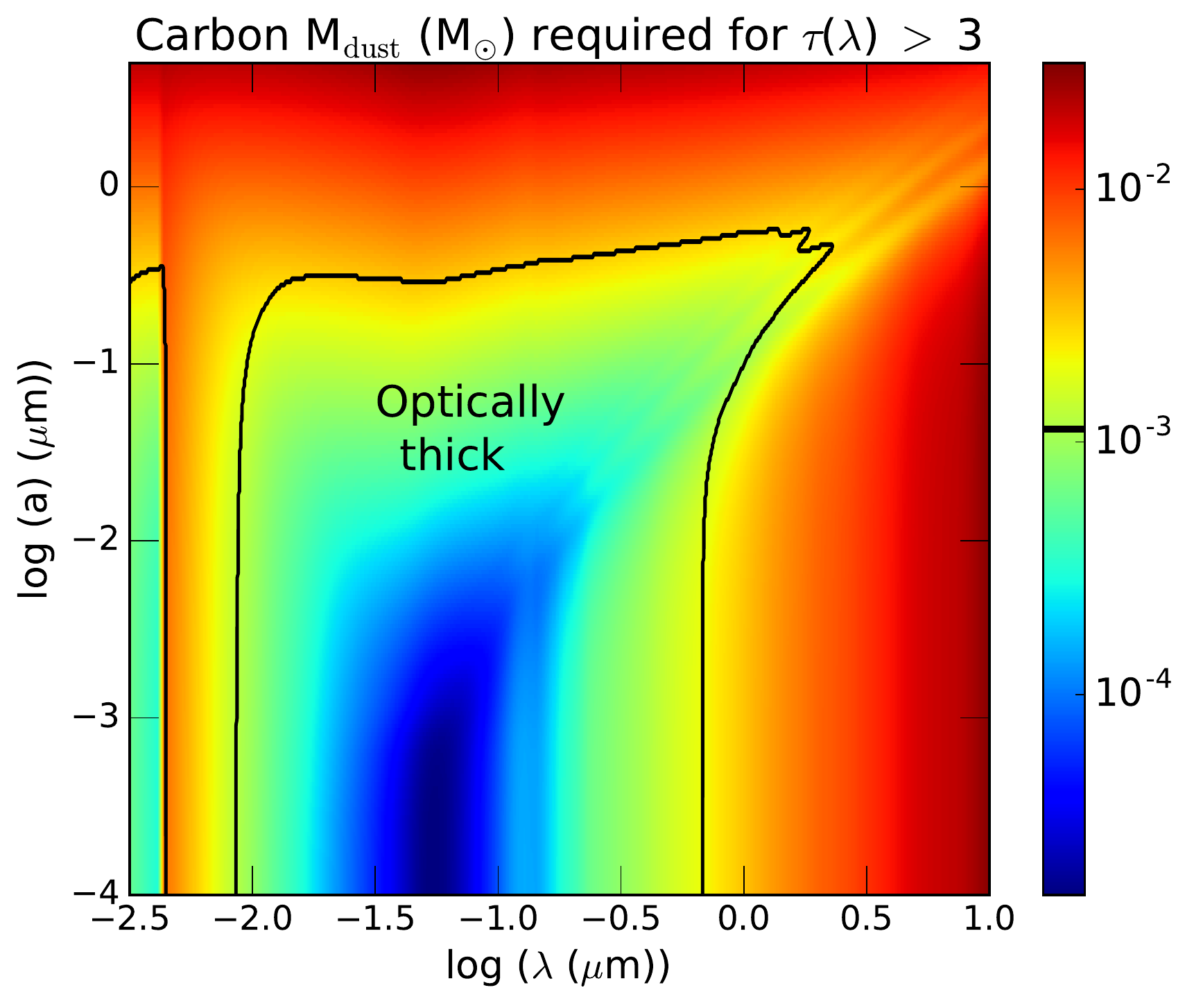}
\caption{\label{fig:dm_kappa} \footnotesize{The required minimum dust mass to achieve $\tau$($\lambda$)$>$3 is shown as a function of wavelength and grain radius for silicate (left) carbon (right) dust. The minimum optical depth $\tau$($\lambda$) that makes the dusty shell opaque to radiation at wavelength $\lambda$ is taken as 3. The colorbar for the mass (in \Ms) is on the right side of each figure. The black contour indicates the enclosed regions where the required minimum dust mass is less or equal to the observed mass of dust (6.5$\times$10$^{-3}$ \Ms\ for silicates and 1.5$\times$10$^{-3}$ \Ms\ for carbon). Therefore, the dusty shell is optically thick only if the grain radii and the wavelength of the incident radiation from the reverse shock are within the limits of the contour.  }} 
\end{figure*}

\begin{figure*}
\centering
\includegraphics[width=3.3in]{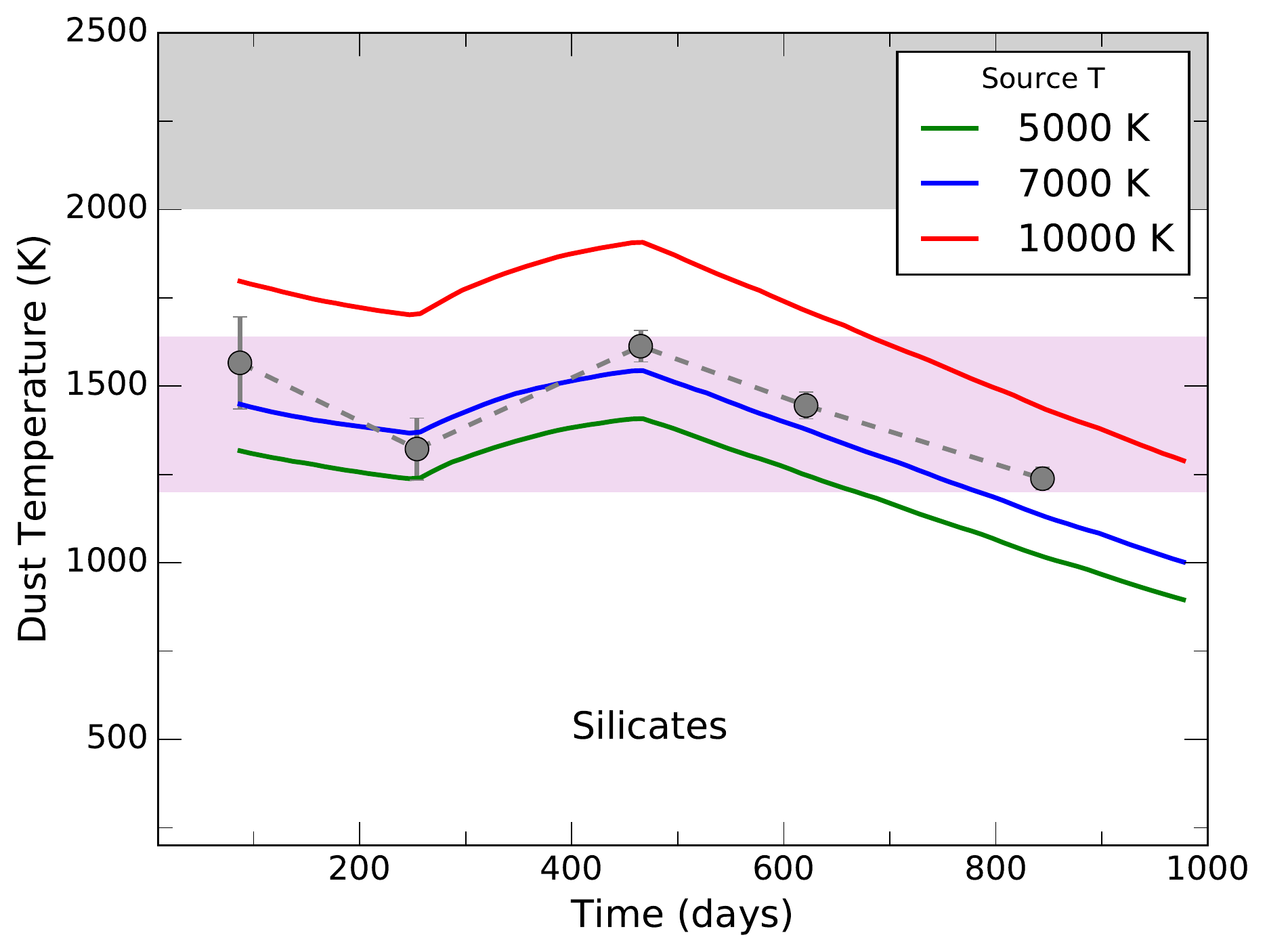}
\includegraphics[width=3.3in]{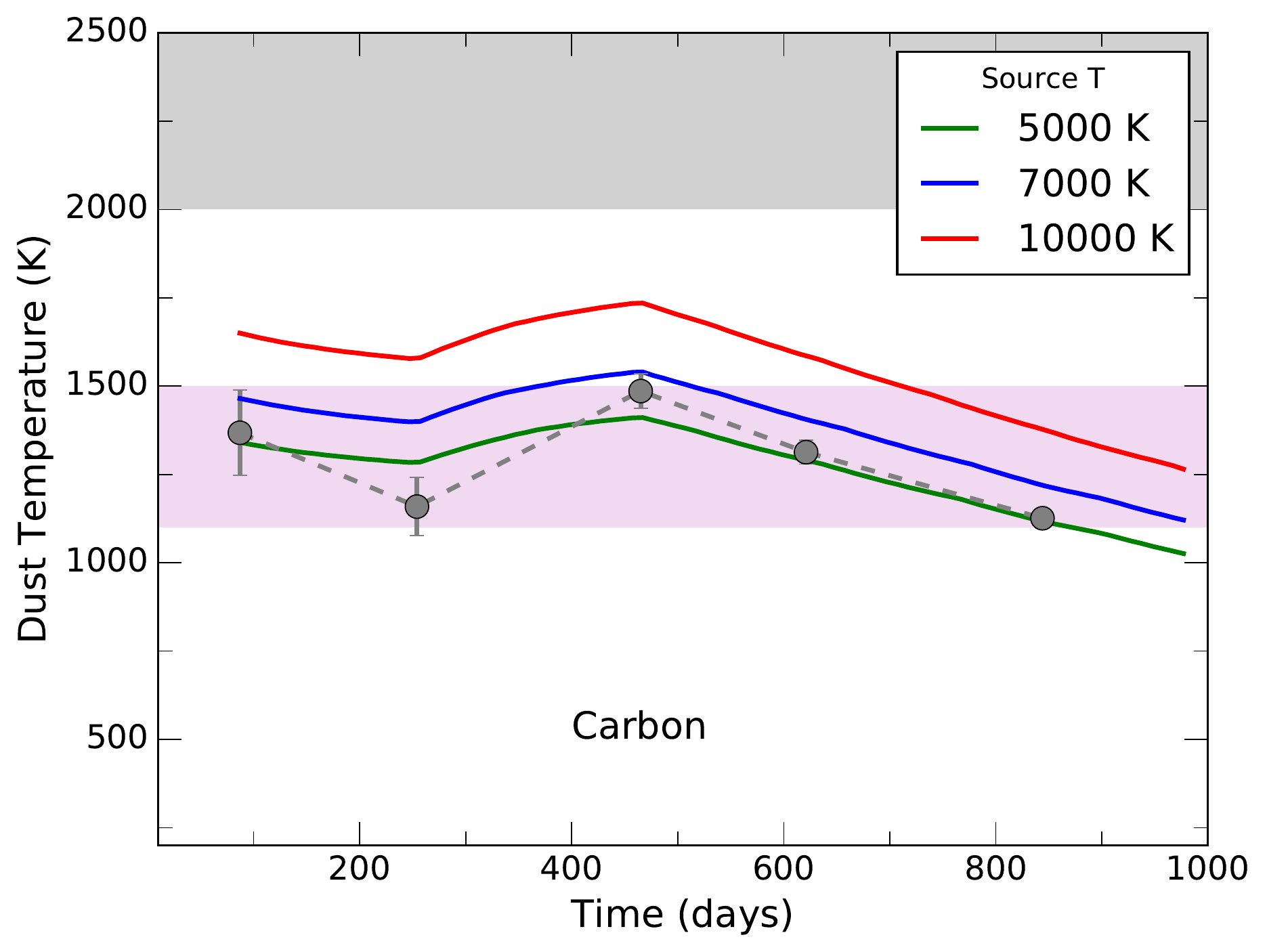}
\caption{\label{fig:revshock_heating} \footnotesize{The figure presents the resultant temperatures for a silicates (left) and a carbon (right) dust grain of 0.01 $\mu$m radii as a function of post-explosion time for various source temperatures, when they are heated by the reverse shock only, assuming $L\rm_{rev} (t)$ = $L\rm_{IR} (t)$. The band in pink represents the range of observed dust temperatures, while the band in gray represents the region where the dust temperature is above the condensation temperature of the grains. The impact of collisional heating by the ambient gas is not shown in the figure. Additionally, the dust being optically thick, the resultant temperature of the entire dusty shell cannot be determined by the equilibrium heating of a single dust grain only. See Section \ref{rev_shock_const} for further explanations. }} 
\end{figure*}


\begin{equation}
\label{vrev_min}
L_{IR} = \frac{1}{2} \times \frac{\rm{d} m}{\rm{d}t} \times (v_{rs}^{\rm{min}})^2 =  2\pi R^2 \rho (v_{rs}^{\rm{min}})^3 \\
\end{equation}

Taking the radius at 2 $\times$ 10$^{16}$ cm and the density at 500 days to be 10$^{-13}$ g cm$^{-3}$ \citep{sar13, noz10}, we get the minimum reverse shock velocity required to be $\sim$ 250 km s$^{-1}$. 

Hard X-rays above 3 keV (4 $\times$ 10$^{-4}$ $\mu$m) have very low absorption cross-section, as also evident from Figure \ref{fig:uvobrem}. Taking $\lambda$ = 4 $\times$ 10$^{-4}$ $\mu$m as the highest energy, the maximum allowed jump temperature (T$\rm _J$ = hc/$\lambda$k) of the shocked gas is 3.6 $\times$ 10$^7$~K. The maximum velocity of the reverse shock is therefore calculated from the relation,

\begin{equation}
\label{vrev_max}
T_{J} = \frac{3}{16} \frac{\mu_m}{k} (v_{rs}^{\rm{max}})^2 \\
\end{equation}

which is the jump condition and $\mu_m$ is the average mass of an atom. In O-rich SN ejecta, the mean molecular weight is around 17 \citep{sar13}. The maximum reverse shock velocity is therefore given by $\sim$ 310 km s$^{-1}$. 


For the sake of generality, we also consider a lower ejecta density of $\sim$ 1.7 $\times$ 10$^{-15}$ g cm$^{-3}$ \citep{utr17} and present an alternative set of parameters. In this case, the lower limit on the shock velocity, $v_{rs}^{\rm{min}}$, calculated using Equation~\ref{vrev_min}, increases to $\sim$ 950~km~s$^{-1}$. Encounter with such velocity shock will lead to the production of some hard X-rays in the shocked ejecta. However, the shocked ejecta would be optically thick enough to cutoff most of the hard X-rays. This explains the absence of hard X-rays in the spectra. The photo-attenuation cross-section of O-rich gas at 3 keV, 10 keV and 30 keV are 5.77$\times$10$^3$, 1.58$\times$10$^2$ and 10.1 barns respectively \citep{vei73}. Assuming around 4 \Ms\ of O-rich gas in the shocked ejecta, the column density is about 10$^{23}$ cm$^{-2}$, which is enough to ensure an optically thick medium.

As the dust is optically thick to the incident radiation from the inner heating source (reverse shock), we have limited informations to characterize the incident spectra. Therefore,  even though we have calculated the the upper and lower limits on the shock velocity, luminosity and the dust masses, the temperature of the dusty shell cannot be determined accurately. 

Figure \ref{fig:revshock_heating} shows the temperatures of a hypothetical 0.01 $\mu$m grain of silicate and carbon when heated by the reverse shock, assuming the reverse shock luminosity to be equal to the IR luminosity at a given time, which is basically the lower limit of the reverse shock luminosity. Three different source temperatures were assumed and the spectra was assumed as a blackbody. The figure shows that the UVO radiation with a temperature ranging between 7000 K and 5000 K fits well with the observed band of temperatures. Radiation in this range comprises of non-ionizing photons, hence it will not be able to ionize the gas efficiently. However, as shown in Figure \ref{fig:dm_kappa}, the dusty shell made up of 0.01 $\mu$m grains is not thick to radiation around 0.4-0.6 $\mu$m, which corresponds to a blackbody temperature in this range. So a more likely scenario is provided by a source at higher temperature, which heats the gas and collisionally heats the dust. Additionally, the dust being optically thick, it can shield the remainder of the gas from the strong flux of radiation generated by the reverse shock. Lastly, the dust is an efficient coolant, hence once the dust is formed in the dense shell it further helps to cool down the ambient gas.



\begin{table*}
\centering
\caption{Dust shell properties for selected days}
\label{finalmass} 
\begin{tabular}{c c c c c c c c }
\hline \hline
Epoch & Post-shock & \multicolumn{3}{c}{M$\rm_{sil}$ (\Ms)} &  \multicolumn{3}{c}{M$\rm_{car}$ (\Ms)}  \\
(day) & shell (\Ms) &  Max.\tablenotemark{1}  & Observation &  $\tau$(U) & Max.\tablenotemark{1} & Observation  & $\tau$(U)  \\
 \hline
 465 & 8.2 & 3.5 $\times$ 10$^{-2}$ & 4.7 $\times$ 10$^{-3}$ & 1.7 &  1.1 $\times$ 10$^{-3}$ & 8.8 $\times$ 10$^{-4}$ & 3.7  \\
 621 & 9.9 & 4.2 $\times$ 10$^{-2}$ & 6.5 $\times$ 10$^{-3}$ & 2.4 &  1.3 $\times$ 10$^{-3}$ & 1.4 $\times$ 10$^{-3}$ & 5.8 \\
 844 & 11.7 & 5.0 $\times$ 10$^{-2}$ & 9.5 $\times$ 10$^{-3}$ & 3.5 &  1.6 $\times$ 10$^{-3}$ & 2.2 $\times$ 10$^{-3}$  & 3.1  \\
 \hline
\end{tabular}
\tablenotetext{1}{The maximum dust mass is purely a stoichiometric quantity, calculated using the solar mass fractions and the CNO abundances given by \cite{fra14}. It ignores the steps of formation of intermediate molecules during the process.} 
\end{table*} 

\subsection{Estimated dust masses}
\label{est_dustmass}

The total amount of dust at a given time depends on several intercorrelated physical and chemical processes, which we do not address in details in course of this current study. The constraints on the dust masses can be derived from the following: (a) the fit to the IR emission (b) the maximum mass of metals that can condense (c) the minimum mass that can make the shell optically thick.

Table \ref{finalmass} presents the upper limit on the dust masses that can be synthesized in the post-shock gas at a given epoch, purely from stoichiometric estimates. The upper limit of the silicate mass is larger than the observed masses, hence silicates can be a major dust component in the post-shock shell. In case of carbon dust however, the maximum atomic carbon present in the gas is smaller than the mass that is required to fit the observations. Hence, carbon dust cannot be the only dust type, which is present in the post-shock gas. 



Yet, absence of silicate features at late times \citep{wil15} makes carbon a more likely dust species. This dichotomy can be resolved, if the dust comprises of a combination of both, the silicate and carbon dust.


In a dusty shell like this one, where the inner radius is $R_0$ and mass of dust is M$_d$, the optical depth is given by, 

\begin{equation}
\tau (\lambda) = \frac{M_d \ \kappa(a, \lambda)}{4\pi R_0^2} \\
\end{equation}

In Figure \ref{fig:dm_kappa} we calculate the mass of dust (silicates and carbon) required in order to make the post-shock shell thick to incident radiation. In terms of optical depth ($\tau$), $\tau_{\lambda}$ $\ge$ 3 is assumed as the necessary condition for optical thickness. The dust masses are shown in the grid of the grain radii and wavelength of incident radiation. 

The contours of the dust masses (of silicates and carbon) that has been derived from observations are also presented in Figure \ref{fig:dm_kappa}. It is evident from the figure that the grains cannot be larger than 0.1~$\mu$m and also the incident spectra of the reverse shock should have a peak at wavelength less than 0.3~$\mu$m. Therefore, the reverse shock spectra cannot be dominated by the UVO only. It is also confirmed by the U-band optical depths listed in Table \ref{finalmass}. 

Importantly, this also confirms that the pre-existing dust alone cannot account for the late time IR emission. This is because, the total mass of pre-existing dust, which is present at later radii than the evaporation radius, cannot be large enough to make the region optically thick. That will lead to a violation of the maximum possible dust to gas mass ratio.

On the other-hand, the soft X-rays at wavelength $\lambda$, given by $\lambda<0.1$ $\mu$m,  are very efficient in heating and ionizing the gas. Strong ionizing radiation from the reverse shock is capable of collisionally destroying the dust grains that have formed in the post-shock CSM. Hence the allowed luminosity of the reverse shock should have the maxima of the spectral distribution between 0.1-0.3~$\mu$m. 

\section{Summary and discussions}
\label{summary}


In this paper, we have studied the origin and evolution of the IR emission in SN~2010jl. 

Our model has combined the results from the study of UVO and IR spectra by \cite{fra14, gal14, mae13} and \cite{jen16} and the analysis of X-ray data by  \cite{cha15} and \cite{ofe14a}. The morphology of the ejecta, adopted in this study, is consistent with Hydrogen column densities estimated from X-rays analysis. The temperature of the spectra representing the optical and the IR light curve, derived from the best-fit scenarios, are also akin to the estimates by \cite{fra14}. 


The study confirms that formation of new dust in the CSM or in the ejecta is not feasible as early as a few weeks after detection, as previously reported by \cite{gal14}. Dust synthesis commences in the post-shock CSM only after day 380 from the time of detection. Also dust is unlikely to be present in the ejecta at the early times. So the IR emission must be attributed to the circumstellar dust which has survived from the pre-explosion era.

Eventually at later times, the pre-existing dust gradually gets destroyed by SN shock encounter and then newly formed dust in the dense post-shock shell acts as the prime source of IR emission. Therefore, the IR observations by $Spitzer$ of SN~2010jl at days 465, 621 and 844 can be attributed to the new dust formed in the post-shock shell. To explain the IR excess at day 87 and day 254, we shall address the case of IR echo from the pre-existing dust in the associated paper \citep{dwe18}.


Dust formation in the circumstellar shell of type IIn supernovae are controlled by radiative transfer processes. Downstream radiation from the shocked gas is responsible for directly heating the dust as well as passively heating the dust through ionizing the surrounding gas. 

The presence of strong ionizing photons has a two-fold impact on the chemical pathways: (a) it hinders the earliest epoch of dust formation (b) it also leads to the formation of the warm dense shell, so the gas phase species get enough time to react via kinetic processes. Therefore, the high post-shock density does not automatically ensure early dust formation, as previously estimated by \cite{gal14}. In absence of the downstream radiation, the gas is likely to cool from 10$^4$~K to $\sim$100~K in a span of less than a day, leaving very little time to the chemical reactions to proceed kinetically. 

In this study, a special emphasis has been given to the progenitor-CSM morphology in the pre-explosion era. As the findings suggest, in order to explain the IR spectra of SN~2010jl, a low density region, extending up to a few times 10$^{16}$ cm and lying between the pre-explosion star and the surrounding CSM, is absolutely necessary. In SNe where a dense CSM almost superposes the progenitor star, early formation of new dust becomes an unlikely scenario, both in the CSM and as well as in the ejecta. Therefore, in addition to the intensity of the radiation field, the shape of the outer circumstellar matter also determines the efficiency of dust formation. 

In brief, the processes or phenomena that control the dynamics of dust formation in such environments can be summarized by the following points. 
The dust formation is facilitated by: 
\begin{itemize}[noitemsep]
\item[--] The high density of the post-shock gas 
\item[--] The short cooling time  
\item[--] Formation of the warm gas reservoir, which does not cool efficiently  
\item[--] A large inner radius of the CSM   
\end{itemize}
On the other hand, the formation and growth of dust in such environment is impeded by: 
\begin{itemize}[noitemsep]
\item[--] The X-rays and UVO from the forward shock 
\item[--] The radiation from the reverse shock 
\item[--] Collision with the ambient hot and ionized gas
\end{itemize}



Even though type IIn SNe have a relatively larger progenitor in general, the final dust yield is influenced by strong circumstellar interactions. The CSM constitutes of materials ejected in the form of the metal-depleted winds in the pre-explosion era. Therefore, assuming an average mass of the CSM to be $\sim$ 10 \Ms, an upper limit of about $\sim$ 0.1 \Ms\ of dust can be considered to form in the CSM. Hence, in the CSM, dust formation might transpire earlier than the ejecta, yet the final mass may not exceed 0.1 \Ms. In case of SN~2010jl the dust mass ranges between 10$^{-3}$ to 10$^{-2}$ \Ms.


This study has developed the first model that addresses the scenario of dust formation in circumstellar shells which are associated with strong ionizing radiation from the supernova forward and reverse shock. Even though the study is based on SN~2010jl, the effect of radiation on the post-shock gas is globally applicable to any Type IIn SNe to study dust formation in such environments. 

There are some analytical steps in the model where some first order approximations or simplifications were required to be imposed. Important to note, the effective cooling mechanism of a gas parcel which is at low temperature (T $<$~4000~K) and high densities, is not very well understood. The cooling rate is expected to be controlled by the combined impact of metal cooling, H$^{-}$ cooling and molecular cooling. However, there are not many available tools that can quantify the processes. Therefore, in our analysis using CLOUDY there are some inherent approximations on the cooling rates. Secondly, the shock equations were solved using a plane parallel geometry approximation. In the limit where the region of shocked ionized gas to be within $\Delta R$ and $\Delta R/R \ll$ 1, the approximation is justified to the first order. 

This is a 1-D spherically symmetric model, which has scope of further development. \cite{and11a} has adopted a bipolar geometry for SN~2010jl in order to simultaneously explain the continuum emission and the optical depths. However the shock equations deal with microscopic properties of the shocked gas, so they do not alter to a great extent, depending on the structure of the shell. Therefore, the inferences drawn from this study related to the radiation induced dust formation remains justified.

\acknowledgements
We acknowledge NASA's 16-ATP2016-0004 grant for supporting this project.  Also we want to acknowledge the guidance and feedbacks from Dr. John Raymond, Dr. Jon Slavin and Dr. Tim Kallman who have helped us thoroughly during the course of this study. We thank the anonymous referee for the valuable suggestions that have helped us in finalizing the manuscript of the paper.

\end{document}